\documentclass[journal=jacsat,manuscript=review,layout=twocolumn]{achemso}

\usepackage[version=3]{mhchem} 
\usepackage{amsmath}
\usepackage{amssymb}
\usepackage{lipsum}
\usepackage{mathtools}
\usepackage{cuted}
\usepackage{color}
\usepackage{graphicx}

\SectionNumbersOn

\newcommand{\br}{\boldsymbol{r}} 
\newcommand{\bR}{\boldsymbol{R}}

\newcommand{\bnabla}{\boldsymbol{\nabla}}
\newcommand{\bFtilde}{\boldsymbol{f}}
\newcommand{\bk}{\boldsymbol{k}}
\newcommand{\bA}{\boldsymbol{A}}
\newcommand{\bAh}{\boldsymbol{\hat{A}}}

\newcommand{\be}{\boldsymbol{\epsilon}}
\newcommand{\bJ}{\boldsymbol{J}}

\newcommand{\imagi}{\mathrm{i}}

\def\ket#1{| #1 \rangle}
\def\braket#1#2{\langle #1 | #2 \rangle}
\def\brakett#1#2#3{\langle #1 | \,#2\, | #3\rangle}

\renewcommand{\d}{\,\mathrm{d}} 

\author{Michael Ruggenthaler}

\affiliation{Max-Planck-Institut f\"ur Strukture und Dynamik der Materie, Luruper Chaussee 149, 22761 Hamburg, Germany}

\alsoaffiliation{The Hamburg Center for Ultrafast Imaging, Luruper Chaussee 149, 22761 Hamburg, Germany}

\email{Michael.Ruggenthaler@mpsd.mpg.de}

\author{Dominik Sidler}

\affiliation{Max-Planck-Institut f\"ur Strukture und Dynamik der Materie, Luruper Chaussee 149, 22761 Hamburg, Germany}

\alsoaffiliation{The Hamburg Center for Ultrafast Imaging, Luruper Chaussee 149, 22761 Hamburg, Germany}

\email{Dominik.Sidler@mpsd.mpg.de}

\author{Angel Rubio}

\affiliation{Max-Planck-Institut f\"ur Strukture und Dynamik der Materie, Luruper Chaussee 149, 22761 Hamburg, Germany}

\alsoaffiliation{The Hamburg Center for Ultrafast Imaging, Luruper Chaussee 149, 22761 Hamburg, Germany}

\altaffiliation{Center for Computational Quantum Physics, Flatiron Institute, 162 5th Avenue, New York, New York 10010, USA}

\email{Angel.Rubio@mpsd.mpg.de}

\title[Polaritonic chemistry from first principles]
{Understanding polaritonic chemistry from ab initio quantum electrodynamics}

\abbreviations{QED, QEDFT, QED-CC, PTA, TBAF}

\begin{document}
	
	\begin{tocentry}
		\begin{center}
			\includegraphics[width=5cm,height=5cm,keepaspectratio]{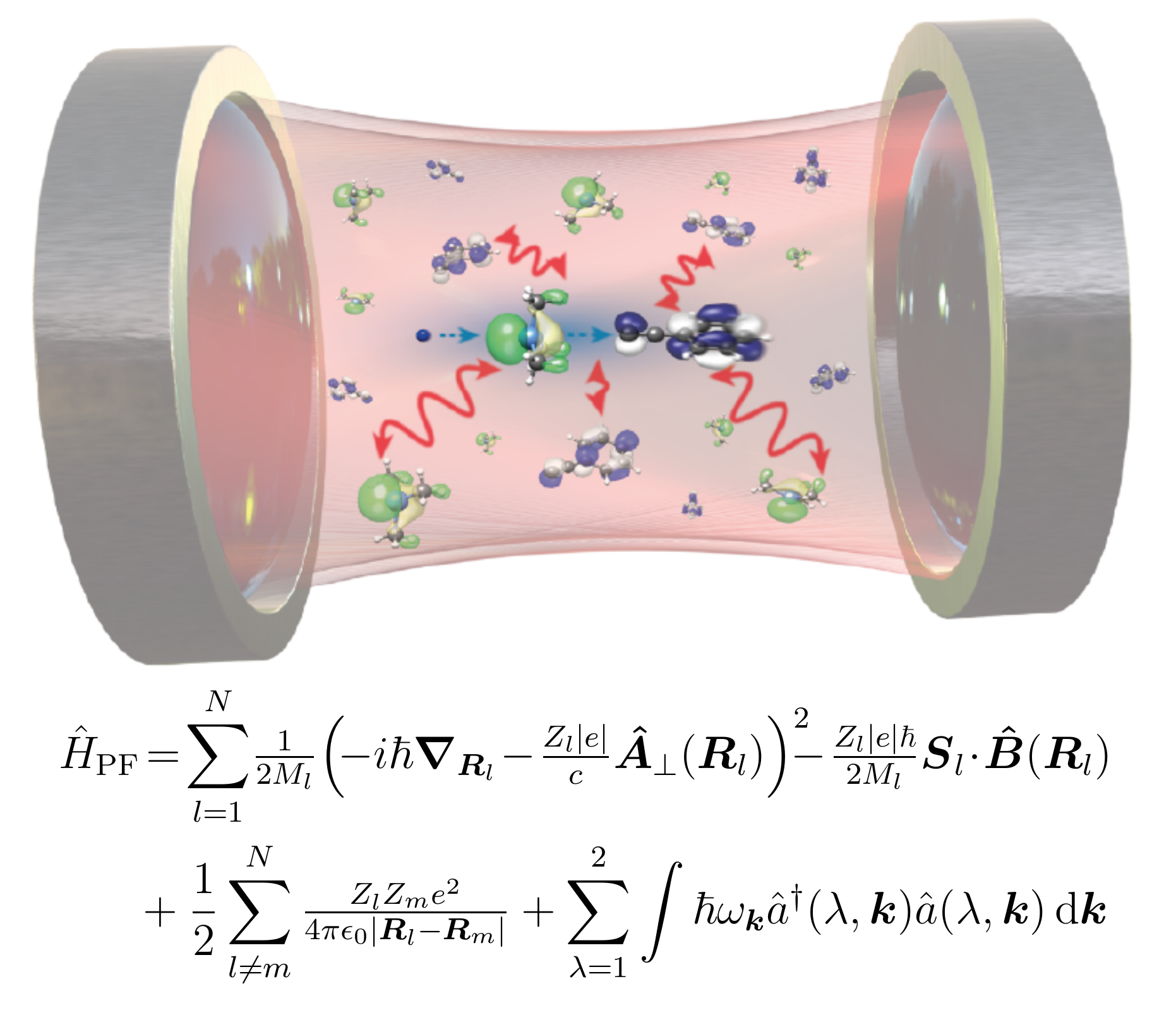}
		\end{center}
	\end{tocentry}
	
	\begin{abstract}
		In this review we present the theoretical foundations and first principles frameworks to describe quantum matter within quantum electrodynamics (QED) in the low-energy regime, with a specific focus on polaritonic chemistry. Having a rigorous and fully quantized description of interacting photons, electrons and nuclei/ions, from weak to strong light-matter coupling regimes, is pivotal for a detailed theoretical understanding of the emerging fields of polaritonic chemistry and cavity materials engineering. At the same time, the use of rigorous first principles avoids ambiguities and problems stemming from using approximate models based on phenomenological descriptions of light, matter and their interactions, and provides a way to systematically derive consistent low-energy models that are fully gauge invariant and mimic the first principles results. By starting from fundamental physical and mathematical principles, we first review in great detail non-relativistic QED, which allows to study polaritonic systems non-perturbatively by solving a Schr\"odinger-type equation. The resulting Pauli-Fierz quantum field theory serves as a cornerstone for the development of (in principle exact but in practice) approximate computational methods, such as quantum-electrodynamical density functional theory, QED coupled cluster or cavity Born-Oppenheimer molecular dynamics. These methods do not depend on phenomenological models of chemical systems, but instead they treat light and matter on equal footing. At the same time, first principles QED methods have the same level of accuracy and reliability as established methods of computational chemistry and electronic structure theory. After an overview of the key-ideas behind those novel ab initio QED methods, we explain their benefits for a better understanding of photon-induced changes of chemical properties and reactions. Based on results obtained by ab initio QED methods we identify the open theoretical questions and how a so far missing mechanistic understanding of polaritonic chemistry can be established. We finally give an outlook on future directions within polaritonic chemistry and first principles QED and address the open questions that need to be solved in the next years both from a theoretical as well as experimental viewpoint.
	\end{abstract}
	
	\tableofcontents
	
	\section{Introduction}
	\label{sec:introduction}
	\textit{"Until the beginning of
		the 20th century, light and matter have been treated as different
		entities, with their own specific properties [...]. The development
		of quantum mechanics has enabled the theoretical description of
		the interaction between light-quanta and matter."}
	
	\noindent
	\\
	M. Hertzog in Ref.~\cite{Hertzog_review}
	\\
	\\

	Chemistry investigates, very broadly spoken, how matter arranges itself under different conditions (temperature, pressure, chemical environment, ...) and how these arrangements lead to various functionalities and phenomena. The basic building blocks of chemical systems, as we understand them today, are the various atoms of the periodic table of elements. Combining these basic building blocks then leads to the formation of molecules and solids, and the arrangement of the atoms determines much of the emerging properties of these complex matter systems. Light, or more generally, the electromagnetic field, usually appears in this context in two distinct capacities: Firstly, as an external (classical) agent that drives the matter system out of equilibrium. External driving is then used to either spectroscopically investigate matter properties, such as when recording an absorption or emission spectrum~\cite{svanberg2012,kuzmany2009,cowan1997,ruggenthaler2018quantum}, or to force the matter system into a different (transient) state~\cite{chu2004beyond,kohler2005driven,mulser2010high,stefanucci2013,basov2017towards}. Secondly, as a (quantized) part of the system~\cite{Greiner_1996,ryder_1996,greiner2003quantum}, such as in the case of the longitudinal electric field between two charged particles, which gives rise to the Coulomb interaction and determines how the atoms are arranged.
	
	Light as an external, classical probe and control field is widely used in chemistry nowadays. However, the potential to employ the quantized light field as part of the system to modify and probe chemical properties has only began to be explored in the last years~\cite{ebbesen2016}. In order to achieve control over the internal light field one can use photonic structures, such as optical cavities~\cite{skolnick1998strong,raimond2001manipulating,vahala2003optical,hugall2018plasmonic}, and in this way control the local electromagnetic field of a molecular system~\cite{haroche1989cavity}. The resulting re-structuring of the electromagnetic modes has very fundamental consequences, since it changes the building blocks of light: the electromagnetic vacuum modes and with this the notion of photons in quantum electrodynamics~\cite{keller2012quantum,buhmann2013dispersion}. Keeping in mind that the interaction between charged particles is mediated via the exchange of photons~\cite{Greiner_1996,ryder_1996,greiner2003quantum}, it becomes clear that such modifications can in principle influence the properties of atomic, molecular and solid-state systems. Even more so, if we realize that the basic building blocks of matter (electrons, nuclei/ions, atoms, ...) are themselves hybrid light-matter systems~\cite{spohn2004dynamics,craig1998molecular} that depend on the \textit{photonic environment} (see also discussion after Eq.~\eqref{eq:hydrogenatom}).

	Although optical cavities have been used in atomic physics and quantum optics routinely since several decades to interrogate and change the behavior of (an ensemble of) atoms~\cite{grynberg2010, scully1997}, it came as a surprise to many that cavities could also influence complex chemical and solid-state processes~\cite{torma2014strong,ebbesen2016,basov2016polaritons,kimble2018matter,bloch2022strongly,basov2021polariton}. The main reason being that in quantum optics, or more precisely in cavity~\cite{haroche2006exploring,walther2006cavity,haroche2013nobel} and circuit~\cite{clerk2020hybrid,carusotto2020photonic} quantum electrodynamics (QED), which focus on the properties of the photons and a limited set of matter degrees of freedom, often ultra-low temperatures and ultra-high vacuua are needed in order to observe the influence of the changed electromagnetic vacuum modes. Such very specific external conditions are not often considered in chemistry and materials science, and hence it was assumed that there would be no observable effect on chemical and material properties upon changing the photonic environment at ambient conditions. Yet there is by now a multitude of seminal experimental results that show that indeed the restructuring of the electromagnetic environment by optical cavities can influence chemical and material properties at ambient conditions, even if there is no external illumination and the effects are driven mainly by vacuum and thermal fluctuations (for an overview see various reviews, e.g., Refs.~\cite{ebbesen2016,kena2016polaritonic,Ribeiro2018polariton,feist2018polaritonic,flick2018strong,dovzhenko2018light,schneider20182Dpolaritons, ruggenthaler2018quantum,gargiulo2019plasmonic,frisk2019ultrastrong,Hertzog_review,hirai_review,herrera2020molecular,nagarajan2021chemistry,garcia2021manipulating,nitzan2022polaritons}). We here only highlight, as exemplifications, changes in energy and charge transport~\cite{coles2014polariton,orgiu2015conductivity,zeng2016transport,tal2018exciton,faist2019landau}, the appearance of exciton-polariton condensates at room temperature~\cite{su2020observation,dusel2020room} and the modification of the phases of solids~\cite{wang2014phase,thomas2021ferromagentism}. In the following we will focus on changes in chemical properties of (finite) molecular systems upon modifying the photonic environment and do not go into detail on changes observed and induced in extended solid-state systems. 
	
\begin{figure}
		\begin{center}
			\includegraphics[width=\linewidth]{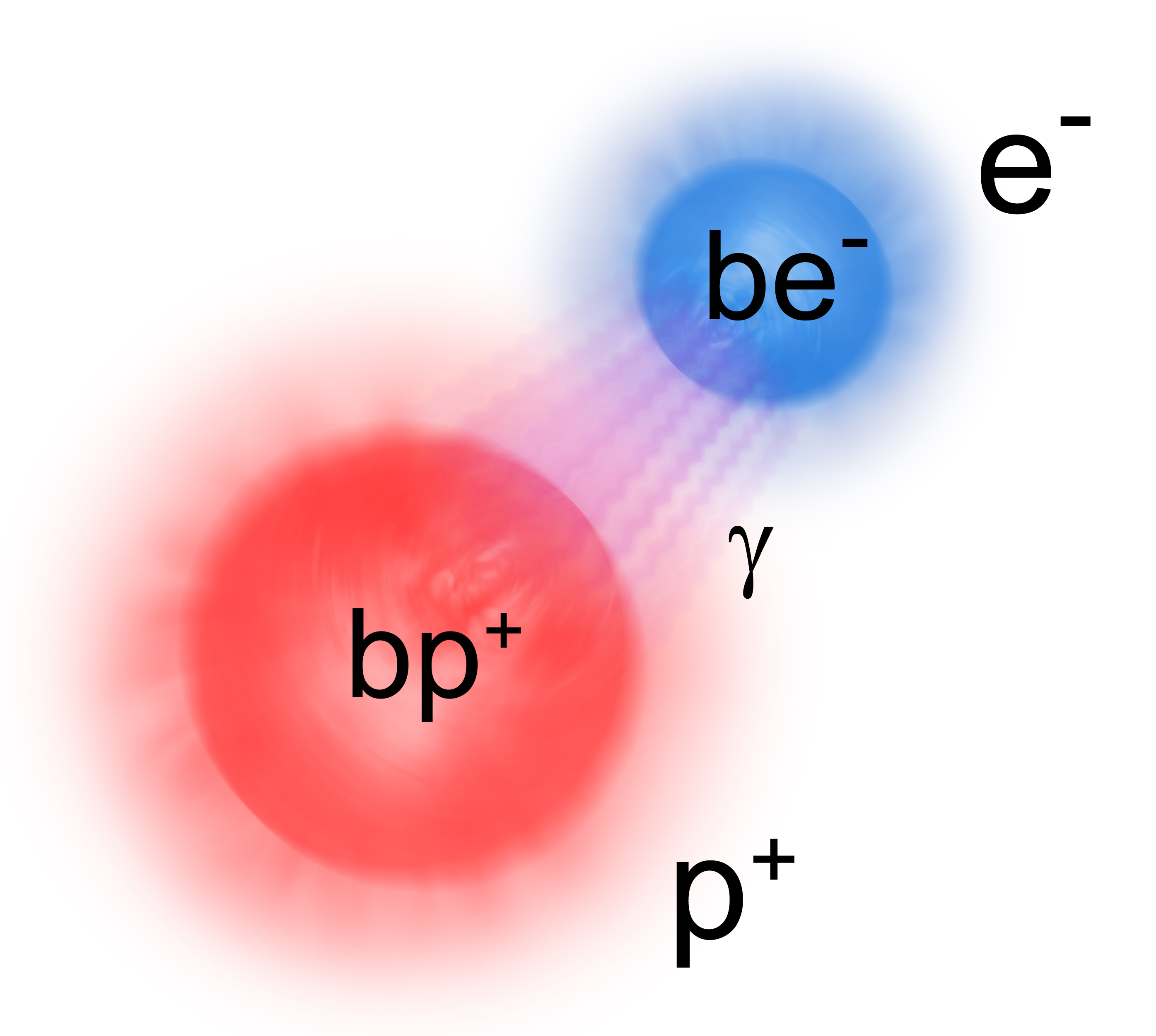}
			\caption{Sketch of the QED perspective on coupled light-matter systems, e.g., a hydrogen atom. In QED the bare (pure matter) proton (bp$^+$) and bare electron (be$^{-}$) are a reminiscence of the (mathematically necessary) smallest length scale (energetically an ultra-violet cutoff) that can be resolved. The observed (dressed or physical) proton ($p^{+}$) and electron (e$^{-}$) include the contributions from the (virtual) photon field $\gamma$, which describes the electromagnetic self-interaction of charged particles. The photons, at the same time, describe the electromagnetic interaction between the electron and proton and lead to the appearance of a bound hydrogen atom. From a QED perspective the distinction between light and matter depends on the energy scale that we look at, the chosen reference frame and the chosen gauge (see discussion in Sec.~\ref{sec:lightandmatter}). Considering one aspect without the other leads to inconsistencies, and for a consistent description always both (quantum light and quantum matter aspects) have to be treated at the same time.}
			\label{fig:lightandmatter}
		\end{center}
	\end{figure}

	This new flavor of chemistry, which uses the \textbf{modification of the photonic environment as an extra control knob}, has been named QED or polaritonic chemistry~\cite{flick2017atoms,feist2018polaritonic}. The latter notion is derived from the quasi particle \emph{polariton}, which is a mixed light-matter excitation~\cite{basov2016polaritons} (see also Fig.~\ref{fig:polaritonsplitting}), and whose appearance in absorption or emission spectra is often assumed to be a prerequisite for observing changes in chemical properties. Polaritonic chemistry is a highly interdisciplinary field, with often conflicting perspectives on the same physical concepts. From a (quantum) optics perspective, for instance, the role of light and matter is reversed as compared to chemistry: One uses matter to either interrogate or change the properties of the electromagnetic field. This clash of perspectives (and their resolution via QED, see also Fig.~\ref{fig:lightandmatter}) makes it a scientifically very rewarding field of research, since it constantly challenges one's basic conceptions. A plethora of theoretical methods from (quantum) optics and (quantum) chemistry are employed and combined to capture and understand the emerging novel functionalities when changes in the electromagnetic environment lead to strong coupling between light and matter~\cite{ruggenthaler2018quantum, frisk2019ultrastrong}. 
	
	While (quantum) optics methods are geared to capture details of the electromagnetic field and photonic states~\cite{grynberg2010, scully1997}, the (quantum) chemical methods are naturally focused on a detailed description of the matter system~\cite{szabo2012modern,haile1992molecular,cramer2013essentials}. Currently employed phenomenological combinations of such methods are able to capture certain effects, but fail in important situations, such as to describe (even only qualitatively) the observed changes in ground-state chemical reactions under vibrational strong coupling~\cite{climent2020sn, climent2021reply,martinez2018can}. On a first glance, owing to the complexity of the systems under study (a large number of complex molecules in solvation at ambient conditions strongly coupled to an optical cavity with many photonic modes), this might not come as a surprise, since already the accurate theoretical description of a single complex molecule in vacuum and at zero temperature is highly challenging~\cite{szabo2012modern}. Even simple working principles of polaritonic chemistry, that single out the most important ingredients to control chemistry via changed electromagnetic environments, remain elusive so far. On a second, more careful glance, however, there might be a more fundamental reason for why currently employed phenomenologically combined approaches are not able to describe some of the experimentally observed effects: Our most fundamental description of how light and matter interact, QED~\cite{Greiner_1996,ryder_1996,cohen1997photons,greiner2003quantum}, does not allow for a strict distinction between light and matter (see also Fig.~\ref{fig:lightandmatter}).   
	
	Indeed, if we reconsider the basic building blocks of matter from a QED perspective, we realize that already electrons and atoms are hybrid light-matter objects themselves and their properties depend on various assumptions. Take, for instance, the hydrogen atom as described by the non-relativistic Schr\"odinger equation in Born-Oppenheimer approximation in SI units (used throughout this review)
	\begin{align}\label{eq:hydrogenatom}
		\left(-\frac{\hbar^2}{2 m_{\rm e}} \bnabla^2 - \frac{e^2}{4 \pi \epsilon_0 | \br |} \right) \Psi(\br) = E \Psi(\br),
	\end{align}  
	where $\hbar$ is the reduced Planck's constant, $m_{\rm e}$ is the physical mass of the electron, $e$ is the elementary charge and $\epsilon_0$ is the permittivity of the free electromagnetic vacuum. However, from a QED perspective the electron of a hydrogen atom has a mass that depends on the structure of the electromagnetic vacuum surrounding the atom and also the Coulomb attraction depends on the form of the surrounding electromagnetic vacuum. Indeed, the physical mass of the electron has two contributions
	\begin{align}
		m_{\rm e} = m + m_{\rm ph},
	\end{align} 
	where the bare mass $m$ depends on how the electromagnetic vacuum modes decay when going to higher and higher frequencies (ultra-violet regularization) and the photon-induced mass $m_{\rm ph}$ comes from the energy due to the interaction of a moving electron with the photon field (see discussion in Sec.~\ref{subsec:paulifierzproperties} and \ref{subsec:approximatepaulifierz} for more details). In addition, the form as well as the strength of the Coulomb interaction is determined solely by the structure of the vacuum modes (see discussion in Sec.~\ref{subsec:approximatepaulifierz} for more details). In other words, what we call a hydrogen atom, is defined with respect to a specific \textit{photonic environment}, i.e., in this case the free electromagnetic vacuum. Similarly, the photonic environment dictates how a laser or thermal radiation interacts with matter. Hence it becomes clear that when we restructure the electromagnetic environment with the help of an optical cavity or other setups~\cite{vahala2003optical,torma2014strong,frisk2019ultrastrong}, we might need to rethink what are the basic building blocks of matter, which statistics they obey, how they interact among each other and how they couple to external perturbations.
	
	Admittedly, having in mind the many other aspects that might have an influence in QED chemistry~\cite{sidler2022perspective} (see also Sec.~\ref{sec:polaritonicchemistry}), such fundamental considerations might seem on a first glance like a theoretical nuisance. However, it is important to realize which assumptions are made and which internal inconsistencies arise when phenomenologically combining methods from (quantum) optics and (quantum) chemistry or electronic structure theory. Especially, since we do not yet have simple and reliable rules for how polaritonic chemistry operates, what are the basic factors that determine the observed changes and how to control them. Furthermore, in recent years theoretical methods have been devised that avoid the common \emph{apriori} division into light and matter, allowing approximate solutions to QED in the low-energy regime directly~\cite{craig1998molecular, spohn2004dynamics}. These first principles QED methods~\cite{ruggenthaler2018quantum} have already provided important insights into certain aspects of polaritonic chemistry and strong light-matter coupling for molecular and solid-state systems.   
	
	We will therefore focus in this review on the basic theory and ab initio description of coupled light-matter systems under the umbrella of QED in the low-energy regime. Such fundamental considerations allow us to address several important (and often very subtle) topics that arise in the context of describing polaritonic chemistry and materials science and that are decisive to find the main physical mechanisms observed in experiment. The first main question to answer is how to devise a (physically and mathematically) \textbf{consistent theory of interacting light and matter} that treats all basic degrees of freedom of the low-energy regime, i.e., photons, electrons and nuclei/ions, on the same quantized and non-perturbative footing. We will give the basic principles and a concise derivation of such a theory in Sec.~\ref{sec:lightandmatter} and discuss the resulting Hamiltonian formulation for fundamentally \textit{polaritonic quantum matter} in Sec.~\ref{sec:PauliFierz}. The next important topic that arises is how the \textbf{gauge choice} influences what we call light and what we call matter. This topic has a direct impact on consistently combining approximate models from quantum optics and quantum chemistry. As we discuss in more detail at the end of Sec.~\ref{subsec:paulifierzproperties}, this topic has provoked many debates and gauge-inconsistencies can even predict wrong and unphysical effects. The next main question is how to find approximations that allow a reduction of complexity and a straightforward combination of different theoretical methodologies without introducing too many uncontrolled assumptions. We will discuss this in Sec.~\ref{subsec:approximatepaulifierz} and specifically highlight the \textbf{long wavelength approximation and its implicit assumptions}. Sometimes the implicit assumptions of this common approximation lead to misunderstandings and can therefore be a barrier for new people in the field of QED chemistry and materials sciences. A further important issue is how changing the photonic environment leads to \textbf{modified vacuum and thermal fluctuations},  specifically when considering changes of chemical properties at ambient conditions. We highlight under which conditions the modified vacuum or thermal fluctuations become important in Sec.~\ref{subsec:restructuring} and might induce non-canonical equilibrium conditions for the matter sub-system. A final question to address in polaritonic chemistry is then the difference between single-molecule strong coupling, also called local strong coupling, and collective strong coupling. We discuss the topic of \textbf{local/collective strong coupling} in Sec.~\ref{subsec:collectivity}, and highlight how an effective single-molecule picture suggests itself.

	Despite the internal complexity and depth of this review, we try to keep it structured modularly and the different sections largely self-contained. This will help the reader, allowing her/him to, for instance, skip the first few sections, which detail the theoretical foundations of ab initio QED, and jump directly to the later sections which focus more on polaritonic chemistry. Yet a better understanding of many arguments (as highlighted above) necessitate detailed discussions and hence we have provided many cross links between various sections. In Sec.~\ref{sec:lightandmatter} we give a concise introduction into QED with a focus on the description of the electromagnetic field. In Sec.~\ref{sec:PauliFierz} we introduce the basic Hamiltonian of ab initio QED, discuss its many important properties and provide its most commonly employed approximations. In Sec.~\ref{sec:firstprinciples} we discuss various first principles QED methods. In Sec.~\ref{sec:polaritonicchemistry} we discuss polaritonic chemistry from an ab initio perspective. Finally, in Sec.~\ref{sec:outlook} we give a conclusion and outlook on how to employ the \textit{photonic environment} as an extra control knob to influence chemical and material properties.

	\section{A theory of light and matter: quantum electrodynamics}
	\label{sec:lightandmatter}
	
	\textit{"In a hydrogen atom an electron and a proton are bound together by photons (the quanta of the electromagnetic field). Every photon will spend some time as a virtual electron plus its antiparticle, the virtual positron [...]"}
	
	\noindent
	\\
	G. Kane in Ref.~\cite{Kane_scientific}
	\\
	\\
	QED is a cornerstone of modern physics, and Feynman, Tomonaga and Schwinger were awarded the Nobel prize in physics in 1965 for their contributions to this theory~\cite{nobelprize1965}. It tells us on the most fundamental level how light and charged particles interact and how their coupling leads to the emergence of the observable electrons/positrons and photons~\cite{Greiner_1996,ryder_1996,greiner2003quantum}. The beauty of QED is that it can be derived from a few very basic principles. However, it is also plagued by several mathematical issues that restrict the applicability of full QED to perturbative high-energy scattering processes~\cite{ryder_1996,greiner2003quantum}. Yet, in certain limits, most notably when the charged particles are treated non-relativistically, QED allows for a beautiful and mathematically well-defined formulation that is very similar to standard electronic quantum mechanics~\cite{spohn2004dynamics}. The resulting non-relativistic QED theory in Coulomb gauge will form the foundation of ab initio QED chemistry and will be discussed in Sec.~\ref{sec:PauliFierz}. But before, we will briefly summarize how QED can be derived from basic principles.
	
	
	\subsection{Relativistic origins}
	
	There are different formulations of the basic equations of QED as well as various different ways to derive them~\cite{ryder_1996, Greiner_1996, greiner2003quantum,baez2014introduction,scharf2014finite}. Let us follow here a route that highlights that both sectors of the theory, that is, the light and the matter parts, follow from the same reasoning and that the coupling between the sectors enforces a strong consistency between the light and matter sector. As a first step we want the matter as well as the light sector to individually obey special relativity in the form of the energy-momentum relation~\cite{ryder_1996,Greiner_1996}
	\begin{align}\label{eq:energymomentum}
		E^2 = m^2 c^4 + p^2 c^2.
	\end{align}  
	This relation can be derived from the assumption of a highest possible velocity $c$ which we call the speed of light in vacuum. We note that Eq.~\eqref{eq:energymomentum} implies that we think about the flat (Euclidean) space $\mathbb{R}^3$ or its extension including time, the Minkowski space~\cite{ryder_1996, Greiner_1996}. Its \textbf{homogeneity}, i.e., that no point is special, and its \textbf{isotropy}, i.e., that no direction is special, are very important, since these symmetries determine the basic building blocks of our theories. These symmetries are connected directly to the position-momentum and energy-time uncertainty relations~\cite{Greiner_1996,teschl2014,blanchard2015mathematical}, i.e., the translations in space are connected to momentum operators and the translations in time to the energy operator. Thus the basic building blocks are (self-adjoint realizations of) the momentum $-\imagi \hbar \bnabla$ and position $\br$ operators and the energy $\imagi \hbar \partial_t$ and time $t$ operators. And the basic wave functions describing matter and light, respectively, should obey Eq.~\eqref{eq:energymomentum}, but with the substitution $E \rightarrow \imagi \hbar \partial_t$ and $p \rightarrow -\imagi \hbar \bnabla$. Just using the resulting second-order equation to determine the basic wave functions leads, however, to several problems~\cite{Greiner_1996, greiner2003quantum,thaller2013dirac}. A possible way out is to recast the second-order equation in terms of a first-order Hamiltonian equation, i.e., an evolution equation for the energy. Following Dirac's seminal idea, we can use for spin-$1/2$ particles the four-component \textbf{Dirac equation}
	\begin{align}\label{eq:dirac}
		\imagi \hbar \partial_t \psi(\br t) = \left[ -\imagi \hbar c \boldsymbol{\alpha} \cdot \bnabla + \alpha_0 m c^2   \right] \psi(\br t),
	\end{align}
	where the vector of matrices $\boldsymbol{\alpha}$ and $\alpha_0$ are the $4 \times 4$ Dirac matrices~\cite{Greiner_1996, greiner2003quantum,thaller2013dirac}. Applying the Dirac equation twice we recover the operator form of Eq.~\eqref{eq:energymomentum} as intended. Eq.~\eqref{eq:dirac} is then used to describe the matter sector of QED. If we use a vector of spin-$1$ matrices $\boldsymbol{S}$ instead, we find the \textbf{Riemann-Silberstein equation}~\cite{silberstein1907,oppenheimer1931,bialynicki1994wave,gersten1999maxwell}
	\begin{align}\label{eq:riemannsilberstein}
		\imagi \hbar \partial_t  \bFtilde(\br t) = -\imagi \hbar c \boldsymbol{S}\cdot \bnabla \bFtilde(\br t),
	\end{align}   
	for a three-component wave function $\bFtilde$ with zero mass and the necessary side condition
	\begin{align}\label{eq:sidecondition}
		\nabla \cdot \bFtilde(\br t) = 0.
	\end{align}
	This side condition ensures that the wave function $\bFtilde$ has only two transverse degrees of freedom, as to be expected for free electromagnetic fields, which have two independent polarizations. Eqs.~\eqref{eq:riemannsilberstein} and \eqref{eq:sidecondition} are then used to describe the electromagnetic sector of QED and recover the usual Maxwell equations in the classical limit, as discussed below. 
	
	
	\subsection{Quantizing the light field}
	\label{subsec:quantizinglight}

	The main issue with these two relativistic equations is that, since they are first order, they necessarily have besides positive- also negative-energy eigenstates. This is an issue that becomes immediately clear from the Riemann-Silberstein wave function $\bFtilde$, which should be a quantum version of the electromagnetic energy expression in terms of the electric field $\boldsymbol{E}(\br t)$ and magnetic field $\boldsymbol{B}(\br t)$, i.e.,
	\begin{align}\label{eq:electromagenticenergy}
		E_{\rm ph} = \frac{\epsilon_0}{2} \int \left( \boldsymbol{E}^2(\br t) + c \boldsymbol{B}^2(\br t) \right) \d \br,
	\end{align} 
	with strictly positive eigenenergies. To resolve this issue, we follow a further seminal idea of Dirac. We re-interpret the single-particle equations as actually being equations for two particles. That is, the positive-energy states are the particles and the negative-energy states are the corresponding anti-particles~\cite{Greiner_1996, ryder_1996, thaller2013dirac}. For the photon, we find that it is its own anti-particle, where positive-energy states are associated with positive helicity and negative-energy states with negative helicity~\cite{greiner2003quantum,ryder_1996, greiner2003quantum}. To translate this idea into a mathematical prescription we perform a second quantization step. In more detail, we use the distributional eigenstates of the respective equations (plane waves with momentum $\bk$ times the corresponding Dirac spinors for matter, or times circular polarization vectors for light), define creation and annihilation field operators for particles and anti-particles, and effectively exchange the meaning of creation and annihilation for the anti-particles such that the energy becomes manifestly positive~\cite{Greiner_1996, ryder_1996,thaller2013dirac}. In the case of the electromagnetic field quantization the respective field operators obey, due to being spin-$1$ particles, the (bosonic) equal-time commutation relations
	\begin{align}
		\left[\hat{a}(\lambda', \bk' ), \hat{a}^{\dagger} (\lambda, \bk)\right] = \delta_{\lambda' \lambda} \delta^{3}(\bk - \bk').
	\end{align}
	Here we interpret $\lambda =1$ as having positive helicity and $\lambda = 2$ as having negative helicity~\cite{Greiner_1996, ryder_1996,greiner2003quantum}. With this we find the quantized form of Eq.~\eqref{eq:electromagenticenergy} to be 
	\begin{align}\label{eq:photonhamiltonian}
		\hat{H}_{\rm ph} = \sum_{\lambda =1}^{2} \int \hbar \omega_{\bk} \hat{a}^{\dagger}(\lambda, \bk) \hat{a}(\lambda, \bk) \d \bk, 
	\end{align} 
	where $\omega_{\bk} = c |\bk|$ (dispersion of the light cone) and we have discarded the trivial and unobservable, yet infinite vacuum contribution $\sum_{\lambda =1}^{2} \int \hbar \omega_{\bk} \d \bk/2$, i.e., we have assumed normal ordering~~\cite{Greiner_1996, ryder_1996,greiner2003quantum}.

	In this very condensed derivation of the quantized electromagnetic Hamiltonian (we do not give further details of the electronic part of relativistic QED, because we will consider non-relativistic charged particles only) we have made some important implicit choices that need to be highlighted. Firstly, we have used a quantization procedure based on the vector potential to arrive at the standard expression of Eq.~\eqref{eq:photonhamiltonian}. Since the Riemann-Silberstein momentum operator is proportional to the curl, i.e., $-\imagi \boldsymbol{S}\cdot \bnabla \equiv \bnabla \times$, its distributional eigenfunctions are also distributional eigenfunctions for the static vector-potential formulation of the homogeneous Maxwell equation (see also Eq.~\eqref{eq:freemaxwellcoulomb2})
	\begin{align}\label{eq:statichomogeneousmaxwell}
		-\bnabla^2 \bA(\br) \!=\! \left( \bnabla\times\bnabla\times - \bnabla \bnabla \cdot \right) \bA(\br) \!= \! \bk^2 \bA(\br).
	\end{align}
	Note that the longitudinal part is zero by construction for the decomposition of the vector Laplacian, due to the side condition of Eq.~\eqref{eq:sidecondition}, i.e. only the transverse part (first term) becomes non-trivial. The quantization in terms of the vector potential is an important choice, since in the context of the Riemann-Silberstein formulation one often uses a quantization procedure based on the electric and magnetic fields instead~\cite{bialynicki1996v, bialynicki1998exponential,keller2012quantum}. We will comment on this and further connections to classical electrodynamics a little below. Furthermore, since we have only considered the transverse eigenfunctions of Eq.~\eqref{eq:statichomogeneousmaxwell}, we have implicitly chosen the \textbf{Coulomb gauge}, i.e., $\bnabla\cdot \bAh_{\perp}(\br) = 0$. Consequently, the electromagnetic vector potential 
	
	\begin{strip}
		\begin{align}\label{eq:quantizedvectorpotential}
			\bAh_{\perp}(\br) = \sqrt{\frac{\hbar c^2}{\epsilon_0 (2 \pi)^3}}\sum_{\lambda =1}^{2}\int \frac{1}{\sqrt{2 \omega_{\bk}}} \left( \hat{a}(\bk,\lambda) \exp(\imagi \bk \cdot \br) \be(\bk,\lambda) + \hat{a}^{\dagger}(\bk,\lambda) \exp(- \imagi \bk \cdot \br) \be^{*}(\bk,\lambda) \right) \d \bk,
		\end{align} 
	\end{strip}
	
	given here in units of Volts, to agree with relativistic notation~\cite{Greiner_1996,ryder_1996,jestadt2019light}, has only the two physical transverse components. If we had chosen a different gauge instead, we would have to take care of unwanted longitudinal and time-like degrees of freedom by employing quite intricate technical methods, such as Gupta-Bleuler or ghost-field methods~\cite{Greiner_1996,srednicki2007quantum,keller2012quantum}. The main drawback of the Coulomb gauge is that it is not explicit Lorentz covariant, i.e., if we perform a Lorentz transformation to a new reference frame the Coulomb condition is violated in general~\cite{Greiner_1996}. However, since we usually have a preferred reference frame for our considerations, i.e., the lab frame, this is a minor restriction in practice. The second point we want to mention is that we have so far chosen, in accordance to the distributional eigenfunctions of Eq.~\eqref{eq:riemannsilberstein}, circularly polarized vectors $\be(\br,\lambda)$~\cite{Greiner_1996, ryder_1996, srednicki2007quantum}. But for the quantization of the electromagnetic field we can equivalently choose any other \textbf{polarization} vectors that obey
	\begin{align}
		\be(\bk,\lambda) \cdot \bk = \be(\bk,1) \cdot \be(\bk,2) = 0,
	\end{align} 
	and are normalized, i.e., $\be^{*}(\bk,\lambda)\cdot \be(\bk,\lambda)=1$. Indeed, in the following, we will assume the standard choice of linearly-polarized vectors if nothing else is stated, because the linearly- and the circularly-polarized representation are connected by a canonical transformation that leaves everything invariant. For the following theoretical considerations, it is sufficient to  overload the meaning of $\hat{a}(\bk,\lambda)$ and $\be(\bk,\lambda)$ to correspond to the respective linearly-polarized objects as well. The only formal difference is that we can take $\be(\bk,\lambda)$ outside the brackets in Eqs.~\eqref{eq:quantizedvectorpotential} and \eqref{eq:electricquantized}, since in this case it is a real-valued three-dimensional vector. We note that in certain cases the linear polarization will be important, e.g., for the derivation of the length gauge Hamiltonian of Eq.~\eqref{eq:paulifierzlengthgauge}. We will come across an electromagnetic field given in terms of circularly-polarized (also called chiral) modes only at the very end, i.e., in the outlook presented in Sec.~\ref{sec:outlook} .  
	
	Going back to the Riemann-Silberstein Eq.~\eqref{eq:riemannsilberstein}, we recognize that there is a well-known classical equation associated with it, in contrast to the Dirac equation. Indeed, if we re-interpret the three-component wave function and give it the units of an energy wave function, i.e., $\sqrt{C V/ m^3}$ where $C$ is Coulomb, $V$ Volts and $m$ meters, we can associate
	\begin{align}\label{eq:RiemannSilbersteinclassical}
		\boldsymbol{F}(\br t) = \sqrt{\frac{\epsilon_0}{2}} \left( \boldsymbol{E}(\br t) + \imagi c \boldsymbol{B}(\br t) \right).
	\end{align} 
	Using this (classical) Riemann-Silberstein vector, Eqs.~\eqref{eq:riemannsilberstein} and \eqref{eq:sidecondition} become the four \textbf{Maxwell equations} without sources~\cite{silberstein1907,bialynicki1994wave,gersten1999maxwell}
	\begin{align}
		\tfrac{1}{c^2} \partial_t \boldsymbol{E}(\br t) & = \bnabla\times \boldsymbol{B}(\br t), \label{eq:maxwellwithoutsources1}
		\\
		\partial_t \boldsymbol{B}(\br t) & = - \bnabla\times \boldsymbol{E}(\br t), \label{eq:maxwellwithoutsources2}
		\\
		\bnabla\cdot \boldsymbol{E}(\br t) & = 0, \label{eq:maxwellwithoutsources3}
		\\
		\bnabla \cdot \boldsymbol{B}(\br t) & = 0.\label{eq:maxwellwithoutsources4}
	\end{align}
	In this re-interpretation of Eq.~\eqref{eq:riemannsilberstein}, the operator $-\imagi \hbar c \boldsymbol{S}\cdot \bnabla$ does no longer refer to an energy but rather to power, since we can cancel the $\hbar$ on both sides of Eq.~\eqref{eq:riemannsilberstein}. Further, the energy of Eq.~\eqref{eq:electromagenticenergy} is given by the norm of the Riemann-Silberstein vector
	\begin{align}
		E_{\rm ph} = \int \boldsymbol{F}^{*}(\br t) \cdot \boldsymbol{F}(\br t) \d \br.  
	\end{align}

	To connect the classical Maxwell equations back to the above second quantization procedure we note that the vector potential representation of Eqs.~\eqref{eq:maxwellwithoutsources1}-\eqref{eq:maxwellwithoutsources4} in an arbitrary gauge is
	\begin{align}
		&- \bnabla^2 \phi(\br t) - \tfrac{1}{c} \partial_t \left(\bnabla \cdot \bA(\br t)\right) =0 \label{eq:maxwellgeneralgauge1} \\
		&\left(\tfrac{1}{c^2}\partial_t^2 - \bnabla^2 \right)\bA(\br t), \nonumber 
		\\
		& \quad+ \bnabla \left(\bnabla\cdot\bA(\br t) + \tfrac{1}{c} \partial_t\phi(\br t)\right) = 0 \label{eq:maxwellgeneralgauge2},		
	\end{align}
	where the four potential vector is given by $(\phi(\br t), \bA(\br t))$ and we have the association
	\begin{align}
		\boldsymbol{E}(\br t) &= - \bnabla \phi(\br t) - \tfrac{1}{c}\partial_t \bA(\br t) \label{eq:electricpotential},
		\\
		\boldsymbol{B}(\br t) &= \tfrac{1}{c} \bnabla \times \bA(\br t).\label{eq:magneticpotential}
	\end{align}
	Choosing now the Coulomb gauge, i.e., $\bnabla \cdot \bA_{\perp}(\br t) = 0$, the above equations become
	\begin{align}
		&\bnabla^2 \phi(\br t) = 0, \label{eq:freemaxwellcoulomb1}
		\\
		&\left(\tfrac{1}{c^2} \partial_t^2 - \bnabla^2\right) \bA_{\perp}(\br t) = 0. \label{eq:freemaxwellcoulomb2}
	\end{align}
	The only zero solution of Eq.~\eqref{eq:freemaxwellcoulomb1} is $\phi(\br t) = 0$, and all zero solutions of Eq.~\eqref{eq:freemaxwellcoulomb2}, i.e., freely propagating Maxwell fields, can be constructed with the help of the distributional eigenstates of Eq.~\eqref{eq:statichomogeneousmaxwell}~\cite{Greiner_1996}. The Coulomb gauge is a maximal gauge, since it removes all gauge ambiguities (compare Eqs.~\eqref{eq:maxwellgeneralgauge1} and \eqref{eq:maxwellgeneralgauge2}) that would still be allowed in other gauges. We further note that we recover the classical equations from the above vector-potential-based second-quantized formulation by using the Heisenberg equations of motions~\cite{Greiner_1996}, where $\boldsymbol{\hat{B}}(\br) = \tfrac{1}{c} \bnabla \times \bAh_{\perp}(\br)$ and

	\begin{strip}
		\begin{align}\label{eq:electricquantized}
			\boldsymbol{\hat{E}}_{\perp}(\br) = \sqrt{\frac{\hbar}{\epsilon_0 (2 \pi)^3}}\sum_{\lambda =1}^{2}\int \frac{\imagi \omega_{\bk}}{\sqrt{2 \omega_{\bk}}} \left( \hat{a}(\bk,\lambda) \exp(\imagi \bk \cdot \br) \be(\bk,\lambda) - \hat{a}^{\dagger}(\bk,\lambda) \exp(- \imagi \bk \cdot \br) \be^{*}(\bk,\lambda) \right) \d \bk.
		\end{align} 
	\end{strip}
	
	Finally we mention that one can also do a second quantization based on the interpretation of Eq.~\eqref{eq:RiemannSilbersteinclassical} without resorting to the vector potential formulation~\cite{bialynicki2013role}. This has the advantage that the resulting basic objects of the theory are gauge-independent. On the other hand, as we will see next, the coupling between light and matter is based on the gauge principle, and hence at that point usually the vector potential formulation appears again. 
	
	
	\subsection{Coupling light and matter}
	
	Let us next couple the two sectors of the theory. Not surprisingly, there are again various ways to derive how photons and quantized charged particles couple~\cite{Greiner_1996, ryder_1996, srednicki2007quantum, keller2012quantum, craig1998molecular, spohn2004dynamics}. We will here use a further symmetry argument to couple light and matter. The Dirac and the Riemann-Silberstein equations are intimately connected to symmetries. One specifically important symmetry is connected to the \textbf{local conservation of charge} (or probability if we do not include the elementary charge $|e|$ in the arguments below). Indeed, from Eq.~\eqref{eq:dirac} we find that the Dirac charge density $\rho(\br t) = -|e| \psi^{\dagger}(\br t)\psi(\br t)$ and the Dirac charge current $\bJ(\br t) = -|e| c \psi^{\dagger}(\br t) \boldsymbol{\alpha} \psi(\br t)$, where and $-|e|$ is the charge of the electron, obey the continuity equation
	\begin{align}\label{eq:continuityequation}
		\partial_t \rho(\br t) - \bnabla \cdot \bJ(\br t) = 0.
	\end{align}
	This equation guarantees that locally charge cannot be destroyed or created, it can only flow from one point to another. Since in the above equation the phase of the wave function becomes irrelevant, we realize that this conservation law holds even if we change the phase of the wave function $\psi(\br t) \rightarrow \psi(\br t) \exp(\imagi \chi(\br t))$. In order to enforce that this phase change does not affect any physical observable, we have to replace $ \imagi \partial _t \rightarrow \imagi \partial_t + (\partial_t \chi(\br t))$ and $- \imagi \bnabla \rightarrow -\imagi \bnabla  - (\bnabla \chi(\br t) )$ in Eq.~\eqref{eq:dirac}. One therefore interprets the resulting linearly-coupled fields $(\partial_t \chi, \bnabla \chi)$ as having no physical effect on the charged particle. Indeed, if we determine the Maxwell energy that such fields would correspond to, we find that the four vector potential $(-\tfrac{\hbar}{|e|} \partial_t \chi(\br t), -\tfrac{\hbar c}{|e|} \bnabla \chi(\br t))$ leads to zero physical fields (compare to Eqs.~\eqref{eq:electricpotential} and \eqref{eq:magneticpotential}) and thus to zero energy (compare to Eq.~\eqref{eq:electromagenticenergy}). The phase of the wave function therefore corresponds to the gauge freedom of the electromagnetic field. This suggests that we should couple a general (non-zero) electromagnetic field in the same linear (minimal) manner, i.e.,  
	\begin{align}
		\imagi \hbar \partial_t &\rightarrow \imagi \hbar \partial_t + |e| \phi(\br t), \label{eq:timegauge}
		\\
		-\imagi \hbar \bnabla &\rightarrow - \imagi \hbar \bnabla + \frac{|e|}{c} \bA(\br t). \label{eq:spatialgauge}
	\end{align}
	This adapted derivative is then called a gauge-covariant derivative~\cite{Greiner_1996, ryder_1996, srednicki2007quantum}. All of this can be formalized much more elegantly in a Lagrangian representation of the problem, where the gauge-covariant derivative makes the local charge conservation explicit~\cite{Greiner_1996, ryder_1996, srednicki2007quantum}.

	Let us next see what that prescription entails for light. For this we look at the (still classical) light-matter interaction energy expression that we recover from the above prescription which is
	\begin{align}\label{eq:interactionenergy}
		E_{\rm int} = -\tfrac{1}{c}\int \bJ(\br t)\! \cdot\! \bA(\br t) \d \br +\!\! \int\!\! \rho(\br t) \phi(\br t) \d \br.
	\end{align}
	Varying this energy expression with respect to the four vector potential we can derive the corresponding contributions to the Maxwell equation~\cite{Greiner_1996}. If we choose the Coulomb gauge we thus find compactly
	\begin{align}
		-\bnabla^2 \phi(\br t) &= \tfrac{\rho(\br t)}{\epsilon_0}, \label{eq:maxwellcoulomb1}
		\\
		\left(\tfrac{1}{c^2} \partial_t^2 - \bnabla^2\right) \bA_{\perp}(\br t) &= \mu_0 c \bJ_{\perp}(\br t),\label{eq:maxwellcoulomb2}
	\end{align}
	where due to the inner product in Eq.~\eqref{eq:interactionenergy} only the transverse part of the charge current contributes. We have thus derived the Maxwell equations including sources that obey the continuity of Eq.~\eqref{eq:continuityequation}. For completeness and later reference we further give the Maxwell equations without vector potentials as
	\begin{align}
		\bnabla\times \boldsymbol{B}(\br t) -\tfrac{1}{c^2} \partial_t \boldsymbol{E}(\br t) & = \mu _0 \bJ(\br t), \label{eq:maxwellwithsources1}
		\\
		\partial_t \boldsymbol{B}(\br t)+  \bnabla\times \boldsymbol{E}(\br t) & = 0, \label{eq:maxwellwithsources2}
		\\
		\bnabla\cdot \boldsymbol{E}(\br t) & = \tfrac{\rho(\br t)}{\epsilon_0}, \label{eq:maxwellwithsources3}
		\\
		\bnabla \cdot \boldsymbol{B}(\br t) & = 0.\label{eq:maxwellwithsources4}
	\end{align}
	
	If we next assume that the only sources for the electromagnetic fields are the (quantized) charged particles, the longitudinal part of the fields, i.e., those corresponding to $\phi(\br t)$ in Eq.~\eqref{eq:maxwellcoulomb1}, can be expressed purely in terms of the charge density itself, i.e., the Hartree potential
	\begin{align}
		\phi(\br t) = \int \tfrac{\rho(\br' t)}{4 \pi \epsilon_0 | \br - \br'|} \d \br'.\label{eq:Coulombpotential}
	\end{align} 
	If we combine this longitudinal interaction energy with the longitudinal contribution in $E_{\rm ph}$ we obtain the well-known Coulomb interaction between the (quantized) charged particles~\cite{Greiner_1996}. So upon second quantization of the electromagnetic field, the longitudinal contributions in Coulomb gauge are only affected by the quantization of the particles and we are left by just replacing $\bA_{\perp}(\br t) \rightarrow \bAh_{\perp}(\br)$ (in the Schr\"odinger picture~\cite{Greiner_1996}).

	Before we give the basic Hamiltonian of non-relativistic QED in the next section, we want to highlight the intimate relation between the geometry of (real) space, the light and the matter sector, the gauge choice and the interaction. Changing any of these ingredients needs to be accompanied with a careful re-evaluation of the basic theory. Firstly, we highlight that if we restrict to only a part of $\mathbb{R}^3$, we need to carefully re-evaluate the basic symmetries in the theory. This is relevant for practical implementations of non-relativistic QED and derivation of corresponding approximate models. For instance, a box with periodic boundary conditions, where all three edges have the same length, keeps all the basic symmetries intact. One finds that the resulting theory, where the plane waves solutions of the various differential operators become normalizable eigenfunctions, converges to the free-space formulation that we have discussed so far. One therefore often uses these two settings interchangeably. Already just choosing other boundary conditions, for instance, zero boundary conditions, might imply subtle differences (see also Sec.~\ref{subsec:approximatepaulifierz}). We further note that both basic equations, i.e., Eqs.~\eqref{eq:dirac} and \eqref{eq:riemannsilberstein}, are based on the same differential operators and hence share the same (distributional) eigenfunctions. This consistency is highlighted again in the gauge principle of Eqs.~\eqref{eq:timegauge} and \eqref{eq:spatialgauge}, where the differential operator and the fields obey the same boundary conditions. Thus changing the modes of the light field independently from the matter can violate, for instance, the basic gauge principle and the Maxwell equations. We will comment on this also later in Sec.~\ref{subsec:approximatepaulifierz}. Finally, the gauge choice influences what we call matter and what we call light. This can be nicely seen from the fact that in Coulomb gauge the longitudinal and time-like photons are absent and subsumed in the Coulomb interaction between the charged particles. This will be further discussed in Sec.~\ref{subsec:paulifierzproperties}.

	
	\section{The Pauli-Fierz quantum-field theory} 
	\label{sec:PauliFierz}
	
	\textit{"The claimed range of  validity of the Pauli-Fierz Hamiltonian is flabbergasting. To be sure, on the high-energy side, nuclear physics and high-energy physics are omitted. On the long-distance side, we could phenomenologically include gravity on the Newtonian level, but anything beyond that is ignored. As the bold claim goes, any physical phenomenon in between, including life on Earth, is accurately described through the Pauli-Fierz Hamiltonian [...]."}
	
	\noindent
	\\
	Herbert Spohn in Ref.~\cite{spohn2004dynamics}
	\\
	\\
	We have discussed above how the (quantized) electromagnetic field can be deduced and how it can be coupled to a quantized matter description. Yet, if we treat matter on the same relativistic level as light, we encounter various conceptual and mathematical issues. Performing a second quantization of also the Dirac equation and coupling it to a second-quantized Maxwell equation via the above gauge-coupling prescription, leads to several divergences~\cite{baez2014introduction,ryder_1996, srednicki2007quantum,derezinski2013mathematics}. Full QED treats these divergences by regularizing and then renormalizing scattering theory~\cite{ryder_1996, greiner2003quantum, srednicki2007quantum}. The simplest realization of a regularization introduces several energy cutoffs in the theory (largest and smallest energy scales for the different particles and their interactions), and it is then shown that the results of perturbative calculations do not depend on how the cutoffs are removed upon renormalizing the theory. In the following, however, we go beyond perturbation theory and consider, for instance, spatially and temporally resolved how a molecule changes during a chemical reaction. In other words, we solve a Schr\"odinger-type equation that gives us access to such processes.
	
	
	\subsection{Non-relativistic QED}
	
	Indeed, within the last decades tremendous progress has been made to reformulate QED as a non-perturbative theory in several limits~\cite{spohn2004dynamics,miyao2009spectral,hidaka2015self,takaesu2009}. The most important situation for our purpose is the \textbf{non-relativistic limit for the matter sector (while keeping the photon sector fully relativistic)}, which allows for a mathematical formulation that is similar to standard electronic quantum mechanics~\cite{teschl2014,blanchard2015mathematical}. So instead of the Dirac equation we are mainly interested in the electronic part of matter and assume that the electrons have small  momenta (with respect to relativistic scales). In other words, we discard the positrons and replace the Dirac momentum by the non-relativistic momentum and hence assume that the electrons are well described by the Schr\"odinger equation. Because this also implies \textbf{matter particle conservation} (no electron-positron pair creation is possible anymore) we do not need to second-quantize the matter sector. This avoids many of the mathematical pitfalls of full QED that arise from working with (mathematically problematic) field operators~\cite{derezinski2013mathematics,thirring2013quantum}. The resulting Hamiltonian, where light and matter couple via the exact minimal coupling prescription from above, is the generalized \textbf{Pauli-Fierz Hamiltonian}~\cite{spohn2004dynamics, jestadt2019light}  
	
	\begin{strip}
		\begin{align} \label{eq:paulifierzhamiltonian}
			\hat{H}_{\rm PF}\! & = \! \sum_{l=1}^{N_e}\tfrac{1}{2m}\left(-i \hbar \bnabla_{\br_l} \! + \! \tfrac{|e|}{c}\boldsymbol{\hat{A}}_\perp(\br_l) \right) ^2 \!+\! \tfrac{|e| \hbar}{2 m} \boldsymbol{\sigma}_l \! \cdot \! \boldsymbol{\hat{B}}(\br_l) + \frac{1}{2}\sum_{l \neq m}^{N_e} \tfrac{e^2}{4 \pi \epsilon_0 |\br_l-\br_m|}    \nonumber \\ 
			& +  \sum_{l=1}^{N_n}\!\tfrac{1}{2M_l}\! \left(\!-i \hbar \bnabla_{\boldsymbol{R}_l}\!-\! \tfrac{Z_l|e|}{c}\boldsymbol{\hat{A}}_\perp(\boldsymbol{R}_l) \right)^2 \!\!\!\!- \! \tfrac{Z_l |e| \hbar}{2 M_l} \boldsymbol{S}_l \! \cdot \! \boldsymbol{\hat{B}}(\boldsymbol{R}_l) + \frac{1}{2}\sum_{l \neq m}^{N_n} \tfrac{  Z_l Z_m e^2}{4 \pi \epsilon_0 |\boldsymbol{R}_l-\boldsymbol{R}_m|} \nonumber \\
			&- \sum_{l}^{N_e} \sum_{m}^{N_n}   \tfrac{Z_m e^2}{4 \pi \epsilon_0 |\br_l-\boldsymbol{R}_m|}  
			+ \sum_{\lambda=1}^{2}\int \hbar \omega_{\bk} \hat{a}^{\dagger}(\lambda,\bk)\hat{a}(\lambda, \bk) \d \bk,
		\end{align} 
	\end{strip}
	
	Here, the first line describes the electronic sector of the theory and its interaction induced by the Coulomb-gauged photon field, where $\boldsymbol{\sigma}$ is a vector of spin-$1/2$ Pauli matrices. The second line is an addition to QED, which would only consider electrons, positrons and photons. We include the nuclei (or more generally ions) as effective quantum particles with an effective mass $M_{l}$, an effective charge $Z_l |e|$ and an effective spin $S$, which gives rise to a vector of spin matrices $\boldsymbol{S}_l$. We do, however, not consider the internal structure of nuclei, which consist of protons and neutrons. The last line describes the longitudinal interaction between the nuclei/ions and the electrons as well as the energy of the free electromagnetic field. It is commonly assumed that this generalized (including also the nuclei/ions) Pauli-Fierz Hamiltonian should be enough to capture most of the physics that happens at non-relativistic energies. Specifically it should be able to describe the situations that arise in QED chemistry and cavity materials engineering. We note, however, that in contrast to the introductory quote by Herbert Spohn, already for simple problems the non-relativistic matter description might not be sufficient. For instance, the color of gold would be much less appealing without relativistic corrections, in many cases spin-orbit interactions can be decisive and often core electrons need to be treated relativistically to find accurate results~\cite{pyykko2012relativistic,reiher2014relativistic}. Semi-relativistic extensions of Eq.~\eqref{eq:paulifierzequation} exist~\cite{miyao2009spectral,konenberg2011existence,hidaka2015self} and adding further corrections seems possible. We will disregard these important details in the following, since they will not lead to qualitative changes in the low-energy regime, and just want to mention that investigating which extra terms need to be included might give indications on how to approach the high-energy problem non-perturbatively. Work along those lines, based on relativistic ab initio QED formulations~\cite{ruggenthaler2011time, ruggenthaler2014quantum,repisky2020respect}, is already in progress. 
	
	
	\subsection{Mathematical properties of the theory}
	
	\label{subsec:paulifierzproperties}
	
	Before we go on, we need to make some comments with regard to this Hamiltonian and discuss some mathematical details that are important for a better understanding of non-relativistic QED. Firstly, while the \textbf{Hilbert space} of the electrons and nuclei/ions are the usual anti/symmetric tensor products of square-integrable Hilbert spaces as in quantum mechanics~\cite{teschl2014, blanchard2015mathematical,spohn2004dynamics}, the space of the photons is a \textbf{symmetric Fock space}~\cite{spohn2004dynamics}. It is build by defining first a single-photon momentum space, i.e., a photon wave function is defined by $\bk$ and the two polarization directions $\lambda$, and from this all symmetric combinations are constructed. This Fock space is different to the very common way of constructing the space of photons, where for each point in momentum or real space a quantum harmonic oscillator is introduced. Such a construction leads to a non-separable Hilbert space~\cite{thirring2013quantum} and thus to a formally different theory. Next, in order for the Hamiltonian $\hat{H}_{\rm PF}$ to be well-defined, the contributions of the photon modes need to be regularized when approaching very high momenta and frequencies. That is, one needs to introduce a form function $\varphi(|\bk|) \rightarrow 0$ for $|\bk| \rightarrow \infty$ with which to regularize the field operators $\hat{a}(\bk,\lambda)$ and $\hat{a}^{\dagger}(\bk, \lambda)$~\cite{spohn2004dynamics}. The simplest way to do so is to introduce a sharp \textbf{cutoff}, which is also called an ultra-violet cutoff, in the mode integrals. Since we have assumed that the particles have non-relativistic momenta, a common choice for the cutoff is the rest mass energy of the particles. An infra-red cutoff, as needed in relativistic QED, is, however, no longer necessary~\cite{spohn2004dynamics}. The interaction between charged particles and photons leads to a stable theory with a finite amount of soft ($\omega_{\bk} \rightarrow 0$) photons (at least for the ground state)~\cite{spohn2004dynamics}. The explicit interaction with the photons, on the other hand, makes it necessary in general to work with \textbf{bare electronic and nuclear/ionic masses} $m$ and $M_l$, respectively. That is, the masses in Eq.~\eqref{eq:paulifierzhamiltonian} are not the observable masses that one uses in quantum mechanics. The physical masses of the particles in quantum mechanics are recovered from non-relativistic QED by tracing out the photon part which leads, e.g., for the electronic mass to $m_{\rm e} = m + m_{\rm ph}$~\cite{bethe1947lamb,hainzl2002mass,spohn2004dynamics} as also highlighted in the introduction. Here the photon contribution $m_{\rm ph}$ is due to the electromagnetic energy that is created by the charged particle itself. When considering the dispersion of a free particle in non-relativistic QED, we realize that the bare mass is necessarily smaller than in quantum mechanics, i.e., $m_{\rm ph} >0$. This is because the free charged particle generates extra energy due to coupling to the photons when having non-zero momentum and is thus effectively slowed down, i.e., the electron is dressed by the photon field (see also Fig.~\ref{fig:lightandmatter} for an artistic view on dressed particles in QED). We will give an explicit expression for the photonic mass (of single particles in the dipole approximation) and comment on further implications of this mass renormalization below in Sec.~\ref{subsec:approximatepaulifierz}. Irrespective of the specific choice of (the positive and finite) bare mass, however, the Pauli-Fierz Hamiltonian has some very nice properties. It is \textbf{self-adjoint}~\cite{Hiroshima2002,spohn2004dynamics}, which guarantees that we can uniquely solve the corresponding static and time-dependent \textbf{Schr\"odinger-type equations}
	\begin{align}\label{eq:paulifierzequation}
		\imagi \hbar \partial _t \ket{\Psi(t)} = \hat{H}_{\rm PF} \ket{\Psi(t)},
	\end{align}  
	and hence we have access to all possible observables. By this we mean that we can calculate the expectation value of all operators, e.g., positions, momenta, kinetic or potential energies (or distribution-valued operators~\cite{thirring2013quantum}, e.g. densities, current densities or kinetic-energy densities) that share the same domain as the Pauli-Fierz Hamiltonian. Furthermore, the Pauli-Fierz Hamiltonian is bounded from below and thus we can use the usual energy minimization principle to find a possible ground state of the coupled light-matter system. Indeed, it can be shown that any system that has a ground state in quantum mechanics, i.e., without coupling to the quantized electromagnetic field, also has a ground state in non-relativistic QED~\cite{Bach1995,bach1998,Bach1999,Fefferman1997,Hiroshima2001,hidaka2010}. This is exactly the property we need in order to be able to discuss the equilibrium properties of a coupled light-matter system. An important difference, however, is that all excited states turn into resonances in non-relativistic QED, i.e., excited states are no longer eigenstates but have a \textbf{finite lifetime}~\cite{Bach1995,Bach1999,derezinski2001,loss2007}. This feature, which is also termed spontaneous emission, is completely missing in standard electronic quantum mechanics, where excited states have the unphysical property to be infinitely long-lived. Indeed, if one just looks at the spectrum of the Pauli-Fierz Hamiltonian, one will usually just find one eigenstate, i.e., the ground state, and then a continuum above the ground state. Thus the spectrum alone does not provide much insight into the properties of the coupled light-matter system~\cite{Bach1999,derezinski2001,spohn2004dynamics}. On the other hand, due to the inclusion of the continuum of photon modes and all the nuclear/ionic degrees of freedom, we have included all dissipation and decoherence channels that are physically present for the subsystems of the total light-matter system and no artifical external baths or non-Hermitian terms need to be added to mimic those processes. In other words, despite the theory being self-adjoint, i.e., closed, the infinite amount of degrees of freedom include also the physical bath degrees of freedom by radiating light from the molecules to the far field and hence being lost to the molecular subsystem. So we can conclude that we have found a fully non-perturbative and consistent theory of light and matter, which answers the first fundamental question from the introduction.

	One final important comment addresses the possibility of working with a \textbf{different gauge}, which relates to the second fundamental question of the introduction. Performing a gauge transformation on the Pauli-Fierz Hamiltonian is far from trivial since the choice of gauge alters the structure of the underlying Hilbert spaces. This becomes even more problematic because the introduced ultra-violet cutoff does not commute in general with the gauge fixing, i.e., exact gauge equivalence is usually lost once a cutoff has been introduced. We will find one notable exception in the case of the dipole approximation of the Pauli-Fierz Hamiltonian below in Sec.~\ref{subsec:approximatepaulifierz}. Furthermore, to the best of our knowledge, only the Pauli-Fierz Hamiltonian in Coulomb gauge has been shown to have all the above desirable properties. Using other gauges to quantize the theory needs careful considerations, as novel problematic terms and divergences arise~\cite{andrews2018perspective,schaefer2020relevance}. In addition one has to note that for other gauges, e.g., the Lorentz gauge, the Coulomb interaction is mediated directly via the (time-like and longitudinal) photons. Consequently even a "quantum-mechanical calculation" that takes into account only the longitudinal Coulomb interaction needs infinitely many quantized modes that need to fulfill certain consistency conditions, such as enforced by the Gupta-Bleuler method~\cite{Greiner_1996,keller2012quantum}. Therefore,  the Coulomb gauge seems to be the most relevant and practical gauge on a non-perturbative Hamiltonian level, and it connects seamlessly with standard quantum mechanics, which is implicitly always assuming the Coulomb gauge~\cite{Greiner_1996,ryder_1996,spohn2004dynamics}. Consequently it is important to choose the Coulomb gauge if combining phenomenologically models of the quantized light field with standard theoretical approaches to quantum matter. This avoids implicit gauge inconsistencies such as double counting the longitudinal interaction between charged particles.  
	
	
	\subsection{Approximations}    
	
	\label{subsec:approximatepaulifierz}

	Non-relativistic QED allows to work with (polaritonic) wave functions $\ket{\Psi}$ of the fully coupled light-matter system~\cite{spohn2004dynamics,ruggenthaler2018quantum,jestadt2019light}, which makes it very similar to standard quantum mechanics. However, the corresponding wave function does not only depend on $N_{e}$ electronic and $N_n$ nuclear/ionic coordinates anymore, but also on a full continuum of photon modes as well. Thus even for a single particle in free space, a wave function solution of Eq.~\eqref{eq:paulifierzequation} is practically unfeasible. Note furthermore that we might need to describe the photonic structure as part of the quantum system in minimal coupling, e.g., the mirrors of a optical cavity are described with the Pauli-Fierz Hamiltonian as well. As will be discussed below, just approximating the cavity structure with a different level of theory runs the risk of introducing severe inconsistencies. So, how can we make the Pauli-Fierz theory applicable? A first slight simplification is found by realizing that we can \textbf{discretize the photon continuum}, and consider then a continuum limit~\cite{glimm1970lambda,arai1997existence}. A good enough discretization (for our setup a very large quantization box with periodic boundary conditions) is virtually indistinguishable from a real continuum. However, this does not really resolve the problem of the still humongous amount of coordinates in the wave function. Therefore, one has to cut back drastically on the amount of coordinates if one is interested in a non-perturbative solution of the Pauli-Fierz Hamiltonian. For perturbative approaches many alternative strategies exist such as to subsume the continuum of modes in a mass-renormalization from the start, i.e., one works with the physical masses of the particles, and everything else is taken into account by, e.g., Wigner-Weisskopf theory~\cite{milonni2013quantum}. We will focus here on the non-perturbative approaches that are necessary for the strong-coupling regime of polaritonic chemistry, which we are interested in in the following. 
	
\begin{figure}
		\begin{center}
			\includegraphics[width=\linewidth]{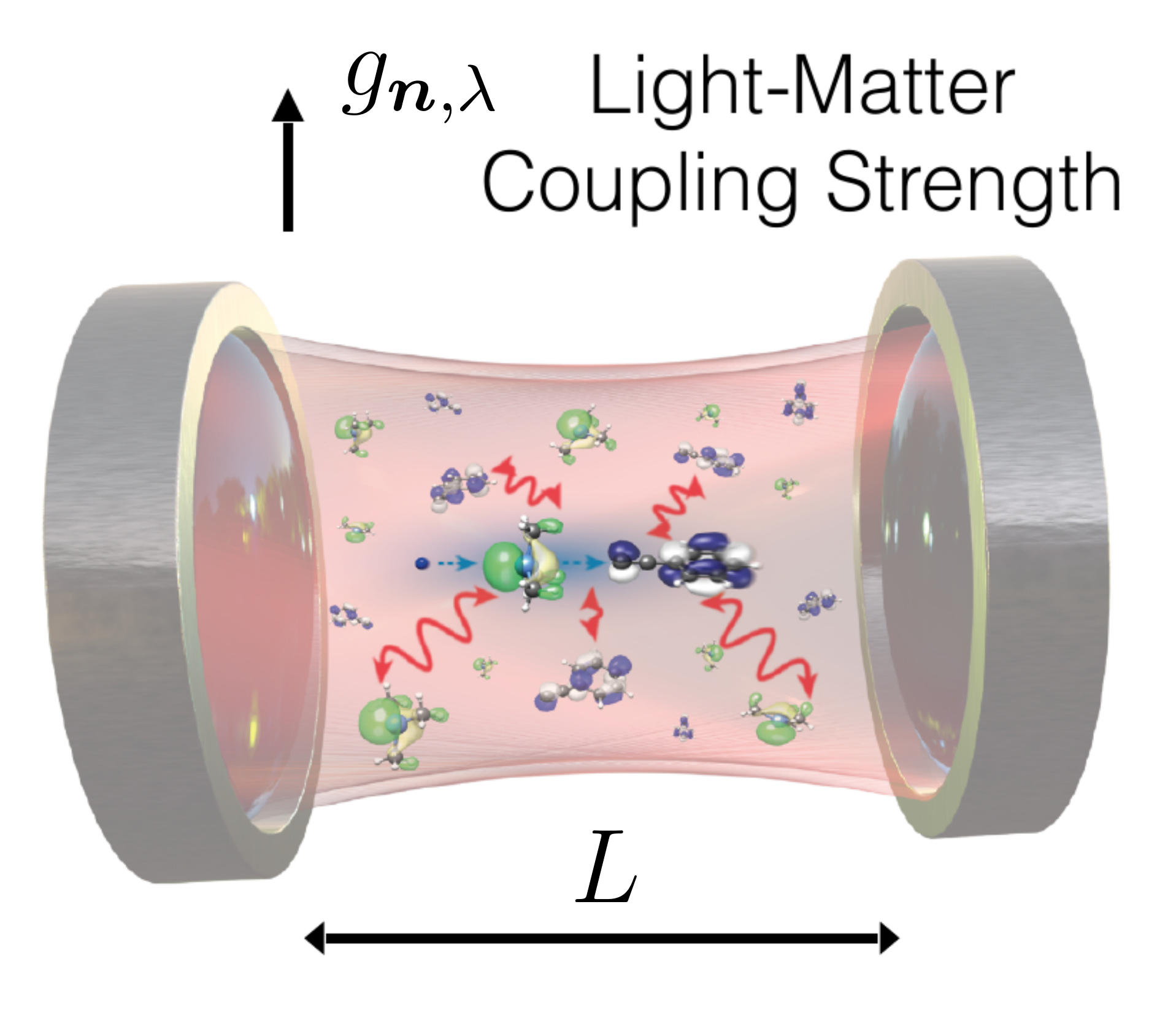}
			\caption{Common Fabry-P\'erot cavity setup. If we assume that the molecules of interest are far removed from the cavity mirrors and localized around the center, one can approximate the main cavity frequencies due to the mirror distance $L$ by $\omega_{n} = c \tfrac{\pi}{L} |n|$, where we have subsummed the effect of the continuum of free-space modes (perpendicular to $L$) in the effective/observed mass of the particles. The coupling strength $g_{n, \lambda}$ for the two independent polarization directions $\lambda$ then increases with $\sqrt{1/L}$ if we keep the low-energy (continuous free-space) modes fixed and take their effect into account by the physical mass of the particles (see Sec.~\ref{subsubsec:dipoleapproximation}).}
			\label{fig:cavitysetup}
		\end{center}
	\end{figure}

	One feasible approximation is to use conditional-wave function approaches~\cite{flick2017cavity, schafer2018ab,villaseco2022exact} to disentangle the different degrees of freedom. In other words, we could apply a Born-Oppenheimer-type of approach, i.e., evolve the nuclei/ions quantum mechanically on a potential-energy surface that is provided by the electrons and/or photons~\cite{flick2017cavity,schafer2018ab}. One then needs to choose whether to group the photons with the electrons or nuclei/ions (see also Sec.~\ref{subsec:photoniondynamics}) and one needs to ensure that there is no double-counting due to the coupling of the photons to both matter degrees of freedom. Further, one needs to take into account that the photons also mediate new couplings between the electrons,  the nuclei/ions and between the electrons and the nuclei/ions. In the dipole approximation (see discussion in Secs.~\ref{subsec:approximatepaulifierz} and \ref{subsec:photoniondynamics}) such extended Born-Oppenheimer approaches have already been investigated and employed in practise. An even further simplification would then be to treat the nuclei/ions classically, which leads to a coupled Ehrenfest-Pauli-Fierz problem~\cite{jestadt2019light,hoffman2019multieherenfest,chen2019ehrenfest}. This option will be discussed in a little more detail in Sec.~\ref{subsec:photoniondynamics}. Another possibility to reduce our problem size is to disentangle different parts of the problem by position, which employs the real-space nature of the Pauli-Fierz Hamiltonian. For instance, we can imaging a common cavity setup, where metallic surfaces constitute the optical cavity and we have the matter system of interest in the middle of this cavity (such as in Fig.~\ref{fig:cavitysetup}). 
	If the surfaces are far enough from the molecular system of interest, the mirrors of the cavity can be described with an effective theory that accounts for changes in the local mode structure of the electromagnetic field instead of describing the (macroscopic) cavity as part of the (cavity+molecular) system. Such a procedure is commonly done, for instance, in macroscopic QED, where the modes of some photonic structures are quantized based on linear-response theory~\cite{buhmann2013dispersion,scheel2009macroscopic}. Such an approximation procedure can lead, however, to severe problems. Keeping in mind our discussion about the necessary \textbf{consistency between light and matter} in Sec.~\ref{sec:lightandmatter}, where we saw that the mode structure of both sectors are the same, we can break various exact relations, such as energy and momentum conservation, if we change the (Fourier) mode structure of light and matter independently. An instructive example is found if we take periodic boundary conditions for matter but zero boundary conditions for light (say in $x$ direction similar to Fig.~\ref{fig:cavitysetup}) to simulate a cavity structure. In this case the gauge principle of Eq.~\eqref{eq:spatialgauge} tells us that just adding $\exp(\imagi \tfrac{2 \pi}{L} x)$ to the wave function on $x \in [0,L]$ corresponds to a pure gauge, and the resulting pure gauge field is proportional to $\tfrac{2 \pi}{L}$, i.e., a constant field. The Maxwell energy with zero boundary conditions of a constant field is, however, infinite. This can be seen either by a basis expansion or by realizing that a self-adjoint differential operator always knows about the boundary conditions and hence interprets that the constant field drops instantaneously to zero at the boundary, which is not differentiable~\cite{ruggenthaler2015existence, berry1996quantum}. Such issues are avoided once we make the dipole-coupling approximation, where the mode consistency between light and matter becomes irrelevant, and we can indeed replace the cavity by a local modification of the electromagnetic modes. We discuss this in more detail below in Sec.~\ref{subsubsec:dipoleapproximation}.

	A different type of simplification follows from a clever basis choice, such as the eigenfunctions of the uncoupled problem, and then to assume that only a few such matter and light states contribute significantly to the solution of the Pauli-Fierz equation. This is a very common way in quantum optics~\cite{grynberg2010,haroche1989cavity}, but it clearly needs already a very detailed understanding or intuition of the subsystems and the physics involved in the light-matter coupling. Moreover, one needs also knowledge about the representation of these states in the original basis of the Pauli-Fierz Hamiltonian to model the proper coupling among the new (many-body) states and the potentially complex photonic states. This knowledge is commonly not available. The many-body methods needed for large systems do not provide the states directly. We will also discuss this issue below in the context of first-principle methods of the Pauli-Fierz Hamiltonian (see Sec.~\ref{sec:firstprinciples}). To circumvent the issues of having the many-body states available, again the dipole approximation comes in handy, since dipole transition moments are readily available for many different systems from various theoretical ab initio methodologies.    
	
	\subsubsection{Cavity as modification of local mode structure: dipole approximation}   
	
	\label{subsubsec:dipoleapproximation}  
	
	For a straightforward simplification of the Pauli-Fierz problem one usually goes directly to the \textbf{dipole approximation} thanks to its many desirable properties. The basic assumption implies that all relevant modes of the electromagnetic field have a wavelength $2 \pi / |\bk| $ that is much larger than the extend of the localized matter system. This clearly requires that we need to adjust the cutoff to low enough frequencies. Indeed, for most calculations one usually reduces the number of modes to only a few effective ones~\cite{latini2022modes}. We will discuss resulting implications below. Following the above assumption, we replace $\bAh_{\perp}(\br) \rightarrow \bAh_{\perp}(0) \equiv \bA_{\perp}$ in Eq.~\eqref{eq:paulifierzhamiltonian}, where we have also assumed implicitly that the matter system is localized (center of charge) at the origin of the coordinate system. An alternative way to arrive at the same approximation is to assume $\exp(\imagi \bk\cdot \br) \approx 1$ in Eq.~\eqref{eq:quantizedvectorpotential}. Besides becoming problematic when the wavelength of the considered modes becomes comparable with the size of the matter system or when retardation effects become important, we also discard in the dipole approximation any direct influence due to the magnetic part of the quantized photon field on the spin degrees of freedom. We further note that we do not use a multi-center dipole approximation, as often assumed in perturbative or phenomenological approaches, where different particles see different fields~\cite{craig1998molecular}, since this would a priori violate the fundamental indistinguishability criterion of quantum particles. Only upon interacting with an environment we can attain distinguishability and classicality, which is discussed in Sec.~\ref{subsec:chemicalreaction}. The resulting (single-center) Hamiltonian is then often also called to be in \textbf{velocity gauge}, which is just the dipole-approximated Coulomb-gauged Pauli-Fierz Hamiltonian. Its form highlights a few important properties that make the dipole approximation so versatile. While Eq.~\eqref{eq:paulifierzhamiltonian} is \textbf{translationally and rotationally invariant} only in the full configuration space of light and matter~\cite{spohn2004dynamics}, in the dipole approximation the Hamiltonian is translationally and rotationally invariant also with respect to the matter subsystem~\cite{rokaj2018light,schaefer2020relevance}. Thus, we find the nice and practical feature that the Pauli-Fierz Hamiltonian can be expanded in the usual matter-only Bloch states in dipole approximation~\cite{schafer2018ab,rokaj2019quantum}, in contrast to the full minimal coupling Hamiltonian. Hence, one usually works in the dipole approximation for solid-state systems. How to properly include beyond dipole contributions remains an active topic of research~\cite{rokaj2019quantum,rokaj2022polaritonic}. 
	
	Specifically in the context of symmetries it is important to highlight that there is a second, unitarily equivalent form of the dipole-approximated Pauli-Fierz Hamiltonian. In more detail, upon performing a unitary transformation $\exp(\tfrac{\imagi}{\hbar c}\bAh_{\perp}\cdot \bR)$, where $\bR = -\sum_{l=1}^{N_{e}} |e| \br_l + \sum_{l=1}^{N_n}Z_l|e|\bR_l$ is the total dipole operator, and a swapping of conjugate photon variables\cite{tokatly2013time,rokaj2018light,schaefer2020relevance}, one finds the \textbf{length gauge Pauli-Fierz Hamiltonian}~\cite{craig1998molecular,tokatly2013time,andrews2018perspective,rokaj2018light,jestadt2019light,schaefer2020relevance}
	
	\begin{strip}
		\begin{align}\label{eq:paulifierzlengthgauge}
			\hat{H}_{\rm PF}'\! = - \sum_{l=1}^{N_e}\tfrac{\hbar^2}{2m}\bnabla^2_{\br_l} &+ \frac{1}{2}\sum_{l \neq m}^{N_e} \tfrac{e^2}{4 \pi \epsilon_0 |\br_l-\br_m|} - \sum_{l}^{N_e} \sum_{m}^{N_n}   \tfrac{Z_m e^2}{4 \pi \epsilon_0 |\br_l-\boldsymbol{R}_m|}  -  \sum_{l=1}^{N_n} \tfrac{\hbar^2}{2M_l} \bnabla_{\boldsymbol{R}_l}^2 + \frac{1}{2}\sum_{l \neq m}^{N_n} \tfrac{  Z_l Z_m e^2}{4 \pi \epsilon_0 |\boldsymbol{R}_l-\boldsymbol{R}_m|} \nonumber \\
			& 
			+ \sum_{\alpha=1}^{M_p} \left[ -\tfrac{\hbar^2}{2} \tfrac{\partial^2}{\partial q_\alpha^2} + \tfrac{\omega^2}{2} \left(q_{\alpha} - \tfrac{g_{\alpha}}{\omega_{\alpha}} \be_\alpha\cdot \bR  \right)^2 \right],
		\end{align}
	\end{strip} 
	
	Here we have already assumed a discretized continuum of $M_p$ modes (given in terms of displacement coordinates $q_{\alpha}$ in units of $\sqrt{J} s$) labeled by $\alpha$, where each $\alpha$ is associated with a specific frequency $\omega_{\alpha}$, coupling strength $g_{\alpha}$ and polarization $\be_\alpha$. In the free space case with a quantization volume $L^3$ these quantities would be associated with $\bk_{\boldsymbol{n}}= 2 \pi \boldsymbol{n}/L$, $\alpha \equiv(\bk_{\boldsymbol{n}}, \lambda)$, $\omega_{\alpha}= c|\bk_{\boldsymbol{n}}|$ and $g_{\alpha} = \sqrt{1/\epsilon_0 L^3}$, where $\boldsymbol{n} \in \mathbb{Z}^{3}_0$. However, now we can adapt the frequencies, coupling strengths and polarizations to match a given cavity structure without breaking fundamental symmetries, since the actual spatial mode structure and the momentum matching (no momentum is transferred in the dipole approximation) is no longer important. For a simple example see Fig.~\ref{fig:cavitysetup}. 
	On a first glance the form of Eq.~\eqref{eq:paulifierzlengthgauge} seems to break the above discussed symmetries and has an unusual self-interaction term proportional to $(\be_\alpha\cdot \bR)^2$. This seeming conundrum can be resolved by carefully analyzing the unitary transformation~\cite{rokaj2018light,schaefer2020relevance} and realizing that one has changed explicitly the conjugate variables of the photonic theory and mixed light and matter. Indeed, $q_{\alpha}$ does not correspond to a pure photonic quantity anymore, but is connected to the auxiliary displacement field of the \textbf{macroscopic Maxwell equations}. The macroscopic Maxwell equations are equivalent to the microsocopic Maxwell equations discussed in Sec.~\ref{sec:lightandmatter}, yet use the auxiliary displacement and magnetization fields that stem from a division of the charge currents and densities into bound and free ones. For completeness and for later reference let us briefly consider how these auxiliary quantities arise. We thus first define
	\begin{align}
		\rho(\br t) &= \rho_{\rm bound}(\br t) + \rho_{\rm free}(\br t),
		\\
		\bJ(\br t) & = \bJ_{\rm bound}(\br t) + \bJ_{\rm free}(\br t),
	\end{align}
	and then introduce the polarization $\boldsymbol{P}(\br t)$ and magnetization $\boldsymbol{M}(\br t)$ due to the bound matter by
	\begin{align}\label{eq:boundcurrent}
		\bJ_{\rm bound}(\br t) & = \bnabla\times\boldsymbol{M}(\br t) + \tfrac{\partial \boldsymbol{P}(\br t)}{\partial t},
		\\
		\rho_{\rm bound}(\br t) &= -\bnabla\cdot\boldsymbol{P}(\br t). \label{eq:bounddensity}
	\end{align}
	We note that these equations are equivalent to Eqs.~\eqref{eq:maxwellwithsources1} and \eqref{eq:maxwellwithsources3}. If we then make a corresponding division in the electromagnetic fields
	\begin{align}
		\epsilon_0 \boldsymbol{E}(\br t) &= \boldsymbol{D}(\br t) - \boldsymbol{P}(\br t),
		\\
		\tfrac{\boldsymbol{B}(\br t)}{\mu _0} &= \boldsymbol{H}(\br t) + \boldsymbol{M}(\br t),
	\end{align}
	and apply these definitions to Eqs.~\eqref{eq:maxwellwithsources1} and \eqref{eq:maxwellwithsources3} we find 
	\begin{align}
		\bnabla \times \boldsymbol{H}(\br t) - \tfrac{\partial \boldsymbol{D}(\br t)}{\partial t} &= \boldsymbol{J}_{\rm free}(\br t),
		\\
		\bnabla \cdot \boldsymbol{D}(\br t)& = \rho_{\rm free}(\br t). 
	\end{align} 
	Consequently, the displacement $\boldsymbol{D}(\br t)$ and magnetization fields $\boldsymbol{H}(\br t)$ describe only the free part of the charges. We note that the homogeneous Eqs.~\eqref{eq:maxwellwithsources2} and \eqref{eq:maxwellwithsources4} are usually obeyed by the bound and free auxiliary fields individually. This formal reshuffling is useful in connecting Maxwell theory to a theory that describes a bound system and its reaction to electromagnetic fields. Thus this formulation is often used in conjunction with approximate (matter-only) linear response theory in terms of constitutive relations~\cite{ehrenreich1966,mochan1985,maki1991}. In our case, where light and matter are treated self-consistently and we have captured the reaction due to (bound) longitudinal fields exactly by using the Coulomb gauge, we are only left with a transverse displacement and polarization fields. In the dipole coupling limit, where the magnetization is disregarded, we therefore find~\cite{abedi2018shedding,rokaj2018light,schaefer2020relevance} that $\sum_{\alpha} \epsilon_0  g_{\alpha}^2 (\be_\alpha\cdot \bR) \be_\alpha = \boldsymbol{\hat{P}}_{\perp}$ and $\sum_{\alpha} \epsilon_0 \omega_{\alpha} g_{\alpha} q_{\alpha} \be_\alpha = \boldsymbol{\hat{D}}_{\perp}$, such that 
	\begin{align}
		\epsilon_0 \boldsymbol{\hat{E}}_{\perp} = \boldsymbol{\hat{D}}_{\perp} - \boldsymbol{\hat{P}}_{\perp},
	\end{align} 
	is the transverse electric field operator. Thus the last line in Eq.~\eqref{eq:paulifierzlengthgauge} corresponds to the mode-resolved $\boldsymbol{\hat{E}}_{\perp}^2 + c^2 \boldsymbol{\hat{B}}^2$, and quadratic self-interaction terms naturally arise when coupling to light in terms of displacement and magnetization fields. Notice that  also the matter coordinates have now a different meaning, since we have mixed light and matter (as we originally defined with respect to the Coulomb gauge). For instance, the translational symmetry is now found along a combined coordinate, i.e., one shifts not only $\br_l$ and $\bR_l$ but at the same time also all $q_{\alpha}$~\cite{rokaj2018light}. In addition, other observables, e.g., the number of photons~\cite{rokaj2018light,schaefer2020relevance}, have now a different representation too. This issue has spawned a lot of misunderstandings, mainly in connection with what is called a \textbf{superradiant phase} transition~\cite{hepp1973superradiant,bialynicki1979no,nataf2010no,viehmann2011superradiant,jaako2016ultrastrong,de2018cavity,stokes2020uniqueness,andolina2020super}. In more detail, the transverse electric field is by construction zero for any eigenstate, which follows from Eq.~\eqref{eq:electricpotential} in Coulomb gauge. Yet the displacement field expectation value can be non-zero for an eigenstate. This merely means that one has a non-zero polarization, i.e., a non-zero total dipole of the system. However, the non-zero displacement field has been often misinterpreted as being the electric field, which led to the wrong conclusion that one can find radiating ground states, i.e., a photonic instability. Due to the symmetries of the Pauli-Fierz Hamiltonian we know that any ground state of atoms, molecules or solids has, by construction, in total zero transverse electric field expectation value. Nevertheless, one could still have a macroscopic amount of virtual photons in the ground state. A macroscopic amount of virtual photons in the ground state, e.g., in form of a constant macroscopic magnetic field, could alternatively be interpreted as a superradiant phase.
	
	Let us note for completeness that the length gauge form of the Pauli-Fierz Hamiltonian of Eq.~\eqref{eq:paulifierzlengthgauge} can also be derived from the \textbf{Power-Zienau-Woolley gauge} in dipole approximation, assuming that this gauge had the same longitudinal Coulomb interaction~\cite{craig1998molecular,andrews2018perspective,woolley2020PZW}. Yet beyond the dipole situation both gauges are, as discussed above, formally different theories. A further reason for this discrepancy can be found in the fact that no multipole expansion exists for unbounded operators. That is, the common argument that a Coulomb-gauged field can be multi-pole expanded and in this way connected to the Power-Zienau-Woolley gauge only holds perturbatively and not on the level of operators~\cite{schaefer2020relevance}. In the context of working with operators instead of with perturbation theory we note that we have implicitly assumed that we are on $\mathbb{R}^3$ and instead of boundary conditions on the matter wave functions we have imposed normalizability to have self-adjoint operators~\cite{blanchard2015mathematical,spohn2004dynamics}. This is the standard setting of quantum physics~\cite{teschl2014,thaller2013dirac}. If we would restrict the matter domain, e.g., choose genuine periodic boundary conditions in the velocity gauge, the length gauge transformation changes these boundary conditions as well in a non-trivial manner~\cite{rokaj2018light, rokaj2019quantum}, again highlighting subtle differences when working with different gauges.
	
	After these important technical details let us return to the main advantage of the dipole approximation. That is, we can treat the photonic environment implicitly by \textbf{changing the mode structure of the electromagnetic field at the position of the matter subsystem}. In our case we chose the origin as the center of charge. Therefore one can take now the mode structure of a photonic environment , e.g., from a Maxwell calculation or from experiment, and adapt the $\omega_{\alpha}, \be_\alpha$ and $g_{\alpha}$ in Eq.~\eqref{eq:paulifierzlengthgauge} accordingly. We note that one needs to use the corresponding displacement modes instead of the electric modes in the length gauge, i.e. in Eq.~\eqref{eq:paulifierzlengthgauge}. A further important detail is that, in principle, when changing the mode structure, also the induced longitudinal interaction would change. For a better understanding of this aspect, let us first highlight how the usual Coulomb interaction arises based on the free-space mode structure. The Coulomb kernel in Eq.~\eqref{eq:Coulombpotential} is connected to the inverse of the longitudinal modes of the electromagnetic field, i.e., the (distributional) eigenfunctions of $-\bnabla \bnabla\cdot$ from Eq.~\eqref{eq:statichomogeneousmaxwell}~\cite{Greiner_1996}. Due to the high consistency between light an matter (see also Sec.~\ref{subsec:approximatepaulifierz}) we can express the longitudinal interaction simply in terms of the scalar (distributional) eigenfunctions and hence find for Eq.~\eqref{eq:Coulombpotential} the usual Coulomb kernel
	\begin{align}
		\frac{1}{4 \pi |\br - \br'|} = \int \frac{c^2}{\omega_{\bk}^2} \underbrace{\braket{\br}{\bk}\braket{\bk}{\br'}}_{ = \frac{\exp(-\imagi\bk \cdot(\br-\br'))}{(2 \pi)^3}} \d \bk.
	\end{align} 
	Now changing the mode structure will also affect the longitudinal eigenfunctions and with this lead to a modified Coulomb interaction. Thus in Eq.~\eqref{eq:paulifierzlengthgauge} we might need to replace the Coulomb kernels by a modified kernel that takes into account this change of interaction. In certain cases it is argued that this modification would be the main difference to free space~\cite{de2018cavity,schuler2020vacua}. Alternatively, especially for nanoplasmonic cavities, one might instead just take into account one or a few quantized longitudinal modes of the photonic structure explicitly. We will comment on this a little later below.
	
	Changing the mode structure in the dipole approximation does, however, have a few further subtle consequences. Firstly, if we have a (discretized) continuum of modes we will have to work with \textbf{bare masses} as already discussed in Sec.~\ref{subsec:paulifierzproperties}. In dipole approximation the connection between the (single-particle) bare mass $m$ and physical mass $m_e$ is known non-perturbatively as~\cite{hainzl2002mass,rokaj2022free}
	\begin{align}
		m_{\rm e} = m + \tfrac{4}{3 \pi} \left(\tfrac{e^2}{4 \pi \epsilon_0 c \hbar}  \right) \tfrac{\hbar}{c} \; \Lambda,
	\end{align}
	where the term in the parenthesis is the fine structure constant and $\Lambda$ is the ultra-violet cutoff wave number. This already implies that the cutoff should not be chosen too large since else we would need an unphysical negative bare mass, i.e., in dipole approximation non-relativistic QED is not fully renormalizable (for a single electron the energy where this happens is, however, gigantic~\cite{rokaj2022free})~\cite{spohn2004dynamics}. If we change the mode structure, the connection between bare and physical mass will change in general. In most cases of polaritonic chemistry it is, however, tacitly assumed that the changes in the mode structures are not so severe as to modify this completely. Hence one usually subsumes the continuum of modes in the physical mass and only keeps a few "enhanced" modes explicitly in the calculations. Indeed, usually just one mode is kept~\cite{walther2006cavity,frisk2019ultrastrong,haroche2006exploring}. On the other hand, if we  use a discretized continuum we have included radiative dissipation and decoherence. In other words, since we have very many photonic degrees of freedom, the quantum revival time tends to infinity~\cite{averbukh1989fractional,berry1996quantum,ROBINETT20041} and hence we have effectively irreversible processes. This is broken once we use the physical mass of the particles and merely keep a few effective modes. To re-introduce the irreversiblity often artificial baths are included in a few mode calculation. But in principle such open-system approaches are not needed in non-relativistic QED as it would contain all dissipation channels explicitly.

	One last subtle but very important point concerns the \textbf{self-polarization} term $(\be_\alpha\cdot \bR)^2$. While often one might hope that this term, which causes the difference between the electric and the displacement field, is not very important, it turns out that without this term the theory becomes unstable and leads to unphysical results~\cite{flick2017atoms,rokaj2018light,schaefer2020relevance}. Indeed, no basis-set limit exists without self-polarization, i.e., the theory has no eigenstates that could be approximated by a finite basis expansion, and an unphysical coordinate- and gauge-dependence is introduced. Thus the results can become highly unphysical for a finite number of basis states, such as having alleged ground states with non-zero transverse (propagating) electric fields. Physically that is easy to understand, since one could only discard this term if one had a perfectly localized system of the form $\delta(\br)$, which is impossible in quantum mechanics~\cite{holstein1972spreading,blanchard2015mathematical}. Therefore, this assumption is equivalent to a classical particle at the origin of the coordinate system with some internal structure. Consequently, the self-interaction term must be included for a physical quantum theory in length gauge. This statement holds true, of course, also if the mode structure is changed as discussed above. We note that the effect of the dipole self-energy term is often not to change the result of a purely dipolar (perturbative or few-level) calculation but to stabilize it and guarantee a unique basis-set limit. Yet it depends on the specific setup and the quantities under investigation whether a decisive difference between a perturbative/few-level and a full ab initio calculation can be observed~\cite{schaefer2020relevance}. Importantly, also for longitudinal modes that are potentially due to, e.g., a nanoplasmonic cavity, self-polarization terms need to be taken into account. This becomes clear from the fact that in principle also longitudinal interactions can be treated in terms of the auxiliary displacement and magnetization fields (see Eqs.~\eqref{eq:boundcurrent} and \eqref{eq:bounddensity}). However, this leads to several mathematical issues for a full continuum of modes and one therefore usually assumes that such terms can be replaced by the usual Coulomb interaction in free space~\cite{andrews2018perspective}. Yet for individual longitudinal modes, which are changed, e.g., due to a nanocavity, such a procedure is straightforward. Because of the manifest positive energy of the photon field, we must include a self-polarization term, otherwise one could lower the energy indefinitely and no basis set limit is possible~\cite{rokaj2018light,schaefer2020relevance} (see also App.~\ref{sec:dipoleselfenergy} for a simple proof of this fact). In practice this issue can often be circumvented by restricting to a finite simulation box with certain boundary conditions, which then serves the same purpose as a self-interaction term. The size of the box, however, then becomes a parameter of theory and needs to be chosen with care. Which way we ever turn it, a stable quantum theory dictates to include quadratic (beyond linear) terms and the only difference with respect to the transverse case of Eq.~\eqref{eq:paulifierzlengthgauge} is that the quadratic contribution might be different (since non-zero longitudinal fields are physically possible even for static eigenstates). The same condition appears in any other coupled quantum systems, such as  electron-phonon systems~\cite{antonvcik1955theory,gonze2011theoretical}.
	
	Finally, after having assumed the dipole approximation, subsuming the continuum of modes in the physical masses of the particles and keeping only one effective mode (this means integrating over the part of the continuum that has been enhanced and thus deducing an effective single-mode coupling), we arrive at the starting point of most currently employed \textbf{phenomenological models}. Upon reducing the matter state to just two states, i.e., a ground and excited state irrespective of whether one considers electronic, rotational or vibrational excitations, one reaches the \textbf{Rabi model}~\cite{frisk2019ultrastrong}. With these approximations the dipole self-energy term becomes a constant offset, and is therefore often discarded. Making then the rotating-wave approximation one finds the famous \textbf{Jaynes-Cummings model} that is virtually always invoked when discussing QED chemistry~\cite{ebbesen2016,flick2018strong,dovzhenko2018light,ruggenthaler2018quantum,frisk2019ultrastrong,Hertzog_review,herrera2020molecular,nagarajan2021chemistry,garcia2021manipulating}. If one wants to consider an ensemble of two-level systems one then often employs the further approximated \textbf{Dicke or Tavis-Cummings models}. The latter becomes equivalent to an effectively scaled Jaynes-Cummings model~\cite{Ribeiro2018polariton,feist2018polaritonic,herrera2020molecular,nitzan2022polaritons}. The Dicke or Tavis-Cummings models assume that the individual physical systems, e.g., molecules, are so far apart that they do not interact with each other directly but only couple via the cavity mode. Yet in the model the dipole self-energy term, which necessarily arises in the length gauge beyond only two levels, is discarded (perfect localization of the whole ensemble is assumed) and no spatial information of the individual systems is kept. We note that also on this level of approximation the choice and knowledge of the gauge is crucial. If the Dicke or Tavis-Cummings model is interpreted in terms of the length gauge without the dipole self-energy, it is possible to find the unphysical case of non-zero transverse electric field in the ground state. If the Dicke or Tavis-Cummings model is interpreted in terms of the Coulomb gauge, such unphysical results are avoided.
	
	
	\section{First-principles approaches to non-relativistic QED} 
	\label{sec:firstprinciples}
	
	\textit{"To
		better understand the properties of the hybrid states, further
		development of QED chemistry calculation methods, akin to
		those in quantum chemistry, would be extremely valuable."}
	
	\noindent
	\\
	Thomas W Ebbesen in Ref.~\cite{ebbesen2016}
	\\
	\\
	
	If we do not want to rely on the many restrictive assumptions underlying the phenomenological models, which we introduced at the end of the previous section, we need to find alternative approaches to handle the extreme complexity of the Pauli-Fierz Hamiltonian. For this purpose, we will rewrite the problem of non-relativistic QED in convenient ways that allow (in practice approximate) solutions of the general Pauli-Fierz Hamiltonian numerically. This means that we want to solve Eq.~\eqref{eq:paulifierzequation} either for the Hamiltonian of Eq.~\eqref{eq:paulifierzhamiltonian} or of Eq.~\eqref{eq:paulifierzlengthgauge} without using too much apriori knowledge or assumptions, e.g., which matter or light states are assumed to be the most important ones. However, before we continue, we generalize the Pauli-Fierz Hamiltonians even further. This is helpful for several reasons: Firstly, for density functional methods (see Sec.~\ref{subsec:QEDFT}) we need to include external fields to establish the necessary mappings~\cite{ruggenthaler2014quantum,ruggenthaler2015ground}. Secondly, external fields are natural to calculate, e.g., absorption spectra or to investigate how a laser would induce non-equilibrium dynamics. Thirdly, in various approximations, e.g., the cavity Born-Oppenheimer approach (see Sec.~\ref{subsec:photoniondynamics}), internal degrees of freedom become effective external fields and hence it is helpful to see how (and which) external fields are included in the Pauli-Fierz Hamiltonian. Therefore, in the full minimal-coupling Eq.~\eqref{eq:paulifierzhamiltonian} we replace
	\begin{align}
		\bAh_{\perp}(\br) \rightarrow \bAh_{\perp}(\br) + \bA_{\rm ext}(\br t)   	
	\end{align}
	and add the terms
	\begin{align}
		\sum_{l=1}^{N_e}-|e| \phi_{\rm ext}(\br_l t) + \sum_{l=1}^{N_n} Z_{l} |e| \phi_{\rm ext}(\bR_l t),
	\end{align}
	and
	\begin{align}
		- \frac{1}{c} \int \bJ_{\rm ext}(\br t)\cdot \bAh_{\perp}(\br). \label{eq:externalcurrent}
	\end{align}
	This means we now include \textbf{external classical electromagnetic fields} $(\phi_{\rm ext}(\br t), \bA_{\rm ext}(\br t))$ to act directly on the matter subsystem and an \textbf{external classical current} $\bJ_{\rm ext}(\br t)$ to act directly on the photons. In the Pauli-Fierz Hamiltonian we can even define the (fully quantized) laser pulse by the chosen initial state of the photon subsystem. This ambiguity raises interesting questions about how to best describe, for instance, a laser pulse and what are the differences in the descriptions~\cite{welakuh2021down}. We further note that we have here subsumed the zero component of the external charge current, i.e., $\rho_{\rm ext}(\br t)$, in $\phi_{\rm ext}(\br t)$ since in Coulomb gauge we can just use Eq.~\eqref{eq:Coulombpotential} to connect both. Further, due to the Coulomb gauge we could even restrict to only the transverse part of $\bJ_{\rm ext}(\br t)$ in accordance to the quantized field being only transverse~\cite{ruggenthaler2014quantum,jestadt2019light} (compare also to Eq.~\eqref{eq:maxwellcoulomb2}). We note in passing that the moment we consider also external fields we effectively gain a second gauge freedom. The physical results will not depend on the choice of the gauge of the external field and we do not necessarily need to choose the internal and the external fields to have the same gauge. In contrast to the gauge choice of the internal fields (see Sec.~\ref{subsec:paulifierzproperties} for further details), it is straightforward to change the gauge of the classical external fields. Having included such general external time-dependent fields leads to an explicitly time-dependent Hamiltonian $\hat{H}_{\rm PF}(t).$
	
	For the dipole-approximated theory of Eq.~\eqref{eq:paulifierzlengthgauge} in length gauge we add merely
	\begin{align}\label{eq:externalscalar}
		\sum_{l=1}^{N_e}-|e| \phi_{\rm ext}(\br_l t) + \sum_{l=1}^{N_n} Z_{l} |e| \phi_{\rm ext}(\bR_l t),
	\end{align}
	and
	\begin{align}\label{eq:externalcurrentdipole}
		\sum_{\alpha =1}^{M_p}q_{\alpha} j_{\alpha}(t),
	\end{align}
	where the last term corresponds to Eq.~\eqref{eq:externalcurrent}. There are, however, several transformations in between~\cite{tokatly2013time,ruggenthaler2014quantum} and so $j_{\alpha}(t)$ is proportional to the mode-resolved time-derivative of $\bJ_{\rm ext}(\br t)$. And accordingly we find in this case an explicit time-dependent dipole-approximated Pauli-Fierz Hamiltonian $\hat{H}'_{\rm PF}(t)$.

	In the following we want to present different first principles methods for non-relativistic QED. Similarly to ab initio methods in quantum mechanics, every approach has certain advantages and drawbacks. Which one to use will not only depend on the system under study or the investigated effects but also on the level of details, e.g., whether the full wave function should be accessible (at least approximately) or reduced physical quantities suffice. The good thing is that many of these methods have overlapping fields of application and can hence be used to validate results obtained with a different ab initio QED approach~\cite{nielsen2018dressed,haugland2021intermolecular}. All of these approaches are extensions of quantum-mechanical methods, which have been applied successfully in theoretical chemistry and electronic structure theory for many decades. These approaches therefore aim at describing molecular systems coupled to photons on the same level of detail as their quantum-mechanical (matter-only) counterparts. We note that there are many advanced models and theoretical methods for molecular polaritons (see, e.g., Refs.~\cite{herrera2014quantum,cwik2014polariton,cwik2016excitonic,herrera2016cavity,hagenmueller2017cavity,DeLiberato2017virtual,hagenmueller2018cavity,reitz2019langevin,hagenmueller2019enhancement,botzung2020dark,reitz2020molecule,Wellnitz2021quantum,gurlek2021engineering,reitz2022cooperative,Wellnitz2022disorder}) that have a more quantum-optical background and hence are geared more towards photonic observables. They are discussed in detail in various reviews on QED chemistry, e.g., Refs.~\cite{Ribeiro2018polariton,feist2018polaritonic,herrera2020molecular}. An important goal of polaritonic chemistry is that, as discussed in the introduction, these different view points align and (quantum) optics, (quantum) chemistry as well as electronic structure theory achieve beneficial synergies.

	
	\subsection{Quantum-electrodyamical density-functional theory}   
	
	\label{subsec:QEDFT}
	
	Quantum-electrodynamical density-functional theory (QEDFT) follows the seminal ideas originally developed by Kohn, Hohenberg and Sham for the electronic ground state~\cite{dreizler2012density,burke2012perspective} and later by Runge and Gross for the time-dependent situation of electronic quantum mechanics~\cite{ullrich2011tddft,marques2012fundamentals}. While the fundamental theorems for the static and the time-dependent situation use different quantities we want to follow here the more general time-dependent perspective which encompasses the static case as well~\cite{tokatly2005manybodydft,tchenkoue2019force, schafer2021making}. 
	
	The basic idea is to replace the high dimensional wave function as a descriptor of the system by a reduced/collective physical variable. This is an ubiquitous idea in physics. For instance, in classical mechanics the description of a fluid is not based on the humongous phase space of all the individual particles but on density and velocity fields such as in the Navier-Stokes equations. A different example is the use of reduced Green's functions in many-body quantum physics~\cite{fetter2012quantum,stefanucci2013}. The main advantage of a density-functional reformulation is that we can do this reformulation in an exact manner. That is, we want to guarantee that we can recover the exact results of the wave-function formulation, at least in principle. In more technical terms, we want to have a bijective mapping between the set of all \textit{physical} wave functions and the set of collective variables~\cite{dreizler2012density,ullrich2011tddft,ruggenthaler2015existence}. In this way, once we know the values of the collective variables, we can uniquely identify the corresponding wave function and determine all observables from it (see App.~\ref{sec:mappings} for details on the basic QEDFT mappings). The existence of such a mapping can be recast into the question whether one can find a closed set of equations that are deduced from the Hamiltonian description in terms of wave functions and that only include the collective variables. In the case of Eq.~\eqref{eq:paulifierzhamiltonian} we find these two equations that form a closed set to be~\cite{jestadt2019light,ruggenthaler2014quantum}
	\begin{align}
		&\partial_t \bJ(\br t) = \frac{\imagi}{\hbar}\brakett{\Psi(t)}{\left[\hat{H}_{\rm PF}(t), \boldsymbol{\hat{J}}(\br t)  \right]}{\Psi(t)} \label{eq:localforceminimalcoupling}
		\\
		&\qquad \qquad \quad + \brakett{\Psi(t)}{\left(\partial_t \boldsymbol{\hat{J}}(\br t)\right)}{ \Psi(t)} \nonumber
		\\
		&\left(\tfrac{1}{c^2} \partial_t^2 - \bnabla^2\right) \bA_{\perp}(\br t) = \mu_0 c \bJ_{\perp}(\br t),\label{eq:maxwellcoulombqedft}
	\end{align} 
	where $\boldsymbol{\hat{J}}(\br t)$ is the total charge current density operator that is explicitly time-dependent even in the Schr\"odinger picture if we have a time-dependent external vector potential. Eq.~\eqref{eq:localforceminimalcoupling} is a local force equation and Eq.~\eqref{eq:maxwellcoulombqedft} is the Maxwell equation in Coulomb gauge of the internal fields induced by the (transverse part of the) charge current density.

	Of course the problem is that we do not know all the terms on the right-hand side of Eq.~\eqref{eq:localforceminimalcoupling} explicitly in terms of $(\bJ(\br t), \bA_{\perp}(\br t))$. So in practice we have to resort to approximations, similar to the case of standard electronic density functional theories~\cite{burke2012perspective}.  Note, however, that for Eqs.~\eqref{eq:localforceminimalcoupling} and \eqref{eq:maxwellcoulombqedft} gauge and relativistic invariance become much easier to enforce then for the wave-function formulation and indeed on a QEDFT level it might be beneficial to employ this facts for more accurate approximation strategies in the future. Yet here we stay in Coulomb gauge and follow the seminal ideas of Kohn and Sham, who proposed that in order to approximate such complicated momentum-stress and interaction-stress terms we should use an auxiliary system, which is as close as possible to the original problem, yet is still numerically tractable~\cite{dreizler2012density,ullrich2011tddft,ruggenthaler2015existence}. So in practice a system of non-interacting electrons, nuclei/ions and photons is usually solved that generate the same current density and vector potential as the original problem. The resulting (single-particle) polaritonic Pauli-Kohn-Sham equations
	\begin{align}\label{eq:paulikohnsham}
		\imagi \hbar \partial_t \varphi_{k}(\br s t ) =  \left[ \tfrac{1}{2M_k}\! \left(\!-i \hbar \bnabla \!-\! \tfrac{Z_k |e|}{c}\boldsymbol{A}_{\rm KS}(\boldsymbol{r} t) \right)^2 \right.
		\\
		\left. -  \tfrac{Z_l |e| \hbar}{2 M_k} \boldsymbol{S}_k \! \cdot \! \boldsymbol{B}_{\rm KS}(\boldsymbol{r} t) -Z_{k}|e| \phi_{\rm KS}(\br t) \right] \varphi_{k}(\br s t),  \nonumber 
	\end{align}
	are non-linearly and self-consistently coupled to Eq.~\eqref{eq:maxwellcoulombqedft}, where 
	\begin{align}
		&\brakett{\Phi(t)}{\boldsymbol{\hat{J}}(\br t)}{ \Phi(t)} = \bJ(\br t), 
		\\
		&\boldsymbol{A}_{\rm KS}(\boldsymbol{r} t) = \boldsymbol{A}_{\perp}(\boldsymbol{r} t) + \boldsymbol{A}_{\rm Mxc}(\boldsymbol{r} t),
		\\
		&\phi_{\rm KS}(\boldsymbol{r} t) = \phi(\boldsymbol{r} t) + \phi_{\rm Hxc}(\boldsymbol{r} t) +  \phi_{\rm pxc}(\boldsymbol{r} t).\label{eq:meanfieldexchangecorrelationpotential}
	\end{align}
	Here the Pauli-Kohn-Sham wave function $\Phi(t)$ is a tensor product of Slater determinants and permanents (of electrons, nuclei/ions and photons~\cite{jestadt2019light}) of the orbitals $\varphi_k(\br s t)$, where $s$ is the corresponding spin coordinate for particle $k$ with mass $M_k$, charge $Z_k |e|$ and spin matrix $\boldsymbol{S}_k$. That is, for electrons we have $s\in \{1,2\}$, $M_k = m$, $Z_{k}=-1$ and $\boldsymbol{S}_k = \boldsymbol{\sigma}$. If we also treat the nuclei/ions quantum-mechanically we then have further species of (massive) particles~\cite{jestadt2019light}. The Kohn-Sham magnetic field is given by $\boldsymbol{B}_{\rm KS}(\boldsymbol{r} t) = \tfrac{1}{c} \bnabla \times \boldsymbol{A}_{\rm KS}(\boldsymbol{r} t)$ and the Kohn-Sham vector potential contains the \textbf{mean-field exchange-correlation} potential $\boldsymbol{A}_{\rm Mxc}(\boldsymbol{r} t)$. Further, the Kohn-Sham (scalar) potential contains now besides the usual Hartree-exchange-correlation potential $\phi_{\rm Hxc}(\boldsymbol{r} t)$ also a photon-exchange-correlation potential $\phi_{\rm pxc}(\boldsymbol{r} t)$ (see also App.~\ref{sec:mappings} for further details). An accurate approximation of these fields is much easier to establish and one can beneficially use the direct connection of density-functional methods to reduced-density matrix and Green's function theories~\cite{pellegrini2015optimized,melo2016greens,buchholz2019reduced,tokatly2018conserving,buchholz2020light,karlsson2021greens}. As can be seen from Eqs.~\eqref{eq:paulikohnsham}-\eqref{eq:meanfieldexchangecorrelationpotential}, in general we work with current-density functionals in QEDFT. However, for the static case or the dipole-approximated version (see also discussion below), functionals in terms of the density are sufficient. Further we note that while new terms appear that generate novel contributions to the exchange-correlation potentials, e.g., $\phi_{\rm pxc}(\br t)$ in Eq.~\eqref{eq:meanfieldexchangecorrelationpotential} that is explicitly due to the photon-matter coupling~\cite{jestadt2019light,tokatly2013time,ruggenthaler2014quantum}, in principle also the usual density functionals are implicitly modified since they are now generated by light-matter coupled (polaritonic) wave functions~\cite{flick2015kohn,dimitrov2017exact,theophilou2020virial}. Let us also note that solving these non-interacting yet non-linearly coupled equations is far from trivial. This has to do, on the one hand, with the fact that we still have to solve (for the matter subsystems) many non-linearly coupled single-particle Pauli equations and, on the other hand, that the subsystems (electrons, nuclei/ions and photons) have vastly different energy/time and length/momentum scales. This makes the development of special multi-system and multi-scale methods necessary~\cite{jestadt2019light,tancogne-dejean2020octopus}. An important technical aspect, that connects back to the introduction of the Riemann-Silberstein formulation of classical electrodynamics (see Sec.~\ref{sec:lightandmatter}), is to recast everything as first-order equations in time such as to (re)use the same numerical propagation routines~\cite{jestadt2019light,tancogne-dejean2020octopus}. The first-order equations of the different particle species then need to be solved self-consistently, i.e., the full feedback between the different subsystems (electrons, nuclei/ions and photons) is included. Another technical aspect, specifically with respect to the Maxwell's equation, is to simulate free space by working in a finite simulation box and to use perfectly-matched layers~\cite{jestadt2019light,tancogne-dejean2020octopus}. This gives rise to radiative dissipation and decoherence from first principles. Finally, owing to the difference in mass between the nuclei/ions and electrons, one often makes a further approximation and simulates the nuclei/ions by classical statistical methods, e.g., multi-trajectory Ehrenfest methods~\cite{hoffman2019multieherenfest}. It is within this approximation for the nuclei/ions that QEDFT for Eq.~\eqref{eq:paulifierzhamiltonian} has been successfully applied~\cite{jestadt2019light}.

	Of course, in many practical situations, especially in the case of molecular systems, a full minimal-coupling description is not always needed (although it is still desirable to have such high-level solutions even in such cases in order to justify approximations). So one often uses QEDFT in the long-wavelength (dipole) approximation, where Eqs.~\eqref{eq:localforceminimalcoupling} and \eqref{eq:maxwellcoulombqedft} reduce to the corresponding equations for the Hamiltonian of Eq.~\eqref{eq:paulifierzlengthgauge}~\cite{ruggenthaler2014quantum,tokatly2013time,jestadt2019light}. QEDFT can indeed seamlessly connect to this and various other limiting cases~\cite{schafer2021making}. Before we discuss specifically QEDFT in the dipole-coupling limit, we want to highlight a related methodology applicable in an intermediate regime. For two-dimensional materials one can approximate the in-plane and out-of-plane coupling differently. Such an ansatz was considered by the authors of Ref.~\cite{svendsen2021combining}, which investigated two-dimensional materials weakly coupled to a cavity and the arising Purcell effect, i.e., the cavity-induced faster spontaneous emission of photons. They employed macroscopic QED to quantize the field of the cavity and then coupled it with the help of Wigner-Weisskopf theory (only zero or a single photon in each mode and effective particle masses) to (electron-only) density-functional Kohn-Sham wave functions. Since in this approach light and matter are treated separately, e.g., matter is described in Coulomb-gauged density-functional theory while the Maxwell field is quantized in Weyl gauge, extra care has to be taken to not generate unphysical effects.
	
	The separate quantization of light and matter becomes less error prone if we consider the interaction with the transverse electromagnetic modes in dipole approximation (see Sec.~\ref{subsec:approximatepaulifierz} for details). Within dipole-approximated QEDFT~\cite{tokatly2013time,ruggenthaler2015ground,flick2018nuclei,flick2019light,yang2021quantum,welakuh2022frequency} dissipation and decoherence is still included~\cite{flick2019light,wang2021light,welakuh2022frequency} if the discretized continuum of photon modes is kept, and one can thus investigate, e.g., the super-radiance of a collection of molecules or mass-renormalization effects~\cite{flick2019light}. To reduce the numerical costs even further, one can either reduce the mode number to a few (or merely one) effective modes or one can, for example, approximate the photon modes by an instantaneous radiation-reaction potential~\cite{schafer2022shortcut,bustamante2021dissipative}. Most of the results in polaritonic chemistry obtained with QEDFT-related methods employ one of these limits (see Sec.~\ref{sec:polaritonicchemistry}). The radiation-reaction approach is specifically efficient in including simple Markovian dissipation and allows, in combination with linear-response theory, to reach the macroscopic collective-coupling limit and explore its implications for real molecules in the dilute gas limit~\cite{schaefer2022polaritonic}. For plasmonic situations one can either include the plasmonic structure itself or (more approximately) some quantized effective (potentially longitudinal) modes (see also App.~\ref{sec:dipoleselfenergy}) or even just modify the Coulomb interaction (see also Sec.~\ref{subsec:approximatepaulifierz}). We finally note that once we take the coupling to the (now only few) transverse modes of the photonic structure to zero, QEDFT recovers standard (time-dependent) density-functional theory~\cite{ruggenthaler2015existence,jestadt2019light}. Time-dependent density-functional theory is then often sufficient to capture strong-coupling effects to longitudinal modes of plasmonic cavities if the plasmonic nanostructure is treated explicitly~\cite{rossi2019strong,tserkezis2020applicability,jojt2021dipolar,Peller2020quantitative,kuisma2022ultrastrong}. 
	
	All in all, QEDFT is highly versatile and allows to access electronic, photonic and nuclear/ionic quantities and their self-consistent interplay on various levels of approximation. The main disadvantages involve that it is not easy to assess the error of an approximate density-functional for a given level of theory and it is not straightforward to access observables that are not trivially given by the auxiliary Pauli-Kohn-Sham wave functions.  
	
	
	\subsection{Exact results}
	
	While QEDFT is able to treat the different forms of the Pauli-Fierz Hamiltonian efficiently, in one way or another, the results are usually approximate. For validation purposes and elementary insights it would be good to have exact results. However, for coupled light-matter systems not many exact results (analytic or numerical) are available. To the best of our knowledge only for dipole coupling some exact reference results are known, whose main insights are summarized in the following.
	
	Firstly, we assume the dipole approximated light-matter coupling of Eq.~\eqref{eq:paulifierzlengthgauge} and consider a single particle trapped in a harmonic potential. It can be shown~\cite{spohn2004dynamics} that the time-dependent dipole moment of this particle can be computed by just solving the classical equation of motion of the harmonically trapped particle coupled to the Maxwell's equation, instead of solving the full quantum field problem. This example is also a good rationalization of QEDFT, where the coupled Eqs.~\eqref{eq:localforceminimalcoupling} and \eqref{eq:maxwellcoulombqedft} directly reduced to these classical equations for this case. The computed time evolution of the dipole moment allows to access, e.g., the lifetimes of the excited states and absorption/emission spectra. 
	
	Staying with a harmonic potential, recently \textbf{analytically exact results} of the influence of the photon field with many (identical) interacting particles have been presented and implications discussed, e.g., that even for a ground state resonant behavior can be observed~\cite{rokaj2022cavity}. Furthermore, exact analytical results are available for free particles (electrons)~\cite{rokaj2022free}, which have been used to devise approximations within QEDFT~\cite{schafer2021making}. Besides others, interesting effects on the linear response of the system have been highlighted (e.g. the appearance of plasmon-polariton resonances and a decrease of the Drude peak) and mass renormalization effects due to the thermodynamic limit of the photon field have been shown. In both cases the authors have used that in velocity gauge the photon field couples only to the center of charge of the total system directly and that this  then leads to only an indirect modification of the relative degrees of freedom.

	A different example concerns the one-mode approximation. In this case \textbf{numerically exact results} are available for a quantum three-body system coupled to this effective mode. For He, HD+ and H$_2^+$ one can reformulate the 10 dimensional problem in a problem-specific coordinate system and solve for the lowest lying eigenstates~\cite{sidler2020chemistry, sidler2022exact}. This seemingly simple problem already provides a lot of new insights and effects that we will highlight in Sec.~\ref{sec:polaritonicchemistry}. Suffice it to say that already for the simple one-mode case the eigenvalues of the problem, without any further knowledge, loose the simple interpretation they have in standard quantum mechanics (see also the discussion in Sec.~\ref{subsec:paulifierzproperties} concerning the loss of excited states in QED). Because one has access to the lowest-lying eigenstates in this numerically exact approach, one can also calculate the exact thermal (canonical) ensemble and deduce cavity modified thermal properties, which will be discussed in Sec.~\ref{sec:polaritonicchemistry}.
	
	The main drawback of either analytically or numerically exact results is that they are only available for very specific situations and cannot be applied to different, chemically more relevant cases.
	
	
	\subsection{Quantum electrodynamics coupled cluster theory for electronic strong coupling} 
	
	A compromise between generality and accuracy can be found if we restrict to Eq.~\eqref{eq:paulifierzlengthgauge} in the static case from the start and additionally treat the nuclei as external (clamped) quantities. Afterwards quantum electrodynamics coupled cluster (QED-CC) theory~\cite{mordovina2020polaritonic,haugland2020coupled} can be employed for electronic strong coupling conditions, which has become another important first-principle QED method. In contrast to many-body methods such as QEDFT, QED-CC theory tries to approximate the many-body wave function of electrons and photons directly. We note that alternative wave-function-based methods are available (see e.g. Refs.~\cite{rivera2019variational,ahrens2021stochastic}), but we will not elaborate further on those in this review. The exact electron-photon wave function in QED-CC is re-expressed by applying a cluster (excitation) operator $\hat{T}$ on a reference wave function
	\begin{align}\label{eq:QEDCC}
		\ket{\Psi} = \exp(\hat{T})\ket{R},
	\end{align}
	where $\ket{R}$ is usually the tensor product of the electronic Hartree-Fock wave function with the vacuum states of the modes $\alpha$. In addition to standard coupled cluster theory the cluster operator now also contains photonic contributions and reads for a single cavity mode as 
	\begin{align}\label{eq:clusteroperator}
		\hat{T}=\sum_{\mu,n} t_{\mu, n} \hat{b}^{\mu} (\hat{a}^{\dagger})^n.
	\end{align}  
	Here $\hat{b}^{\mu} \in \{\hat{b}^{a}_{i}, \hat{b}^{ab}_{ij},...  \}$ are the electronic excitation operators of rank $\mu$, $n$ is the number of photons in the mode and the unknown parameters (amplitudes) $t_{\mu,n}$ are to be determined. Also, when comparing to standard coupled cluster theory, one might wonder whether the bosonic nature of the photons imply some sort of symmetrization of the mode wave functions. Yet in Eq.~\eqref{eq:paulifierzlengthgauge} the bosonic nature of the photons is made explicit by the quantum harmonic oscillators $\alpha$. This happens because the excitations of a quantum harmonic oscillator $\alpha$ are connected to the number of photons (note that, as discussed in Sec.~\ref{subsec:approximatepaulifierz}, the "length gauge photons" are not the physically observed photons) in this mode $\alpha$, i.e., we can have infinitely many photons (bosons) in one mode. The expression of Eq.~\eqref{eq:QEDCC} for the wave function $\ket{\Psi}$ leads to an expansion in the number of electronic and photonic excitations that, even if truncated early, gives very accurate results provided the exact ground state $\ket{\Psi}$ is dominated by the single reference wave function $\ket{R}$.
	
	The choice of truncation is important and in practice the number of electronic excitations is chosen as two (with potentially perturbative triples) and the mixed electronic-photonic and pure photon excitations in each mode is either one or two~\cite{haugland2021intermolecular,fregoni2021strong,deprince2021cavity}. This truncation allows to perform practical calculations for relatively large systems. Embedding approaches allow to reach larger systems~\cite{pavovsevic2022wavefunction}, where only part of the problem is treated on the QED-CC level and other parts with, e.g., a Hartree-Fock-type approximation. Ultra-strong and deep-strong coupling~\cite{frisk2019ultrastrong}, where many more than just one excitation per mode arise, need a truncation at higher excitations. Here reformulations of the problem in an adapted basis~\cite{schafer2021making, ashida2021cavity} might prove helpful. Further, as can be inferred from Eq.~\eqref{eq:clusteroperator}, to treat many different photon modes can become numerically costly and no extension to full minimal-coupling (see Eq.~\eqref{eq:paulifierzequation} with the nuclei/ions treated classically) has been devised as of yet. However, the reformulation of our QED eigenvalue problem in terms of (unitary) coupled cluster theory allows for a relatively straightforward implementation~\cite{pavosevic2021polaritonic} on noisy intermediate scale quantum devices~\cite{preskill2018quantum,cao2019quantum} employing variational quantum eigensolvers~\cite{peruzzo2014variational,cao2019quantum}. The representation on a quantum computer has the appealing feature that in principle many (entangled) photon modes could efficiently be represented in contrast to classical devices.
	
	
	\subsection{Nuclear/ionic dynamics in the generalized Born-Huang picture}
	
	\label{subsec:photoniondynamics}

\begin{figure*}
		\begin{center}
			\includegraphics[width=0.9\textwidth]{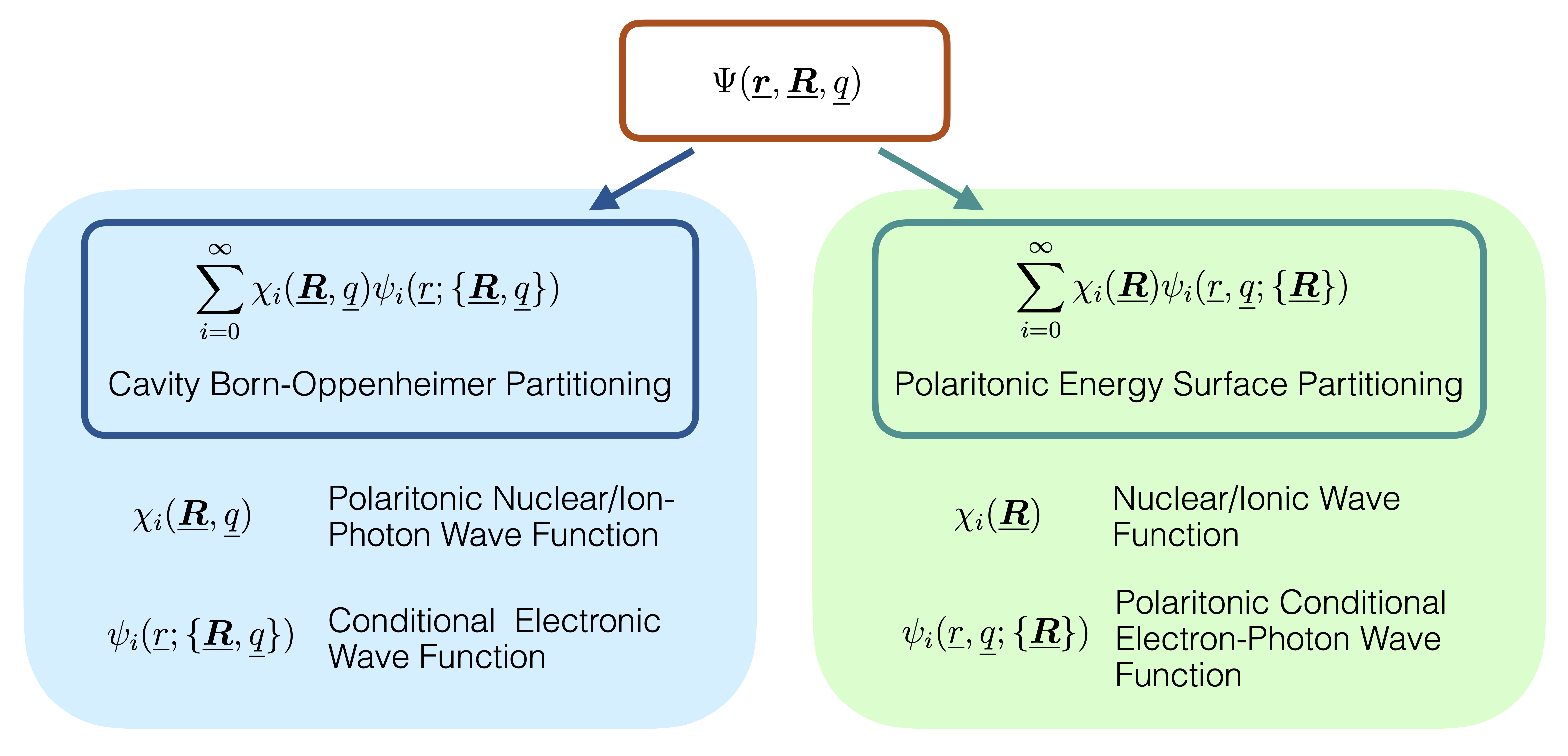}
			\caption{The two main forms of the generalized Born-Huang expansion for coupled light-matter systems discussed in the main text. While the cavity Born-Oppenheimer partitioning is geared towards ground-state chemical reactions, the polaritonic energy surface partitioning is more geared towards photo-chemical processes.}
			\label{fig:bornhuang}
		\end{center}
	\end{figure*}

	If we want to investigate properties of the nuclear/ionic degrees of freedom when strongly coupled to a cavity, we usually take the dipole-coupling approximation and hence start from Eq.~\eqref{eq:paulifierzlengthgauge}. In this case we first perform a \textbf{generalization of the Born-Huang expansion}~\cite{flick2017atoms, flick2017cavity,schafer2018ab}. That is, we re-express the fully correlated wave function of electrons, nuclei/ions and photons
	\begin{align}\label{eq:wfctelectronsnucleiphotons}
		\Psi(\underbrace{\br_1 \sigma_1, ... ,\br_{N_e} \sigma_{N_e}}_{=\underline{\br}}, \underbrace{\bR_1 S_1, ... \bR_{N_n} S_{N_n}}_{=\underline{\bR}},\underbrace{q_1,...q_{M_p}}_{=\underline{q}} )
	\end{align}
	in terms of conditional wave functions. There are now several ways of how to perform this expansion, i.e., which subsystem depends conditionally on the others. While the generalized Born-Huang expansion is exact irrespective of partitioning, the choice of partitioning is important when performing approximations~\cite{flick2017atoms,schafer2018ab}. We will here focus on the two most relevant choices for investigating the nuclear/ionic degrees of freedom (see Fig.~\ref{fig:bornhuang}). We note that there are also alternative schemes, such as exact factorization approaches~\cite{Hoffmann2018exact,Abedi2018exact,lacombe2019pes,martinez2021case,villaseco2022exact}, that we will not go into further detail here. 
	The first choice, which we call the \textbf{cavity Born-Oppenheimer approach}~\cite{flick2017atoms,flick2017cavity}, is to group the photons with the nuclei/ions and to make the electrons depend parametrically on $\underline{\bR}$ and $\underline{q}$. In this case, in order to find the exact solution for
	\begin{align}
		E \Psi(\underline{\br}, \underline{\bR}, \underline{q}) = \hat{H}'_{\rm PF} \Psi(\underline{\br}, \underline{\bR}, \underline{q})
	\end{align}
	via the generalized Born-Huang expansion, we have to solve the equations
	\begin{align}\label{eq:cavitypotentialenergysurfaces}
		E_{i}(\underline{\bR}, \underline{q}) \psi_{i}(\underline{\br}; \{\underline{\bR}, \underline{q}\}) = \hat{H}'_{\rm PF}(\underline{\bR}, \underline{q}) \psi_{i}(\underline{\br}; \{\underline{\bR}, \underline{q}\}),
	\end{align}
	where $\hat{H}'_{\rm PF}(\underline{\bR}, \underline{q})$ is the Hamiltonian of Eq.~\eqref{eq:paulifierzlengthgauge} parametrically dependent on $\underline{\bR}$ and $\underline{q}$ and the kinetic nuclear/ionic and photonic parts are set to zero (treated classically), together with
	\begin{strip}
		\begin{align}\label{eq:cavityBornOppenheimer}
			&E \chi_{i}(\underline{\bR}, \underline{q}) =  \left[\sum_{l=1}^{N_n} \tfrac{\hbar^2}{2M_l} \bnabla_{\boldsymbol{R}_l}^2 + \sum_{\alpha=1}^{M_p} -\tfrac{\hbar^2}{2} \tfrac{\partial^2}{\partial q_\alpha^2} + E_{i}(\underline{\bR}, \underline{q}) \right] \chi_{i}(\underline{\bR}, \underline{q})\nonumber \\
			& 
			+ \sum_{j=0}^{\infty} \int \d \underline{r} \left[ \psi_{i}^{*}(\underline{\br}; \{\underline{\bR}, \underline{q}\})\left( \sum_{l=1}^{N_n} \tfrac{\hbar^2}{2M_l} \bnabla_{\boldsymbol{R}_l}^2 + \sum_{\alpha=1}^{M_p} -\tfrac{\hbar^2}{2} \tfrac{\partial^2}{\partial q_\alpha^2}  \right) \psi_{j}^{*}(\underline{\br}; \{\underline{\bR}, \underline{q}\}) \right] \chi_{j}(\underline{\bR}, \underline{q}).
		\end{align}
	\end{strip}
	Here the last term in Eq.~\eqref{eq:cavityBornOppenheimer} is the non-adiabatic coupling between the polaritonic nuclear/ionic wave functions $\chi_{i}(\underline{\bR}, \underline{q})$. Furthermore, with respect to the usual case without photonic degrees of freedom, also the electronic potential-energy surfaces $E_i(\underline{\bR}, \underline{q})$ are now changed, since they depend explicitly on $\underline{q}$. Thus to distinguish them we call them cavity (Born-Oppenheimer) potential energy surfaces. The cavity Born-Oppenheimer expansion is specifically efficient if we are interested in \textbf{ground-state chemical reactions under vibrational strong coupling}~\cite{sidler2022perspective}. If the ground-state cavity potential energy surface is well-separated from the first excited cavity potential energy surface, we can make the cavity Born-Oppenheimer approximation $\Psi(\underline{\br}, \underline{\bR}, \underline{q}) \approx \chi_0(\underline{\bR},\underline{q})\psi_0(\underline{r};\{\underline{\bR}, \underline{q}\})$~\cite{flick2017cavity}. However, even the resulting simplified equations are far from trivial~\cite{bonini2022vibro} and we discuss various first-principles approaches to approximately solve them below.
	
	A second important partitioning is to choose the electrons grouped with the photons (see Fig.~\ref{fig:bornhuang}), such that the resulting potential energy surfaces $E_i^{\rm pol}(\underline{\bR})$ are \textbf{polaritonic energy surfaces}~\cite{schafer2018ab}. If we partition also the electron-photon conditional wave function we can solve the photonic part analytically. We therefore still have only two coupled equations, one for nuclei/ions on polaritonic energy surfaces and one for electrons, yet the analytic solution of the photons leads to novel (analytically known) non-adiabatic coupling elements among electronic states as well as among nuclear/ionic states. These new analytically known non-adiabatic coupling elements are akin to the couplings in Floquet theory, i.e., they connect states with different number of excited photons~\cite{schafer2018ab,hubener2021engineering}. The here chosen partitioning, which leads to the polaritonic potential energy surfaces, is now specifically efficient if one is interested in \textbf{photo-chemistry}, where the cavity modes are in resonance with electronic excitations and we consider the influence of electronic strong coupling on chemistry. In this case, if we assume that the novel non-adiabatic couplings in the nuclear/ionic sector are negligible and the photons only couple efficiently to the electronic sector, we find the explicit polariton approximation~\cite{schafer2018ab}. In this case the nuclear/ionic degrees of freedom are only indirectly modified by the photon degrees of freedom due to changes in the potential energy surfaces. If we further assume that in the electronic sector the coupling to the cavity modes acts only perturbatively, we recover polaritonic potential energy surfaces as originally introduced in Ref.~\cite{galego2015cavity}.

	Either ways, in order to determine the influence of the cavity modes on the nuclear/ionic subsystem we, in principle, need to solve high-dimensional coupled quantum equations. A similar problem appears also for the usual electron-nucleus/ion dynamics and various approaches have been developed to approximately solve such situations. However, when compared to the traditional electron-nucleus/ion-only problem, the inclusion of the photonic modes implies novel non-adiabatic coupling terms which might become important to faithfully describe certain effects~\cite{kowalewski2016nonadiabatic,csehi2019nonadiabatic,gudem2021controlling,couto2022suppressing}. 
	
	If we assume only a very few nuclear/ionic degrees of freedom to be relevant, one can cut back on the dimensionality of the problem and perform numerically exact simulations~\cite{galego2015cavity,bennett2016novel,kowalewski2016nonadiabatic,gudem2021controlling,couto2022suppressing}. We note, however, that a priori it is not clear whether the same nuclear/ionic degrees of freedom are relevant as outside a photonic structure, e.g., that the cavity can correlated nuclear/ionic degrees of freedom that are largley uncorrelated outside the cavity. Hints towards this issue are highlighted in Sec.~\ref{subsec:chemicalreaction}. For this often an adiabatic to quasi-diabatic basis transformation is performed~\cite{farag2021polariton,badanko2022diabatic,hu2022quasi}, which makes the treatment of non-adiabatic couplings and (potentially cavity-induced~\cite{szodarovsky2018conical,farag2021polariton,csehi2019nonadiabatic}) conical intersections simpler. Such simulations show that the influence of the cavity also on the electronic (non-adiabatic couplings) degrees and a consistent treatment of the dipole self-energy terms (see Sec.~\ref{subsubsec:dipoleapproximation}) can be decisive~\cite{flick2017atoms,gudem2021controlling,couto2022suppressing}. If a strong a priori reduction to merely a few nuclear/ionic degrees of freedom is not possible then one can, for instance, extend the multi-configurational Hartree approach to polaritonic problems~\cite{vendrell2018coherent,triana2018entangled,ulusoy2019nonradiative,csehi2019nonadiabatic,csehi2022competition}. Alternatively, the use of path-integral methods~\cite{mandal2019investigating,mandal2020polariton} and ring-polymer quantization~\cite{chowdhury2021ring,li2022ringpolymer} of light and the nuclear/ionic degrees of freedom allows to investigate higher-dimensional (in terms of photonic and nuclear-ionic degrees of freedom) cases. Simplifying even further, especially in the case of \textbf{thermally driven chemical reactions}, extensions of a semi-classical methods or surface-hopping approaches to coupled nucleus/ion-photon systems are possible~\cite{li2018mixedqm, hoffman2019multieherenfest,chen2019ehrenfest,chen2019ehrenfestII,hoffmann2019benchmarking,fregoni2020tully,antoniou2020role}. Here the use of cavity Born-Oppenheimer potential energy surfaces as introduced in Eqs.~\eqref{eq:cavitypotentialenergysurfaces} and \eqref{eq:cavityBornOppenheimer} seems the best choice to formulate a generalization of molecular-dynamics simulations for coupled cavity-nuclei/ions systems~\cite{li2021collective,sun2022suppression}. One should, however, be careful regarding the treatment of the nuclear/ionic and photonic degrees of freedom. The displacement field dynamics in Eq.~\eqref{eq:cavityBornOppenheimer} can be orders of magnitude faster then the nuclear/ionic dynamics and hence might necessitate the use of adapted Langevin/open-system approaches~\cite{sidler2022perspective}. We will comment in more detail on the physically relevant implications later in Sec.~\ref{sec:polaritonicchemistry}. 
	
	Notice that commonly the free-space electronic surfaces or force-fields are employed instead of cavity potential energy surfaces. This implies a further approximation, since in principle the displacement coordinates also influence the reduced energy eigenvalues. Aside from this it is important to note that (not only for nuclear/ion-photon dynamics) a basis truncation, has to be performed in practice, which can introduce an artificial gauge dependence in such calculations. That is, if we performed a simulation in velocity gauge and one in length gauge (see Sec.~\ref{subsec:approximatepaulifierz} for details) then at the same level of truncation we might find different results~\cite{debernardis2018gauge,li2020electromagnetic}. Only for converged results we should compare different gauges. While one can mitigate such effects between the two relevant (length and velocity) dipole-coupled gauges~\cite{di2019resolution,taylor2020resolution}, we recall (see Secs.~\ref{subsec:paulifierzproperties} and \ref{subsec:approximatepaulifierz}) that for the original minimal coupling Hamiltonian mainly the Coulomb gauge seems practically relevant. If we finally make further assumptions, e.g., that only zero- or one-photon states can be occupied and that we are in a perturbative limit such that we can discard the dipole self-energy terms (see also Sec.~\ref{subsec:approximatepaulifierz}), then we recover common Dicke-type interaction models~\cite{luk2017multiscale,tichauer2021multiscale}.

	Overall we can conclude that to accurately describe the influence of a strongly-coupled photon mode on the nuclear/ionic degrees of freedom we need access to cavity Born-Oppenheimer or polaritonic potential energy surfaces and potentially their non-adiabatic couplings. The usage of potential energy surfaces from a bare matter problem (accessible with standard quantum chemistry software) is a widely applied approximation, which neglects the modifications of the electrons  by the photon field entirely and important effects might be missing. Let us finally note that in quantum chemistry (outside of cavities) the potential energy surfaces are always with respect to a single system undergoing a chemical reaction and the full ensemble of reacting molecules is treated statistically (as is also assumed in transition-state and Marcus theory). However, this approach is no longer straightforward to apply, considering that many molecules are collectively coupled via the cavity modes. In polaritonic chemistry sometimes the concept of a "super-molecule" is invoked, with a potential energy surface that now encompasses the full ensemble. We will comment on this controversial concept that commonly assumes (quantum) coherence among a macroscopic amount of molecules later in Sec.~\ref{subsec:chemicalreaction}.
	
	
	\section{Polaritonic chemistry from first principles} 
	\label{sec:polaritonicchemistry}
	
	\textit{"It has been argued that the Rabi splitting experienced by each
		molecule involved in the collective coupling is not $\hbar \Omega_{R}$ but
		$\hbar \Omega_{R}/\sqrt{N}$. If this were the case, the splitting would be tiny, and
		it is unlikely that any molecular or material property would be
		modified as observed experimentally."}
	
	\noindent
	\\
	Thomas W. Ebbesen in Ref.~\cite{ebbesen2016}
	\\
	\\
	
	Let us now turn to the main focus of this review, the modification of chemical and material properties by strong light-matter coupling. As already highlighted in the introduction, we will present here a perspective on QED chemistry, which is based on first-principles results. For more traditional perspectives on polaritonic chemistry based on various model considerations, we refer the reader to the many reviews available, e.g., Refs.~\cite{Ribeiro2018polariton,feist2018polaritonic,herrera2020molecular} and references therein. In the following we assume that we can capture the observed effects by employing either the Hamiltonian of Eq.~\eqref{eq:paulifierzhamiltonian}, where we describe also the cavity as part of the system, or we can use the approximate Hamiltonian of Eq.~\eqref{eq:paulifierzlengthgauge}, where the cavity is taken into account by modifying the mode structure of the electromagnetic field. The presented results are then obtained by solving the Schr\"odinger-type equations with one of the above described first-principles methods (see Sec.~\ref{sec:firstprinciples}). We want to relate the various results with each other, but at the same time also highlight explicitly the underlying assumptions. Such questions of consistent assumptions turn out to be very important for various reasons as will become clear in the next sections. First of all, QED chemistry is a novel research discipline and many assumptions are still under debate and not yet generally accepted. Moreover, the strong coupling between light and matter can potentially invalidate accepted assumptions of theoretical chemistry, which were successfully applied for decades outside of photonic structures.
	In addition, the increased theoretical complexity of polaritonic chemistry includes many additional ingredients, which makes the choice of reasonable assumptions even more delicate. For example, in most applications we have to account for
	\begin{enumerate}
		\item the chemical complexity of the (individual) molecular system under study,
		
		\item the effect of non-zero temperature,
		
		\item potential chemical effects from the solvent in which the molecular system under study is contained,
		
		\item the self-consistent interaction with the restructured (quantized) electromagnetic field,
		
		\item the collective/cooperative effects due to an ensemble/solvent or by the photonic structure itself.
	\end{enumerate}
	Already without a photonic structure, when only points 1-3 are relevant, the complexity is staggering. Combining the first three points encompasses most issues describing reactivity in theoretical chemistry~\cite{szabo2012modern, haile1992molecular,cramer2013essentials}. Adding the last two points is the origin of the observed changes in chemical properties, but also the origin of even more theoretical complexity. In more detail, they can potentially change the basic ingredients of chemistry, as has been highlighted already in the introduction, which in turn also affect how points 1-3 combine. Let us try to unravel these aspects and their connections a little more from an ab initio perspective in the following.
	
	
	\subsection{Restructuring the electromagnetic field modes}    
	
	\label{subsec:restructuring}

	The first fundamentally new ingredient is that a photonic structure, e.g., an optical cavity or some plasmonic structure~\cite{skolnick1998strong,raimond2001manipulating,vahala2003optical,torma2014strong}, will modify locally the modes of the electromagnetic field from simple plane waves (see also Secs.~\ref{sec:lightandmatter} and \ref{subsec:approximatepaulifierz}) to more complex forms. Of course, this re-structuring is automatically contained in non-relativistic QED if we explicitly include the photonic structure as part of the physical system.

	A nice demonstration of this fact is found in, e.g., Ref.~\cite{jestadt2019light}, where the time-resolved field structure between plasmonic nanospheres is considered. It is also shown how longitudinal and transverse electromagnetic modes are modified at the same time for such very small cavities that are explicitly treated as part of the system (see also Sec.~\ref{subsec:quantizinglight} for the usual free-space distinction). It is no surprise that such near-field effects can have a strong influence on the properties and dynamics of molecules. Physically it is quite simple to understand that the (large) charge densities and currents of the nanospheres lead to a modified electromagnetic mode structure, and that the fluctuations of these charge densities and currents are connected to the fluctuations of the electromagnetic field inside the cavity. Abstracting further, the photon-field fluctuations can be understood as current-current correlators between the charged particles of the cavity and the molecules inside the cavity, in analogy to the arguments that can be made for the Casimir forces~\cite{jaffe2005casimir, buhmann2013dispersion}. This idea also underlies the theory of macroscopic QED, where the photon field fluctuations are expressed in terms of currents obtained from linear-response functions of the cavity material~\cite{buhmann2007dispersion,buhmann2013dispersion}. Furthermore, it is nice to observe that the local photon modes lead to strong radiative dissipation, since exciting them transfers energy from the near to the far field and this energy is effectively lost from the localized (cavity-molecule) system~\cite{jestadt2019light, flick2019light}. One should, however, be aware that strictly speaking the photonic structure does not really generate new photon modes, but the transient nature of the excitation in the cavity material rather leads to quasi modes~\cite{buhmann2007dispersion,buhmann2013dispersion,franke2019quantization,medina2021few}. So it is a theoretical abstraction/simplification to denote the cavity-induced local changes in the electromagnetic field as new modes.

	Keeping this cautionary note in mind, we will still use the (approximate) picture of changed electromagnetic modes due to a photonic structure in the following. This becomes specifically handy, when we want to unite various different physical situations where strong light-matter coupling appears. For instance, often strong coupling is not considered in a nanocavity but rather on a surface and the molecules couple to an evanescent wave, a surface plasmon-polariton, which itself is actually a light-matter hybrid state~\cite{torma2014strong}. Overall the strong-coupling effects in these different physical situations are quite similar~\cite{jestadt2019light,rossi2019strong, buhmann2013dispersion}, at least in a coarse-grained view~(see also Sec.~\ref{subsubsec:dipoleapproximation}). Now, putting one or a few molecules in contact with these modified local electromagnetic modes can have strong effects on the molecules. Such situations are commonly called single-molecule or \textbf{local strong coupling}. It is simple to accept that, for instance, plasmonic near-field modes, which (if excited) can generate very strong local fields, can transiently affect molecular properties or change chemical reactions~\cite{doi:10.1126/science.aao0872,cortes2017plasmonic,de2017plasmonic,schaeverbeke2019single}. An important point is that one does not need to excite these modes externally, but also at equilibrium they can have a strong influence. Indeed, the main interest in the following, as already highlighted as one of the main questions in polaritonic chemistry in the introduction, lies in the equilibrium fluctuations of these modes and their impact on molecular properties. These fluctuations can either be of quantum nature or of thermal nature as we will discuss in the next section.
	
	\subsubsection{Modified fluctuations and fields}

	Assuming that our coupled cavity-molecule system is completely isolated and we consider the coupled ground state (see also Sec.~\ref{subsec:paulifierzproperties} about the existence of ground states in non-relativistic QED), the fluctuations of these quasi modes inside the cavity are purely quantum in nature. If we then focus on the equilibrium molecule inside our photonic structure, any changes with respect to free-space equilibrium can then be attributed to the changed mode structure and its changed \textbf{vacuum fluctuations}. 
	Instead, if we start from an excited state, which then can decay due to (radiative and potentially also vibronic/phononic) dissipation, we expect to observe different dynamics due to the changed mode structure. Nevertheless, in this case the main driving force will be the induced non-zero electromagnetic (near) fields and not so much the coupling to the fluctuations~\cite{flick2015kohn,flick2019light, welakuh2021down,kuisma2022ultrastrong}. Certainly, the dominant mechanism will depend on the amount of energy transferred from the molecular system to the cavity modes. If we now bring our cavity-molecule system in contact with a thermal reservoir, the mode fluctuations will additionally get a thermal component. Depending on the temperature and the energy range of the cavity coupling, e.g., ro-vibrational, vibrational or electronic, the \textbf{thermal fluctuations} can dominate over the vacuum contributions. There is now, however, a simple but important point to be highlighted. While the thermal state of the total system is canonical, this is not necessarily the case anymore for the (nuclear/ionic) dynamics of the strongly-coupled molecular system inside the cavity. Similarly, the thermal cavity mode fluctuations can also be very different to the empty-cavity thermal fluctuations. Only in the limit of  weak coupling between light and matter, we can expect to reach a canonical state for the  molecular subsystem. Such effects have been observed for simple molecular systems coupled to a single cavity mode~\cite{sidler2022exact}.

	A different way to quantify cavity induced modifications is to measure the impact  on the basic molecular building blocks, i.e., on the electrons or on the nuclei/ions. For example, by measuring the dispersion relation and determining its curvature at zero momentum, one can determine the mass of the free particles~\cite{spohn2004dynamics,hainzl2002mass,rokaj2022free}. If the same measurement is performed inside a cavity, the photonic environment will alter the dispersion. In addition, the loss of isotropy in the cavity (think about mirrors that restrict the $x$ direction, as displayed in Fig.~\ref{fig:cavitysetup}) will imply that one has (slightly) different masses in different directions~\cite{spohn2004dynamics}. An example of such a \textbf{mass renormalization} can be found, e.g., in Ref.~\cite{rokaj2022free}. Notice that the mode restructuring can not only affect particle masses, but it can also imply that the longitudinal (Coulomb) interaction between the charged particles of the molecules gets modified (see also Sec.~\ref{subsec:approximatepaulifierz}).

	\subsubsection{Chemical consequences of cavity-restructured modes}
	
	What are now the chemical consequences due to the restructured photonic modes? Considering the impact on the \textbf{electronic sector} first, where (room) temperature effects are usually assumed negligible, one finds that the electronic \textbf{ground state} can get modified appreciably only for quite strong coupling, i.e., when the relevant modes correspond to large local fields if excited~\cite{flick2017atoms,flick2018ab,haugland2020coupled} (This does, however, not mean that perturbative/few-level calculations are correct for small changes, since these changes can strongly vary locally~\cite{schafer2018ab, buchholz2019reduced}). In the common dipole approximation of Eq.~\eqref{eq:paulifierzlengthgauge} this happens very roughly when for some modes $\alpha$ (or the sum of all the enhanced modes) we have $g_{\alpha}^2 \Delta l^2 |e|^2$ comparable to the free-space Coulomb interaction, where $\Delta l$ is the relevant (microscopic) length scale of the localized quantum system~\cite{schaefer2020relevance}. It has to be highlighted (see also Sec.~\ref{subsec:approximatepaulifierz}) that for the combined ground state of the cavity-molecule system no real (propagating) fields are generated but the mode occupation is virtual, i.e., it is the vacuum fluctuations of these modes that lead to changes. The hybridized nature of the ground state in a cavity can not only modify the energy or the ionization potential~\cite{deprince2021cavity,riso2022characteristic} but also the electronic density of the ground state~\cite{flick2018ab,buchholz2019reduced,buchholz2020light,haugland2020coupled}. For a fixed coupling strength the magnitude of these effects also depends on the position of the (clamped) nuclei/ions, i.e., since the relevant lengths scale $\Delta l$ from above is also modified. For instance, if dissociating molecules are considered, a cavity mode can lead to strong effects due to novel long-range correlations~\cite{schafer2019modification} and it can modify Van der Waals interactions substantially~\cite{haugland2021intermolecular}.	
	\begin{figure}
		\begin{center}
			\includegraphics[width=\linewidth]{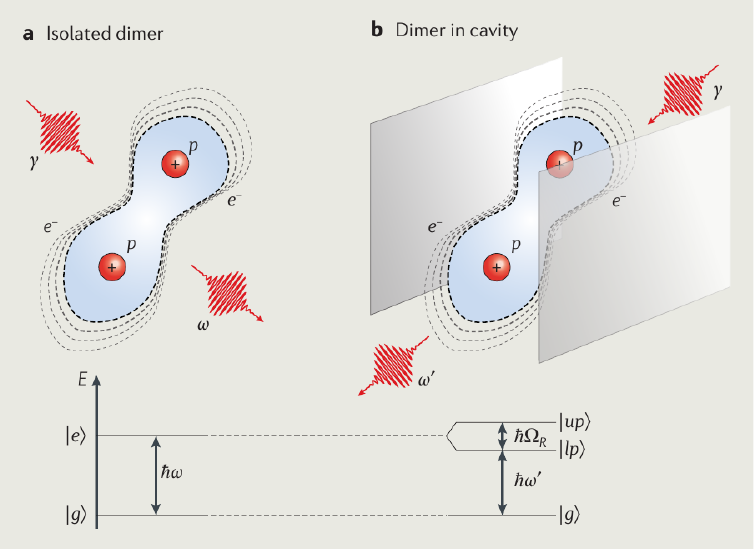}
			\caption{A free-space molecule (a) has specific electronic transitions of frequency $\omega$ from its ground state $\ket{g}$ to some excited state $\ket{e}$. These transitions show up in an absorption (or emission) spectrum where some external probe pulse $\gamma$ interacts with the free-space molecule. If the molecule is placed inside a Fabry-P\'erot cavity (b) with the same resonance frequency $\omega$, one observes that the two degenerate (matter and photon) excitations turn into an avoided crossing. This is due to the coupling between light and matter and instead of one peak one finds now two peaks, i.e., the upper $\ket{\rm up}$ and lower $\ket{\rm lp}$ polaritons, which are split by the Rabi frequency $\Omega_{\rm R}$. From the simple Jaynes-Cummings (for a single molecule) or the Tavis-Cummings (many identical molecules) model (see end of Sec.~\ref{subsec:approximatepaulifierz}) one infers  that the vacuum Rabi splitting depends inversely on the volume of the Fabry-P\'erot cavity, is proportional to the dipole matrix element of the individual molecules and scales with the square root of the number of molecules as well as photons. Reproduced with permission from Ref.~\cite{ruggenthaler2018quantum}, Copyright 2018 Springer Nature.} 
			\label{fig:polaritonsplitting}
		\end{center}
	\end{figure}
	For \textbf{time-dependent and excited state properties} modifications can be observed already for much smaller couplings compared to ground state effects. Notice that (time-dependent) excitations typically also imply a further delocalization with respect to the ground state. For example,  electronic (usually vacuum) Rabi splitting, the hallmark of strong coupling (see Fig.~\ref{fig:polaritonsplitting}), can usually already be observed for coupling regimes where the electronic ground state still remains unaffected~\cite{frisk2019ultrastrong, flick2017atoms, flick2019light}. In most cases, the calculated Rabi splitting shows an asymmetric behaviour~\cite{flick2019light,welakuh2022frequency,yang2021quantum} and one also recovers the super-/sub-radiant features (radiative lifetime is shorter/longer than free-space counterpart) of these polaritonic states~\cite{flick2019light,wang2021light,welakuh2022frequency}. This is a nice consistency check with respect to experimental evidence. To include the radiative losses these time-dependent simulations either need to take into account the continuum of modes for a specific environment~\cite{flick2019light,wang2021light,welakuh2022frequency} (see also Sec.~\ref{subsec:approximatepaulifierz}), consider time-propagation that are shorter than the dephasing times~\cite{schafer2019modification, welakuh2021down} or explicitly include dissipation phenomenologically~\cite{antoniou2020role,felicetti2020uracil,davidsson2020lindblad,schafer2022shortcut}. Specifically interesting for chemistry is the appearance of new non-adiabatic couplings between (excited) electronic surfaces and novel \textbf{conical intersections}~\cite{szodarovsky2018conical,csehi2019nonadiabatic,farag2021polariton,fabri2021born} (see also Sec.~\ref{subsec:photoniondynamics}). 
	
\begin{figure*}
		\begin{center}
			\includegraphics[width=1.0\textwidth]{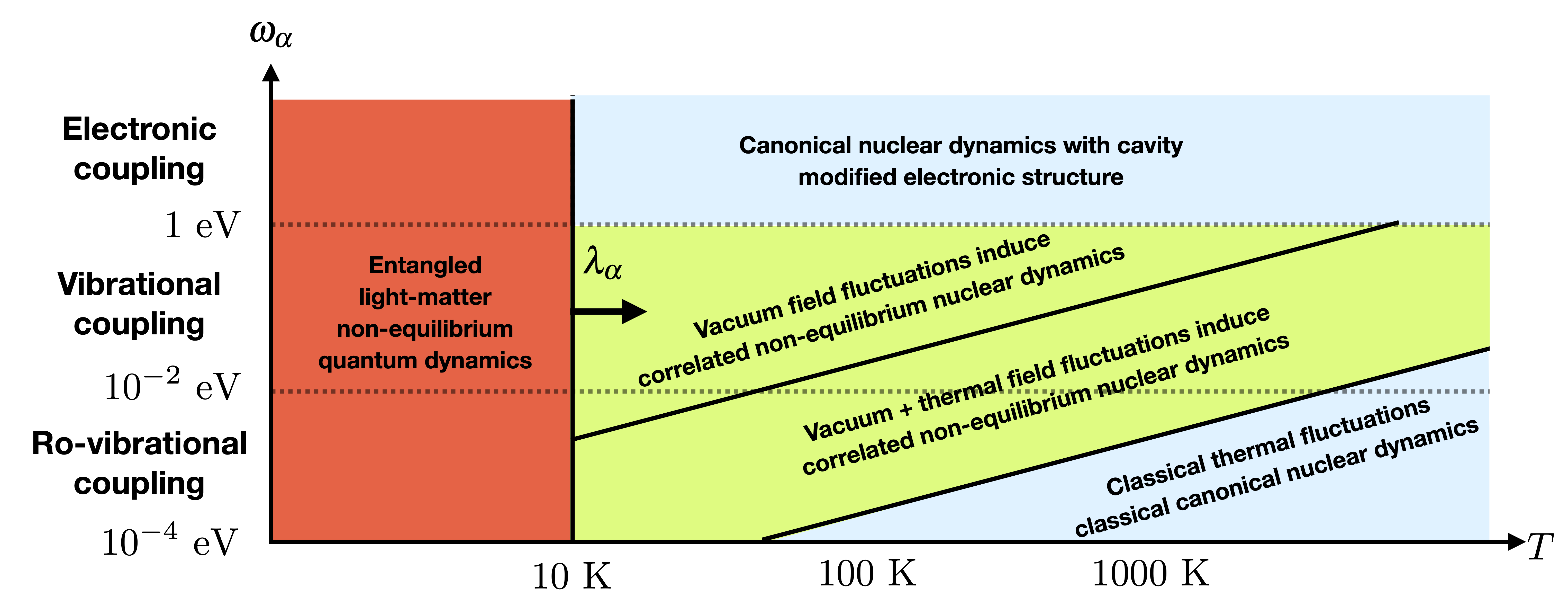}
			\caption{Pictorial sketch of distinguishable thermal (non)-equilibrium regimes emergent under different molecular strong coupling conditions in a cavity. They are inferred from exact quantum thermal equilibrium simulations for one HD$^+$ molecule coupled to a single cavity mode.\cite{sidler2022exact} In more detail, the exact results suggest three different regimes for the dynamics of the nuclei: First,  light and matter remains quantum entangled at low cryogenic temperatures (red). Second, the light-matter entanglement is quickly lost with increasing temperature, however, the field fluctuations remain governed by quantum laws (vacuum and thermal fluctuations), which can drive the nuclei out of classical canonical equilibrium. Third, either at very high temperatures or for electronic strong coupling no direct impact on the nuclear dynamics is expected, which implies that standard canonical equilibrium conditions are preserved (blue). Reproduced with permission from Ref.~\cite{sidler2022exact}.} 
			\label{fig:thermal}
		\end{center}
	\end{figure*}

	For the \textbf{rotational and vibrational} degrees of freedom the effects of (room) temperature can become decisive to describe chemistry. In this case the (energetically) relevant modes of the photonic structure might have a non-negligible thermal occupation. For photo-chemical reactions these modified thermal fluctuations might typically be less important then the new cavity-induced non-adiabatic couplings and conical intersections, but in general the interplay of these cavity-induced effects will alter chemical properties~\cite{sidler2020chemistry,li2021collective,Li2022energy,sun2022suppression,wang2022cavity,wang2022chemical,sidler2022exact}. Notice, however, that  one can observe already very interesting changes in simple photo-chemical reactions, even when disregarding these thermal contributions~\cite{triana2018entangled, antoniou2020role,gudem2021controlling,kowalewski2016nonadiabatic,fregoni2021strong,torres2021photo, schaefer2022polaritonic,Riso2022molecular}. On the other hand, the modification of the thermal fluctuations are expected to be specifically important for ground-state chemical reactions~\cite{ebbesen2016, sidler2022perspective} and many other phenomena of materials in cavities (e.g. quantum phase transitions~\cite{ashida2020quantum,latini2021ferroelectric}). We will discuss this issue in Sec.~\ref{subsec:chemicalreaction} in more detail for a specific case. For the generic situation we want to highlight that the common simplification to describe classically the thermal fluctuations of the (relatively heavy) nuclei/ions is not necessarily appropriate for the fluctuations of the modes even at ambient conditions (depending on the chosen cavity frequency). The mode fluctuations can still have strong non-classical contributions of vacuum and quantum thermal nature~\cite{sidler2022exact} (see also Fig.~\ref{fig:thermal}).
	
	
	\subsection{Collectivity and cooperativity}
	\label{subsec:collectivity}
	
	The second fundamentally novel aspect (as also highlighted in the introduction) that becomes decisive inside a photonic structure is that the cavity can facilitate strong \textbf{collective} or \textbf{cooperative} effects. "Collective" here means that similar physical entities, e.g., the same type of molecules, start to interact with each other via the cavity and potentially synchronize, while "cooperative" means that such a cross talking happens between different physical entities, e.g., solute and solvent. Strictly speaking, any effect that we observe is cooperative, due to the cavity being a different physical entity than the material inside, but this distinction inside the photonic structure is common~\cite{hirai_review, garcia2021manipulating}.  
	
	In order to construct cavities that have a particular strong coupling to molecules, it is often helpful to further fill the cavity with a highly polarizable medium~\cite{ebbesen2016,kena2016polaritonic,schneider20182Dpolaritons,herrera2020molecular,nitzan2022polaritons}. Indeed, in many cases of QED chemistry one simply employs the molecules of interest themselves to increase the coupling effect~\cite{ebbesen2016,kena2016polaritonic,herrera2020molecular,nitzan2022polaritons}. Clearly this cannot be done ad infinitum, since even in gas phase the molecules get densely packed at one point and loose their individual character and hence will respond very differently. Ab initio QED simulations can nicely reproduce this behavior and recover the well-known $\sqrt{N_{mol}}$ increase of the vacuum Rabi-splitting by the number of coupled molecules $N_{mol}$~\cite{sidler2021collective,li2021collective,schafer2022shortcut} (see also Fig.~\ref{fig:polaritonsplitting}). As we will also highlight later, this does not necessitate quantum coherence between the different molecules though. Such collective effects are not only observed for the excited states but also for the collective ground state of molecules~\cite{haugland2021intermolecular}. One of the interesting aspects of the collective coupling situation is the appearance of \textbf{dark states}. That is, the ensemble of molecules can attain a collective state which does not couple to external (dipole) radiation and hence is "dark" for absorption spectra~\cite{Ribeiro2018polariton,sidler2021collective,du2022dark}. These states will only be thermally populated and can modify the relaxation dynamics or they can also act as a thermal reservoir for the coupled "bright" collective states~\cite{ulusoy2019nonradiative,du2022dark,csehi2022competition,cederbaum2022cooperative}. An alternative approach to modify chemistry collectively is by resonantly tuning on the solvent (or highly polarizable plasmonic structures) which yields a density-dependent Rabi splitting with respect to the solvent concentration~\cite{garcia2021manipulating,schutz2020collective,Li2022energy}. The difference is that one hopes that the strongly-coupled solvent either induces strong single-molecule coupling to the solute~\cite{sidler2021collective,schutz2020collective} or that the cooperative behavior of the solvent leads in some other way to observable changes in the solute. To describe theoretically the mesoscopic amount of molecules that is present in experimental ensembles, one usually needs to make some further approximations, e.g., that the molecules (assumed in gas phase) only couple with each other via the cavity in a semi-classical way~\cite{flick2019light,sidler2021collective}. In this way, first-principle simulations are able to also consider the macroscopic limit~\cite{schafer2022shortcut}.
	
	There are now two important observations to be made for collective and cooperative effects. Firstly, once a molecule out of the ensemble is slightly modified, e.g., due to the onset of a chemical reaction, the distinction between collectivity and cooperativity even inside the cavity becomes fuzzy again. Indeed, ab initio simulations have shown that a collectively-coupled ensemble induces strong single-molecule effects on a modified molecule similar to cooperative strong coupling~\cite{sidler2021collective}. Secondly, for phenomenological models it is often argued that the collective effects are quantum in nature and that a robust and collectively  delocalized (over a mesoscopic amount of molecules) polaritonic quantum state is generated~\cite{feist2018polaritonic,Ribeiro2018polariton}(see Sec.~\ref{subsec:chemicalreaction} for more details). From an ab initio perspective a mesoscopic quantum collective mechanism does not seem to be necessary and in certain cases it even becomes  problematic. For instance, the response of the collective system, together with the dark state configurations, can be captured purely semi-classically~\cite{herrera2020molecular,flick2019light, sidler2021collective}. Therefore, the term "state of the ensemble" does not need to imply a quantum state, since also the response of classical dipoles will show such configurations. Furthermore, it has been shown that the quantum entanglement between light and matter vanishes rapidly above zero degrees Kelvin even for simple molecular systems~\cite{sidler2022exact}. On the other hand, at least for the common long wavelength approximations, the light-matter Hamiltonian is not size-extensive~\cite{schafer2019modification,haugland2021intermolecular}. That is, the more molecules are fully quantum coherently coupled, the stronger the effect of the modes becomes (even if these molecules are arbitrarily far apart). As a simple consequence of this, the cavity modes would be strongly blue shifted from the alleged mesoscopic amount of quantum-coherently coupled molecules, which is, however, not observed in experiment~\cite{sidler2022perspective} (see also Sec.~\ref{subsec:chemicalreaction} for an example).
	
	To conclude, ab initio approaches provide access to collective and cooperative coupling regimes and they reproduce the well-known effects from phenomenological models. However, at the same time ab initio results suggest rather a semi-classical mechanism than a fully quantum collective/cooperative origin of the experimentally observed effects at ambient conditions.

	\subsubsection{Chemical consequences of collective coupling}
	
	With these caveats in mind we can ask what chemical consequences can be expected that originate from collectivity or cooperativity? First of all, essentially all previously mentioned effects in Sec.~\ref{subsec:restructuring} can in principle arise (and even be collectively enhanced), since the coupled ensembles can mediate single-molecule strong coupling. However, we can now find additional, non-trivial modifications that emerge specifically due to having ensembles with a large number of molecules. Such effects include, for instance, ensemble-induced changes in lifetimes~\cite{groenhof2019polariton,tichauer2021multiscale,cederbaum2022cooperative}, dark-state-influenced relaxation dynamics~\cite{du2022dark,Li2022energy}, modified inter-molecular interactions~\cite{haugland2021intermolecular,philbin2022fluids} and enhanced transport properties~\cite{Du2018transfer,mauro2021charge,sokolovskii2022transfer}. In addition, how an ensemble changes local molecular properties can have a non-trivial dependence on the number of molecules in the ensemble~\cite{schafer2022shortcut}. Of course, the probably most relevant effect for chemical applications will be the site/bond selective modifications and control of chemical reactions in an ensemble of molecules without external driving, i.e. in thermal equilibrium~\cite{ebbesen2016,sidler2022perspective}. We note that chemistry is local, i.e. the electronic and nuclear structure is modified on a single-molecule or nearest-neighbour level. However, in the case of collective/cooperative strong coupling this prevalent paradigm is challenged, since chemical reactions seemingly become dependent on the total ensemble. For example, \textit{a priori} it is unclear if the reaction mechanism in a cavity is altered due to a quantum-collective state, cavity-mediated inter-molecular interactions, cavity-modified thermal fluctuations or single-molecule strong-coupling effects. To quantify the extend and origin of these modifications is currently one of the main goals of QED chemistry. This understanding will allow to reach a qualitative and quantitative theoretical understanding, and accurate predictions become feasible that can significantly advance experiments and applications of polaritonic chemistry.  
	
	
	\subsection{Cavity-modified chemical reactions}
	
	\label{subsec:chemicalreaction}
	
	As pointed out before, the cavity induced contributions to the chemical complexity offer many tantalizing opportunities, yet make a detailed understanding even more challenging. Their additional interplay with (single-molecule) symmetries~\cite{garcia2021manipulating,riso2022chiral} and external probes~\cite{Li2022energy} is just getting explored and might lead to further very interesting effects. Let us next focus on a specific experiment to reduce the immense amount of possibilities, and thus complexity. This paradigmatic example will highlight how ab initio theory can help to unravel the main mechanisms of cavity-modified chemistry.
	\begin{figure*}
		\begin{center}
			\includegraphics[width=1.0\textwidth]{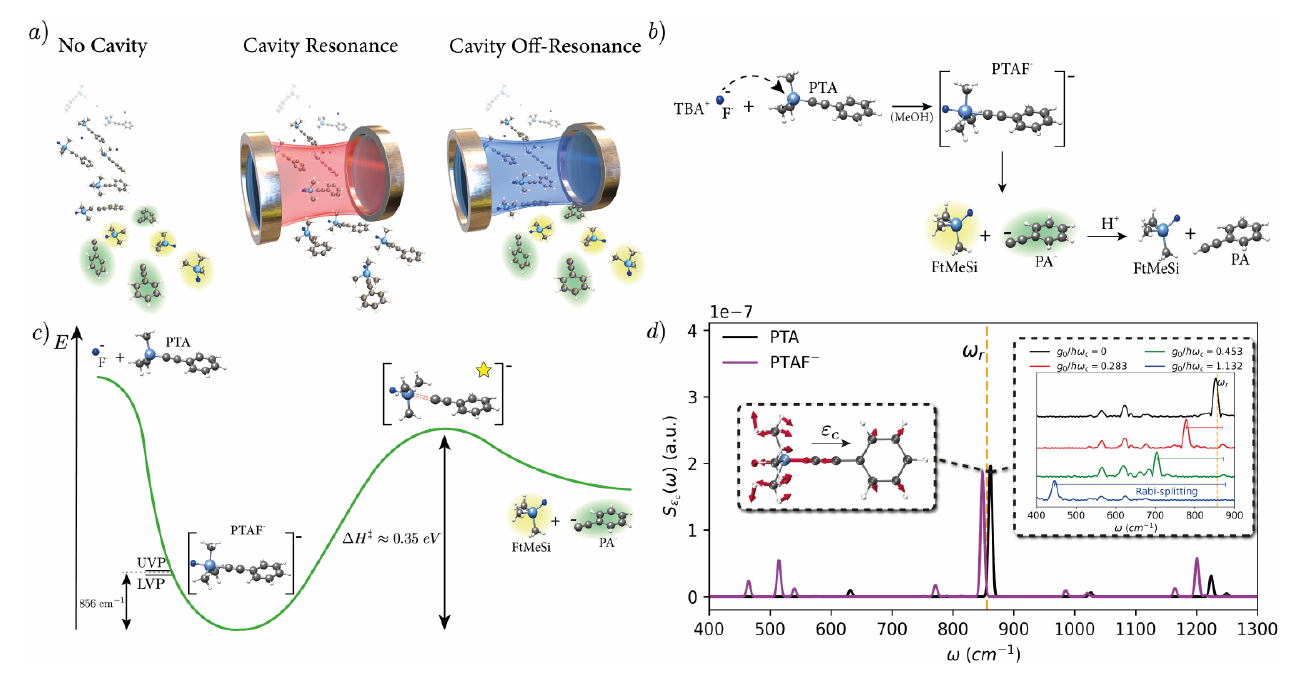}
			\caption{(a) Resonant vibrational strong-coupling can inhibit chemical reactions. (b) Illustration of the reaction mechanism for the deprotection of 1-phenyl-2-trimethylsilylacetylene (PTA), with tetra-n-butylammonium fluoride (TBAF) and (c) energetic of the reaction in (b) in free-space. The successful reaction involves breaking the Si-C bond and thus overcoming a transition-state barrier of 0.35 eV. (d) Vibrational absorption spectrum along the cavity polarization direction illustrating the strong-coupling of the vibrational eigenmode at 856 cm$^{-1}$ with the cavity polarized along for PTAF- (magenta) and the isolated PTA complex (black). The insets show the coupled vibrational mode of PTA and the light-matter hybridization under vibrational strong-coupling. Reproduced with permission from Ref.~\cite{schafer2021shining}}
			\label{fig:experiment}
		\end{center}
	\end{figure*}
	The seminal experiment that we consider in the following is the ground-state deprotection reaction of 1-phenyl-2-trimethylsilylacetylene (PTA) under vibrational strong coupling~\cite{Thomas2016ground}. The PTA molecules are mixed with tetra-n-butylammonium fluoride (TBAF) in methanol. In the ensuing deprotection reaction fluoride ions released from TBAF interact with PTA, forming an intermediate complex, which makes the breaking of the Si-C bond in the PTA molecule more likely (see Fig.~\ref{fig:experiment}). The Fabry-P\'erot cavity is then set on resonance with the Si-C stretching modes at roughly 856 cm$^{-1}$. It is important to note that the cavity is not pumped except of the thermal effects due to ambient conditions. To verify the vibrational strong coupling condition, the transmission spectrum is observed and shows a large Rabi splitting. Eventually, one finds that the deprotection reaction rate is strongly suppressed for the non-pumped resonantly-coupled system, when compared to free-space or off-resonant coupling. The measured suppression is also strongly dependent on the temperature of the total system, such that from fitting simple equilibrium rate models even a qualitative change in the transitions state would be predicted. This observation was interpreted as potential evidence for cavity-induced non-equilibrium effects. For the further interpretation of this experiment one important remark has to be made: The vacuum Rabi splitting of the mixture of molecules, which is the usual way to identify strong coupling situations, depends on the density of PTA molecules and their products alike, since both contain the same Si-C stretching modes. Therefore the Rabi splitting stays constant throughout the minutes-long reaction and is a self-adapting mixture of collectivity and cooperativity. 
	
	Most of the phenomenological interpretations of this experimental result follow a two-step procedure. Firstly, a perfect ensemble of aligned PTA molecules in gas phase with a small single-molecule coupling constant (obtained from the coupling constant of the empty Fabry-Perot cavity) is assumed. Then, by assuming zero temperature, the Dicke or Tavis-Cummings model (see Sec.~\ref{subsec:approximatepaulifierz} for all the other assumptions that go into this model) is used to determine the number of (two-level) molecules that are quantum-collectively coupled to the vacuum of the cavity mode. Based on these assumptions, the phenomenological fit suggests a mesoscopic  number of quantum-collectively coupled molecules on the order of $10^9$ molecules~\cite{galego2015cavity,martinez2018can}. In a second step, concepts of quantum chemistry are applied on this \textit{quantum collective state}, i.e., assuming that a single collective "super-molecule" is formed with many dark states~\cite{feist2018polaritonic,Ribeiro2018polariton}. In contrast to usual quantum chemistry, where only a single molecule and its potential energy surface is considered, the "super-molecule" has now a potential energy surface that is formed by the $10^9$ molecules (for each molecule reduced to the main free-space reaction-coordinate) plus the single-excitation subspace of the cavity mode~\cite{galego2015cavity,feist2018polaritonic}. This new humongous potential energy surface is then assumed to change the chemistry, since now all molecules move in a concerted motion and no longer statistically independently~\cite{galego2017many}. However, this phenomenological combination of quantum optics and quantum chemistry concepts cannot explain (even qualitatively) the experimentally observed findings~\cite{martinez2018can,climent2020sn,climent2021reply}. Moreover, from a rigorous theoretical perspective even a single quantum-mechanical molecule would never attain, e.g., a permanent dipole moment or specific internal structures without coupling to the environment~\cite{anderson1972more}. It is the interaction with the environment that leads to a specific realization of the molecular structure, e.g., a certain orientation of the pyramid of the NH$_3$ molecule. In a "super-molecule" all these (exactly similar) realization of the individual molecular structures are assumed to happen simultaneously and fully quantum-coherently due to coupling to the cavity even at ambient conditions. 
	
	Can now ab initio polaritonic chemistry help to understand this stark discrepancy between theory (based on a simplified model calculation) and experiment and maybe hint at a potential mechanism? Let us first fix the basic level of theory that we deem sufficient and computationally feasible to investigate the PTA experiment theoretically.  We assume that the dipole-coupled Hamiltonian of Eq.~\eqref{eq:paulifierzlengthgauge} with one effective mode and the physical masses of the particles (see Sec.~\ref{subsec:approximatepaulifierz} for more details) is a sufficient framework for describing polaritonic chemistry in a Fabry-P\'erot cavity. From the chosen Hamiltonian, the standard Hamiltonian of quantum chemistry can directly be recovered for zero coupling strength. In addition, the quantum-optical Dicke and Tavis-Cummings model can also be deduced from it. After having made this theory choice, we immediately realize that a fundamental inconsistency arises with the alleged number of quantum-collectively coupled molecules, which are suggested by the Dicke or Tavis-Cummings model (see also Ref.~\cite{tserkezis2020applicability} for related problems with phenomenological models in plasmonic cavities). Not surprisingly, the matter inside the cavity modifies the frequency of the cavity mode, which will be  accounted for in the Pauli-Fierz theory. This means the enhanced refractive index of the filled cavity will shift the bare (empty cavity) frequency towards smaller wave numbers~\cite{Thomas2016ground,schafer2021shining,sidler2022perspective}. However, the assumption of $10^9$ quantum-collectively coupled molecules would lead to a diamagnetic shift of the cavity frequencies, which is an order of magnitude larger than the experimentally observed frequency of 856 cm$^{-1}$. This discrepancy suggests that at the Pauli-Fierz level of theory, we need to restrict quantum coherence to a much smaller length scale (closer to the common understanding of chemistry as being local) and potential collectivity/cooperativity effects on a macroscopic scale will rather be semi-classical in nature. This seems reasonable since the amount of degrees of freedom (translational, rotational, vibrational and electronic) which can lead to decoherence in a real chemical system in solvation is so breathtaking that a quantum coherence at ambient conditions over large distances seems implausible in practice. 
	
	An alternative interpretation arises if one keeps in mind that the observed Rabi splitting is not an absolute, but rather a statistical quantity, i.e., not all molecules contribute with the same amount~\cite{ebbesen2016}. Therefore,  there is no reason to assume that all molecules experience the coupling to the cavity mode in the same way. Indeed, as discussed above, single molecules can experience strong local coupling effects in a collectively/cooperatively-coupled environment~\cite{schutz2020collective,sidler2021collective}. Note again that in the experiment the (constant) Rabi splitting is by construction a mixture of collectivity and cooperativity. Consequently, it seems plausible that a fraction of the PTA molecules in the cavity could feel strong single-molecule effects, specifically in the case that they undergo a chemical reaction. Taking into account that chemical reactions are rare events and that the likelihood of these events is determined by the temperature, this fraction can become decisive for the observed rate change. This setting suggests that the cooperative/collective coupling can effectively be interpreted in terms of a highly-polarizable and strongly frequency-dependent medium in the vicinity of a reacting PTA molecule. Based on this (simplified) ab initio picture, recent QEDFT simulations were able to reproduce the experimental PTA results qualitatively~\cite{schafer2021shining} and they could also reproduce other predictions in connection to solvent effects~\cite{Li2021cavity}. Overall, these simulations suggest that the cavity can correlate various intra-molecular vibrational modes and hence can transfer energy from the bond-breaking stretching modes to other internal motions, thus effectively strengthening the Si-C bond in the PTA experiment. This indicates that restricting to the main cavity-free degree of freedom of a potential energy surface in vibrational strong coupling simulations could miss important contributions (see discussion in Sec.~\ref{subsec:photoniondynamics}). 
	
	Of course, this simple local model, which infers a frequency-dependent polarizable environment from the collective/cooperative ensemble, is not the end of the story. The ab initio simulations also suggest -- again in agreement with the original interpretation of the experiment -- that the cavity might induce non-equilibrium effects. In the context of chemical reactions, this means that the nuclear/ionic system might follow a non-canonical (classical) thermal distribution. In contrast, for the uncoupled, bare matter system the thermal state is usually well-described by a classical canonical distribution. Non-equilibrium dynamics for the coupled matter system is not surprising, since it is a strongly-coupled subsystem, i.e., tracing out the cavity degrees of freedom will usually induce a non-canonical/non-stationary distribution for the subsystem. However, what might be more exceptional is that even for ambient conditions it is not correct to treat the cavity degrees of freedom (particularly the fluctuations) purely classically and assume that the thermal fluctuations are uncorrelated~\cite{sidler2022perspective,sidler2022exact} (see also Fig.~\ref{fig:thermal}). Furthermore, it has been argued that such non-canonical dynamics of classical particles (nuclei/ions) can lead to stochastic resonances~\cite{sidler2022perspective}, which could explain on the ensemble level, why the experiment sees a strong frequency dependence (resonance effect) in the polaritonic reaction rates, without any external periodic driving. At the same time, stochastic resonances are quite delicate and they seem to arise only under very special conditions~\cite{Gang1993stochastic,Gammaitoni1998stochastic}. This could also rationalize why in many experimental situations of strong coupling no changes in chemical properties could be observed~\cite{imperatore2021reproducibility,wiesehan2021negligible}. Therefore, it might not only be intra-molecular re-distribution of vibrational energy that stiffens the Si-C bond, but on resonance one might also find effective inter-molecular energy re-distribution. Indeed, recent ab initio results suggest that inter-molecular forces could be efficiently altered by a cavity~\cite{schafer2019modification,haugland2021intermolecular}. 
	
	All in all, ab initio QED suggests a more nuanced interpretation of the seminal PTA experiment under vibrational strong coupling, i.e., a delicate interplay of local (potentially quantum) effects with collective/cooperative semi-classical effects, which lead to non-canonical thermal distributions. The advantage of this perspective is that it naturally connects to the usual understanding of chemical reactions as a macroscopically statistical process, whose parameters are determined by the microscopic quantum description on the single-molecule level. We note that a similar perspective as originally proposed based on ab initio results (effective local theory, non-canonical equilibrium and intra/inter-molecular energy redsitribution)~\cite{sidler2022perspective} has recently also been promoted based on experimental~\cite{Chen2022cavity,Ahn2022modification} and quantum-optical results~\cite{perez2022collective}. At the same time, ab initio QED also connects directly to the quantum-optical perspective for the photonic quantities. Therefore, if indeed subtle details of the light field, such as the exact spatial form of the cavity modes and their intrinsic lifetimes are important, these details can be re-introduced in a straightforward way.

	
	\section{Conclusion and outlook} 
	\label{sec:outlook}
	
	\textit{"As more researchers
		enter the field, influx of new viewpoints will ensure rapid
		development of polaritonic chemistry concepts and further
		pioneering cross-disciplinary breakthroughs."}
	
	\noindent
	\\
	K. Hirai in Ref.~\cite{hirai_review}
	\\
	\\  
	
	If you followed this review chronologically then it has been a real tour-de-force. It encompasses very basic considerations of relativistic quantum physics (how relativity, symmetries and spin lead to the Maxwell equations and their coupling to matter) in Sec.~\ref{sec:lightandmatter}, the basic Hamiltonian of non-relativistic QED (properties and potential approximations) in Sec.~\ref{sec:PauliFierz}, ab initio QED methods in Sec.~\ref{sec:firstprinciples}, and their  applications on relevant research questions of polaritonic chemistry in Sec.~\ref{sec:polaritonicchemistry}. Clearly, many of the details that were highlighted might not be relevant for a specific experiment in QED chemistry, where the re-structuring of the local electromagnetic modes can modify chemical properties. However, as highlighted in the introduction, in the absence of established simple mechanistic rules for polaritonic chemistry, which challenges the locality assumption prevalent in common chemistry, a re-evaluation of all the intrinsic assumptions in our theoretical modeling is needed. This hopefully helps to select among the existing phenomenological models and combinations of (quantum) optics and (quantum) chemistry approaches the most reliable ones and allows to develop more accurate phenomenological models in the future, in order to get an intuitive understanding of the relevant mechanisms in polaritonic chemistry.
	
	Let us repeat in this context the main aspects of the different sections and their answers to the main questions raised in the introduction, i.e., the basic Hamiltonian, the choice of gauge, the implications of the dipole approximation, cavity-induced changes in vacuum and thermal fluctuations as well as the interplay of local and collective strong coupling. In Sec.~\ref{sec:lightandmatter} we have shown how the light and matter sectors follow from the same basic principles and need to be treated consistently, especially when they interact. On the most basic level one cannot even distinguish between light and matter degrees. Changing one sector can have a strong influence on the other and might even break basic physical principles. This should serve as a guidance on how to carefully re-combing on a phenomenological level theoretical methods describing matter, e.g., quantum chemistry methods, and theoretical tools for photons, e.g., quantum optics methods. In Sec.~\ref{sec:PauliFierz} we have presented the basic Hamiltonians of non-relativistic QED, which form the basis of a consistent and non-perturbative theory of light and matter. We highlighted that the Pauli-Fierz Hamiltonian guarantees the stability of matter, that excited states turn into resonances with finite lifetimes, and that we have to work with bare masses of the charged particles. Furthermore we have discussed that the Coulomb gauge is the natural gauge to work in (at least on the wave-function level), because it guarantees internal consistency between quantum mechanics and quantum optics, and that only in the dipole-coupling limit we can easily replace a photonic structure by a local modification of modes. We have also spelled out the various assumptions (extension of localized matter small when compared to the cavity wavelengths, single charge center to have indistinguishability, small enough frequency cutoff to avoid non-renormalizability, linear and quadratic coupling terms to have stable theory) that go into the dipole approximation. In Sec.~\ref{sec:firstprinciples} we have highlighted the necessity of first principle methods to be able to cope in an unambiguous way with the humongous amount of degrees of freedom (photonic, electronic and nuclear/ionic) of a realistic coupled light-matter system. Depending on the specific question and/or coupled systems, different theoretical methods become more appropriate than others (e.g. QEDFT as a general-purpose approach, QED-CC methods for electronic strong coupling or the cavity Born-Oppenheimer partitioning for cavity-modified ground state chemical reactions). In Sec.~\ref{sec:polaritonicchemistry} we have discussed polaritonic chemistry from an ab initio QED perspective. We have highlighted the two main differences to chemistry outside of cavities, i.e., the self-consistent interaction with the restructured (quantized) electromagnetic field and collective/cooperative effects due to an ensemble/solvent, and presented several results obtained with various ab initio QED methods. Based on these results we have argued for the importance of modified quantum/thermal fluctuations that can induce non-canonical equilibrium conditions for the matter subsystem. Furthermore, simulations suggest that collective ensemble/solvent effects are mainly semi-classical and can be approximated as a frequency-dependent modification of the local polarizable medium. We have then scrutinized a paradigmatic experiment in polaritonic chemistry and found that an interplay of single-molecule coupling with semi-classical non-equilibrium effects is indeed a natural explanation for the observed changes in chemical reactions.  
	
	Clearly, we cannot yet provide a general and universally accepted answer for all the microscopic mechanisms at work when chemical properties are changed by a photonic structure. Many possible other effects, which have not been taken into account in the different ab initio simulations, might be important as well. The most obvious shortcoming is that a macroscopic ensemble of molecules inside a cavity cannot be simulated on a full ab initio level. However, several recent developments~\cite{schafer2022shortcut} make it possible to also get approximate results for the macroscopic case. Yet we believe that statistical and thermal effects dominate on a macroscopic scale at ambient conditions, in analogy to chemistry outside of cavities. Therefore, a semi-classical description should be appropriate to recover the observed effects. This suggest that it will be paramount to develop adapted statistical methods in the future that can faithfully include the contributions of the (quantized) cavity mode. A further issue that is often disregarded for simplicity, is the effect of the solvent on chemical properties. Indeed, there are recent experiments~\cite{garcia2021manipulating,hirai_review} which show that strongly-coupled solvents can have different effects on chemical processes than their uncoupled counterparts. Although this fits into the simplified picture of collective/cooperative coupling as a frequency dependent polarizable surrounding for molecules, actual solvent effects can be much more intricate. Another obvious shortcoming of most considerations so far is the simplified treatment of the cavity as an effective single- or few-mode structure. Specifically for nanocavities, where a few molecules couple to plasmonic excitations, a detailed treatment of the cavity as an active physical entity, which can efficiently dissipate energy, might become crucial. A further aspect that might become important for the specific design of chemical properties is to go beyond the dipole-coupling approximation in our theoretical description. Dipole coupling implies that no momentum is transferred between light and matter and also that we loose locality (at least approximate for non-relativistic particles) and no retardation effects are included. Beyond-dipole contributions can become specifically important once we take into account the exact structure of the modes of a cavity, e.g., when we couple strongly to chiral (circularly polarized) light modes. To disentangle which details are important calls for a combined theoretical and experimental effort. Besides theoretical developments on the ab initio and the model side, new experimental setups and observables need to be identified to unravel the influence of the above highlighted issues. It is clear that merely considering the Rabi splitting is not enough to understand the mechanisms at work and spatially as well as temporally resolved experimental investigations are key for the future development of QED chemistry.
	
	While this list of extra complications might seem like spelling doom for a comprehensive understanding of strongly-coupled light-matter systems, it at the same time opens the door for many still to be discovered chemical effects. We hope that this extensive review does highlight where seemingly small changes in the photonic environment might lead to novel effects. Take, for example, the quantization of the electromagnetic degrees of freedom in Sec.~\ref{sec:lightandmatter}. By following the Riemann-Silberstein approach we saw that it is quite natural to quantize the free vacuum in terms of chiral modes. Based on this perspective it seems possible to use photonic structures to suppress one of the two naturally occurring chiralities such that one can manufacture chiral opical cavities~\cite{hubener2021engineering,genet2022chiral,voronin2022single}. Indeed, recent experimental efforts have demonstrated that the engineering of chiral photonic structures is possible, which brings enantiomeric polaritonic chemistry within reach. Hence, one can hope for enantiomer-selective catalysis controlled by optical cavities. Such enantiomeric reactors would, for example, be a great asset for the efficient synthesis of drug molecules. Thinking one step further: Being able to engineer symmetries of the electromagnetic modes inside a cavity might allow to circumvent common excitation selection rules and steer chemical reactions into completely new directions based on breaking or enhancing intrinsic molecular symmetries. Even more fundamentally, we might be able to engineer the properties of the basic molecular building blocks directly. As we have seen, the vacuum determines the physical masses of electrons and ions (and if we consider full QED also other basic properties~\cite{Greiner_1996, greiner2003quantum, ryder_1996}) and thus how atoms and molecules form and combine. Until now, the influence of the resulting (not necessarily scalar~\cite{spohn2004dynamics}) photonic mass on chemical properties and the potential of inducing relativistic effects remains largely unexplored for molecules. In contrast, such engineering of the photon vacuum is actively being explored in solid-state physics, as a way to influence fundamental properties of matter, e.g., the quantization rule of the quantum Hall effect~\cite{rokaj2022polaritonic,appugliese2022breakdown}. Eventually, we note that what we call atoms and molecules and their interactions is always defined with respect to a given \textit{photonic environment}, and this is exactly what we want to engineer in order to understand, control and develop polaritonic chemistry.

	\begin{acknowledgement}
		
		We acknowledge enlightening discussions with Tal Schwartz, Abraham Nitzan, Sharly Fleischer, Thomas W. Ebbesen, Christian Sch\"afer, Simone Latini, Johannes Flick, Enrico Ronca, Henrik Koch, Fabijan Pavosevic and Markus Kowalewski. This work was made possible through the support of the
		RouTe Project (13N14839), financed by the Federal Ministry of Education and Research (Bundesministerium für Bildung und Forschung (BMBF)) and supported by the European Research Council (ERC-2015-AdG694097), the Cluster of Excellence “CUI: Advanced Imaging of Matter” of the
		Deutsche Forschungsgemeinschaft (DFG), EXC 2056, project
		ID 390715994 and the Grupos Consolidados (IT1453-22).
		The Flatiron Institute is a division of the Simons Foundation
		
	\end{acknowledgement}
	
	\appendix
	
	\section{Necessity of longitudinal dipole self-energy term}
	\label{sec:dipoleselfenergy}
	
	Let us for simplicity and without restriction of generality treat the nuclei/ions clamped and consider (as commonly done) only the electronic interaction between a plasmonic and a molecular system of interest. In this case the longitudinal modes due to a plasmon-molecule interaction are all mediated in Coulomb gauge via 
	\begin{align}
		\hat{W}_{\rm ee} = \frac{1}{2}\sum_{l \neq m}^{N_e} \tfrac{e^2}{4 \pi \epsilon_0 |\br_l-\br_m|}, 
	\end{align}
	as can immediately be inferred from the Pauli-Fierz Hamiltonian of Eq.~\eqref{eq:paulifierzhamiltonian}. It is obvious that for all possible square-integrable wave functions $\Psi$ in the domain of the operator~\cite{blanchard2015mathematical,spohn2004dynamics}, irrespective of statistics and indistinguishability, we have
	\begin{align}
		\brakett{\Psi}{\hat{W}_{\rm ee}}{\Psi} > 0.
	\end{align}
	That is, $\hat{W}_{\rm ee}$ is a positive operator. Assume now that we choose instead of $\hat{W}_{\rm ee}$ an operator of the form~\cite{de2018breakdown,fregoni2021strong}
	\begin{align}
		\hat{W}_{\rm dip} = \frac{1}{2}\sum_{l \neq m}^{N_e} \sum_{\mu=1}^{3}g_l^\mu r_l^{\mu} r_m^{\mu} g_m^{\mu},
	\end{align}
	where $r_m^{\mu}$ are the different Euclidean coordinates of particle $m$ and $g_{m}^{\mu}$ are some arbitrary real constants such that we even allow to break the symmetry of the original Coulomb operator. Obviously this operator is not positive and it can be made arbitrarily negative, i.e., $\hat{W}_{\rm dip}$ is not bounded from below and we can find $\Psi$ such that
	\begin{align}
		\brakett{\Psi}{\hat{W}_{\rm dip}}{\Psi} \rightarrow - \infty.
	\end{align} 
	Not surprisingly, with similar arguments as for the transverse case~\cite{rokaj2018light,schaefer2020relevance}, we find that an Hamiltonian with purely dipolar interaction has no eigenstates and hence does not have an equilibrium solution.
	
	The reason for this unphysical behavior is quite obvious since we have broken the very basic condition that the longitudinal electromagnetic energy is positive. As discussed in Sec~\ref{subsec:approximatepaulifierz}, we have two options that amount to the same physics:  We either keep also quadratic contributions of the form $\sum_{\mu=1}^{3}(g^\mu r^{\mu})^2$ that counter the purely linear coupling, or we restrict to a finite area (finite simulation box with chosen boundary conditions). In both cases the approximate interaction becomes bounded from below and can be made manifestly positive by a finite energy shift. We have thus shown, in yet a different way, that quadratic/counter terms are necessary to have a stable quantum theory and claims to the contrary in the literature arise from a basic misunderstanding of the difference between perturbative/few-level calculations and solving a (necessarily infinite-dimensional~\cite{blanchard2015mathematical}) Schr\"odinger-type quantum equation~\cite{rokaj2018light,schaefer2020relevance}. Let us finally note that in the long-wavelength or dipole limit we can, strictly speaking, no longer distinguish between transverse and longitudinal modes~\cite{jestadt2019light}. Only the physical context of the dipolar approximation provides us with the information which type of field we consider.
	
	
	\section{QEDFT mappings and the effective fields}
	\label{sec:mappings}
	
	Let us first note that QEDFT and its basic mapping theorems are distinct from using (time-dependent) density functional theory and approximately taking into account the interaction with a quantized light field. Indeed, QEDFT is the exact reformulation of the Pauli-Fierz quantum field theory in terms of current densities and vector potentials~\cite{ruggenthaler2014quantum,ruggenthaler2015ground,jestadt2019light}, in analogy to electronic density functional theory being the exact reformulation of the static Schr\"odinger theory with scalar external potentials~\cite{dreizler2012density,burke2012perspective}. Further we note that the term "QEDFT" is used synonymously for the ground-state, the real-time or linear-response time-dependent, minimal-coupling or dipole-coupling density-functional reformulation of the (generalized) Pauli-Fierz field theory~\cite{ruggenthaler2014quantum,ruggenthaler2015ground,jestadt2019light, flick2019light,welakuh2022frequency, flick2018nuclei}. The context unambiguously identifies which specific realization/implementation of QEDFT is meant/used. We use the same convention when referring to "density functional theory"~\cite{ruggenthaler2015existence,tchenkoue2019force}  
	
	To highlight these details let us compare the basic mappings of density functional theory and QEDFT. For simplicity we assume clamped nuclei to connect to the standard case of density functional theories and avoid issues due to localizing finite systems in free space~\cite{kreibich2008multi,sutcliffe2012molecules}. In this case we have either due to the Hohenberg-Kohn~\cite{dreizler2012density}, the Runge-Gross~\cite{ullrich2011tddft} (where we for simplicity assume the ground state as initial state~\cite{maitra2002memory}) or van Leeuwen Laplace-transform~\cite{van2001key} theorems the mappings (note the notation convention from Eq.~\eqref{eq:wfctelectronsnucleiphotons})
	\begin{align}
		v(\br t)  \overset{1:1}{ \leftrightarrow}   \Psi(\underline{\br} t) \overset{1:1}{ \leftrightarrow} n(\br t). 
	\end{align}
	For the static case the parameter $t$ (time) is redundant. This mapping shows that instead of the electronic wave function $\Psi(\underline{\br} t)$ we can express everything in terms of $n(\br t)$, since we can perform a functional-variable transformation within electronic quantum mechanics of the form~\cite{dreizler2012density,ullrich2011tddft,ruggenthaler2015existence}
	\begin{align}
		\brakett{\Psi(t)}{\hat{O} }{\Psi(t)} &= \brakett{\Psi([n],t)}{\hat{O}}{ \Psi([n],t)} \nonumber\\
		&= O([n],t) 
	\end{align}
	for any observable $\hat{O}$. Here the notation $O[n]$ means that the object $O$ is uniquely determined by $n$. We have thus replaced the usual quadratic-form structure of quantum physics in terms of wave functions by \textit{exact} non-linear functionals in terms of the density. But this reformulation has so far no practical relevance, since we do not know how to determine from a given $v(\br t)$ the corresponding $n(\br t)$ without going through the wave function $\Psi(\underline{\br} t)$ first. In the ground-state case the obvious way would be reformulate the minimzation over all wave functions in terms of densities~\cite{dreizler2012density, burke2012perspective}. Yet it is very hard to express the various contributions in terms of the density only. Alternatively one can consider the local ground-state force equilibrium~\cite{tokatly2005manybodydft,ruggenthaler2015existence}. In practice (also for the time-dependent case) one tries to approximate the mapping $v(\br  t)  \overset{1:1}{ \leftrightarrow}  n(\br  t)$ for interacting electrons with the help of an auxiliary mapping that is physically close yet numerically still tractable. The standard choice is to consider the mapping of non-interacting electrons
	\begin{align}
		v_{\rm s}(\br  t)  \overset{1:1}{ \leftrightarrow}   \Phi(\underline{\br} t) \overset{1:1}{ \leftrightarrow} n_{\rm s}(\br  t), 
	\end{align}
	where $\Phi(\underline{\br} t)$ is then (usually) a Slater determinant of single-particle orbitals $\varphi_{k}(\br\sigma t)$, $n_{\rm s}(\br t) = \sum_{k} \sum_{\sigma} |\varphi_k(\br \sigma t)|^2$ and the sub-index "s" refers to "single particle" indicating that the non-interacting many-body Schr\"odinger equation can be recast as single-particle Schr\"odinger equations~\cite{ruggenthaler2015existence}. Assuming now that interacting and non-interacting systems generate the same set of densities, we can thus combine both maps and find~\cite{ruggenthaler2015existence}
	\begin{align}
		v(\br  t)  \overset{1:1}{ \leftrightarrow} v_{\rm s}(\br t). 
	\end{align}
	Thus we have mapped the interacting problem to a non-interacting problem that generates the same density. This new effective potential is then called the Kohn-Sham potential and is denoted as $v_{\rm KS}([v],\br t) = v_{\rm s}([v], \br t)$. Still we did not gain anything, because the mapping from interacting to non-interacting potentials is even harder to approximate. The final step is to once again use the bijectivity between densities and potentials to re-express the Kohn-Sham potential as~\cite{ruggenthaler2015existence}
	\begin{align}
		v_{\rm KS}([v], \br  t) &= v(\br t) + v_{\rm KS}([v],\br t) - v(\br,t) \nonumber \\
		&= v(\br t) + \underbrace{v_{\rm s}([n],\br t) - v([n],\br t)}_{v_{\rm Hxc}([n],\br  t)}. 
	\end{align}
	This turns the linear single-particle Schr\"odinger equations in terms of $v_{\rm KS}([v], \br  t)$ into the well-known non-linear Kohn-Sham single-particle equations in terms of the Hartree-exchange-correlation potential $v_{\rm Hxc}([n],\br  t)$. To approximate the difference between interacting and non-interacting maps in terms of densities, many different highly successful strategies exist.

	In the case of QEDFT we have instead of the Schr\"odinger Hamiltonian the Pauli-Fierz Hamiltonian that does not only dependent on the electronic degrees but also on the (continuum of) photonic degrees of freedom. Since there are some subtle differences in the minimal-coupling QEDFT case between the ground-state and the time-dependent case with respect to the suitable functional variables (although these differences allow to cure old issues of ground-state current-density functional theory)~\cite{ruggenthaler2015ground}, we in the following restrict for simplicity to the dipole-coupled Pauli-Fierz Hamiltonian of Eq.~\eqref{eq:paulifierzlengthgauge} with the external fields of the form of Eqs.~\eqref{eq:externalscalar} and \eqref{eq:externalcurrent}. In this case we have with the identification of $v(\br t) = -|e| \phi_{\rm ext}(\br t)$, the interacting mapping in analogy to the density-functional case~\cite{tokatly2013time,ruggenthaler2014quantum,ruggenthaler2015ground,flick2019light}
	\begin{align}
		\left(v(\br  t),\underline{j}(t)\right)  \overset{1:1}{ \leftrightarrow}   \Psi(\underline{\br},\underline{q},t) \overset{1:1}{ \leftrightarrow} \left(n(\br t), \underline{q}(t) \right), 
	\end{align}
	where $\underline{j}(t)$ and $\underline{q}(t)$ indicates that we have $M_{\rm p}$-long vectors of mode-resolved external currents and displacement coordinates, respectively. The reason why we have now a pair of functional variables $\left(n(\br t), \underline{q}(t)\right)$ is that we can change not only $v(\br  t)$ to consider different physical situations but also adapt $\underline{j}(t)$ to influence the full system of light and matter. Obviously, even if the electronic Schr\"odinger equation and the dipole-approximated Pauli-Fierz Hamiltonian have the same external fields, the expectation values of the same operators give different answers in general and we have access in QEDFT to all photonic observables since we have re-expressed everything in terms of
	\begin{align}
		\brakett{\Psi(t)}{\hat{O}}{ \Psi(t)} &= \braket{\Psi([n,\underline{q}],t)}{\hat{O}}{ \Psi([n,\underline{q}],t)} \nonumber
		\\
		&= O([n, \underline{q}],t). 
	\end{align}
	In analogy we then find adapted effective fields
	\begin{align}
		\left(v(\br  t),\underline{j}(t)\right) \overset{1:1}{ \leftrightarrow} \left(v_{\rm s}(\br  t),\underline{j}_{\rm s}(t)\right),
	\end{align}
	if we choose as auxiliary system non-interacting electrons and photons to generate the same density and displacement field~\cite{tokatly2013time,ruggenthaler2014quantum,ruggenthaler2015ground}. Accordingly we find with the definition of the Maxwell-Kohn-Sham fields $v_{\rm MKS}([v,\underline{j}],\br  t) = v_{\rm s}([v,\underline{j}],\br  t)$ and $\underline{j}_{\rm MKS}([v,\underline{j}],t)=\underline{j}_{\rm s}([v,\underline{j}],t)$ that we have
	\begin{align}
		&v_{\rm MKS}([v,\underline{j}],\br  t) = v(\br  t) + v_{\rm Mxc}([n,\underline{q}],\br  t),\\
		&\underline{j}_{\rm MKS}([v,\underline{j}],t)= \underline{j}(t) + \underline{j}_{\rm Mxc}([n,\underline{q}],t).
	\end{align}
	Here we have denoted the non-linear terms as mean-field-exchange-correlation (Mxc) fields in order to highlight that besides the Hartree term there are now also other mean-field contributions that can be made explicit in the effective fields. Similarly to the electronic density-functional case there are various ways of approximating the difference of these mappings~\cite{pellegrini2015optimized, flick2018ab,schafer2021making,flick2022simple} (note that in the dipole case the Mxc current is exactly the mean-field current, i.e., $\underline{j}_{\rm Mxc}([n,\underline{q}],t) = \underline{j}_{\rm M}([n,\underline{q}],t)$~\cite{tokatly2013time,ruggenthaler2014quantum,ruggenthaler2015ground}), and we can distinguish the contribution due to the Coulomb interaction and due to the direct electron-photon interaction as~\cite{tokatly2013time, ruggenthaler2014quantum,ruggenthaler2015ground,jestadt2019light}
	\begin{align}
		v_{\rm Mxc}([n,\underline{q}],\br  t) = v_{\rm Hxc}([n,\underline{q}],\br  t) + v_{\rm pxc}([n,\underline{q}],\br  t).
	\end{align}
	While in principle $v_{\rm Hxc}([n,\underline{q}],\br  t) \neq v_{\rm Hxc}([n],\br  t)$, in practice one usually employs approximations to $v_{\rm Hxc}([n],\br  t)$ from electronic density functional theories also in QEDFT.
	
	We finally note that one is not restricted to using non-interacting electrons and photons as auxiliary system. Similar to electronic density functional theory, where non-local (generalized) Kohn-Sham systems~\cite{seidl1996generalized,koerzdoerfer2010single,Baer2018time} or strictly-correlated electrons~\cite{seidl2007strict,gori2009density,buttazzo2012optimal} are sometimes used, in QEDFT one can, for instance, use polaritonic orbitals as reference~\cite{nielsen2018dressed,buchholz2020light}. The main drawback of using polaritonic orbitals is that we have mixed statistics that need to be handled with extra care to not generate unphysical results~\cite{nielsen2018dressed,buchholz2019reduced,buchholz2020light}.

	\bibliography{references}
	
\end{document}